\documentclass[a4paper,11pt]{article}

\usepackage[utf8]{inputenc}
\usepackage[T1]{fontenc}
\usepackage[margin=1in]{geometry}
\usepackage{url}
\usepackage{booktabs}
\usepackage{threeparttable}
\usepackage{multirow}
\usepackage{mathtools}
\usepackage{amsfonts}
\usepackage{amssymb}
\usepackage{amsthm}
\usepackage{bm}
\usepackage{graphicx}
\usepackage{algorithm}
\usepackage[noend]{algpseudocode}
\usepackage[short]{optidef}
\usepackage[dvipsnames]{xcolor}
\usepackage{soul}
\usepackage{enumitem}
\usepackage[small, margin=1cm]{caption}
\usepackage{subcaption}
\usepackage{pgfplots}
\pgfplotsset{compat=1.18}
\usepgfplotslibrary{groupplots}
\usepackage{microtype}
\usepackage{natbib}

\usepackage[colorlinks=true,linkcolor=blue,citecolor=blue,urlcolor=blue]{hyperref}
\usepackage[capitalize]{cleveref}
\makeatletter
\providecommand{\theHALG@line}{}
\renewcommand{\theHALG@line}{\thealgorithm.\arabic{ALG@line}}
\makeatother

\definecolor{LinkBlue}{HTML}{1F4E79}
\definecolor{CiteBlue}{HTML}{2F6FA3}
\definecolor{URLBrick}{HTML}{8A3B12}
\hypersetup{
    colorlinks=true,
    linkcolor=LinkBlue,
    citecolor=CiteBlue,
    urlcolor=URLBrick,
    filecolor=LinkBlue,
    pdfborder={0 0 0},
    breaklinks=true,
    bookmarksopen=true,
    bookmarksnumbered=true
}

\makeatletter
\let\@maketablecaption\@makecaption
\makeatother

\newcommand{\arxivtablefit}[2]{#1\setlength{\tabcolsep}{#2}}

\newcommand{\floor}[1]{ \left\lfloor #1 \right\rfloor }
\newcommand{\greedy}{\textsc{Greedy}}
\newcommand{\vargreedy}{\textsc{OrderedGreedy}}

\providecommand{\N}{\mathbb{N}}
\providecommand{\threepart}{\textsc{3-Partition}}
\providecommand{\nonoverlap}{\textsc{NonOverlap}}
\providecommand{\overlap}{\textsc{Overlap}}
\providecommand{\singletest}{\textsc{SingleTest}}
\providecommand{\ksum}{$k$-\textsc{Sum}}
\providecommand{\rksum}{\textsc{Relaxed}-$k$-\textsc{Sum}}
\providecommand{\countis}{\#\textsc{Independent-Set}}
\providecommand{\thresholdis}{\textsc{Threshold-Independent-Set}}
\DeclareMathOperator{\opt}{\text{OPT}}
\DeclareMathOperator{\gain}{\mathcal{G}}
\providecommand{\argmax}{\operatorname*{arg\,max}}

\definecolor{RedactionGray}{gray}{0.55}
\sethlcolor{RedactionGray}
\providecommand{\arxivtablefit}[2]{}
\newcommand{\redacted}[1]{{\color{white}\hl{redacted for anonymity}}}
\renewcommand{\redacted}[1]{#1}

\usepackage{multicol}
\usepackage{wasysym}%
\usepackage{enumitem}%
\usepackage{color}%
\usepackage{forloop}%
\usepackage{stmaryrd}

\newcommand{\QO}{$\Box$}%

\newcounter{qr}

\newcommand{\Qline}[1]{\noindent\rule{#1}{0.6pt}}

\newcounter{ql}

\newlength{\qt}

\newcounter{itemnummer}
\newcommand{\Qitem}[2][]{%
\ifthenelse{\equal{#1}{}}{\stepcounter{itemnummer}}{}
\ifthenelse{\equal{#1}{a}}{\stepcounter{itemnummer}}{}
\begin{enumerate}[topsep=2pt,leftmargin=2.8em]
\item[\textbf{\arabic{itemnummer}#1.}] #2
\end{enumerate}
}

\definecolor{bgodd}{rgb}{0.8,0.8,0.8}
\definecolor{bgeven}{rgb}{0.9,0.9,0.9}
\newcounter{itemoddeven}
\newlength{\gb}
\newcommand{\QItem}[2][]{%
\setlength{\gb}{\linewidth}
\addtolength{\gb}{-5.25pt}
\ifthenelse{\equal{\value{itemoddeven}}{0}}{%
\noindent\colorbox{bgeven}{\hskip-3pt\begin{minipage}{\gb}\Qitem[#1]{#2}\end{minipage}}%
\stepcounter{itemoddeven}%
}{%
\noindent\colorbox{bgodd}{\hskip-3pt\begin{minipage}{\gb}\Qitem[#1]{#2}\end{minipage}}%
\setcounter{itemoddeven}{0}%
}
}

\newcommand{\PaperTitle}{Welfare-Maximizing Pooled Testing}

\newcommand{\PaperAbstract}{%
Pooled testing increases the reach of scarce diagnostic resources, but optimally composing pools for individuals differing in infection risk and the utility they derive from a negative test is combinatorially challenging. We study the problem of maximizing the expected welfare of individuals cleared by a negative result, given a testing budget. Assigning a sample to several pools can raise welfare but is operationally burdensome; we show the restriction to non-overlapping allocations costs at most a factor of two for any budget or population, less under a pool-size cap at high health probabilities, and nothing when no health probability exceeds one-half. Welfare decomposes across non-overlapping pools, whereas evaluating overlapping allocations is \#P-hard for pools of three or more. Finding optimal allocations is NP-hard and admits no FPTAS, with or without overlap, unless P = NP. We provide single-test routines and greedy algorithms with constant-factor guarantees. On real-world data, greedy achieves over 99\% of optimal non-overlapping welfare in milliseconds, against hours for exact benchmarks. In a randomized field experiment at a Mexican research institute, our mechanism conditioned campus access on negative qPCR results. Relative to unrestricted access, we found no statistical evidence of adverse effects on participants’ performance, learning, or mental health.
}

\newcommand{\PaperKeywords}{pooled testing, welfare maximization, approximation algorithms}

\newtheorem{theorem}{Theorem}
\newtheorem{lemma}{Lemma}
\newtheorem{proposition}{Proposition}
\newtheorem{corollary}{Corollary}

\theoremstyle{definition}

\newtheorem{example}{Example}
\newtheorem{problem}{Problem}
\newtheorem{definition}{Definition}

\newtheorem{observation}{Observation}

\allowdisplaybreaks

\title{\PaperTitle}
\author{%
Simon Finster\thanks{Johannes Kepler University Linz, \href{mailto:simon.finster@jku.at}{simon.finster@jku.at}}
\and
Michelle González Amador\thanks{Wageningen University \& Research, \href{mailto:michelle.gonzalezamador@wur.nl}{michelle.gonzalezamador@wur.nl}}
\and
Edwin Lock\thanks{King's College London, \href{mailto:edwin.lock@kcl.ac.uk}{edwin.lock@kcl.ac.uk}}
\and
Francisco Marmolejo-Cossío\thanks{Boston College, \href{mailto:marmolf@bc.edu}{marmolf@bc.edu}}
\and
Evi Micha\thanks{University of Southern California, \href{mailto:evi.micha@usc.edu}{evi.micha@usc.edu}}
\and
Ariel D. Procaccia\thanks{Harvard University, \href{mailto:arielpro@seas.harvard.edu}{arielpro@seas.harvard.edu}}
}
\date{}

\begin{document}

\maketitle

\begin{abstract}
\PaperAbstract
\end{abstract}


\newpage

\section{Introduction}
\label{section:intro}

Diagnostic scarcity is not only a public-health constraint; it is an allocation problem. When an institution has fewer reliable tests than people who need testing, a negative test result becomes more than medical information; it becomes a scarce form of permission: to enter a laboratory, staff a workplace, attend class, use specialized equipment, or provide essential services.  Large-scale testing was central to containment because it enabled targeted isolation rather than blanket shutdowns, yet shortages of laboratory equipment, consumables, and trained personnel made individual testing infeasible at scale, especially in low- and middle-income countries~\citep{kavanagh2020access, dhabaan2020challenges, abera2020establishment}. At the same time, the economic and social costs of broad lockdowns made selective, testing-based access a key instrument for balancing public health with productive activity~\citep{deb2022economic, camera2021economic}. The same access-allocation problem can also arise beyond pandemic reopening, for example when hospitals, schools, universities, shelters, or other congregate institutions face localized outbreaks under finite testing capacity~\citep{de2021containing,national2020reopening}. Pooled testing is the canonical way to expand diagnostic reach under such constraints~\citep{dorfman1943detection}: samples from several individuals are combined and processed as a single test; a negative result clears the entire pool, while a positive result indicates that at least one member is infected and, in the one-stage clearance regime we study, clears no one in that pool.\footnote{Pooled tests can also address privacy concerns, as pooling provides a degree of anonymity to test takers.}

Classical group testing asks how to identify infected individuals using as few tests as possible. Our social planner has a different operational objective: given a fixed testing budget, how should tests be allocated to maximize the expected welfare generated by individuals who can safely be cleared? Prior work has made progress on related allocation problems, including welfare-maximizing pooled testing for populations that differ in infection risk~\citep{Lipnowski2021pooled, Bobkova2023}, group testing with imperfect multi-stage tests~\citep{aprahamian2019optimal}, and data-driven allocation of individual tests~\citep{bastani2021efficient}. A crucial dimension missing from these analyses is that individuals differ not only in infection risk but also in the utility they stand to gain from a negative test result. A pool's expected welfare depends jointly on how likely the pool is to test negative and how valuable it is to clear its members.

This paper gives a welfare and computational analysis of pooled testing under diagnostic scarcity. We first quantify the welfare cost of restricting from overlapping to non-overlapping allocations, which are operationally simpler: the welfare ratio is 
at least $\frac{19}{16}$ with three tests, and never exceeds a factor of $2$ for any testing budget or population, with sharper guarantees in regimes with low or high health probabilities. We then map the computational limits, showing that welfare evaluation is easy for non-overlapping allocations but \#P-hard for overlapping allocations when pool size is at least three, and that finding optimal allocations is NP-hard even in the non-overlapping regime. We develop exact and approximate methods, including single-test optimization routines, greedy algorithms with provable constant-factor guarantees, and a mixed-integer linear programming benchmark for the non-overlapping problem. Finally, we validate the framework in a pilot study at a Mexican research institute, where the mechanism grants scarce in-person access with no evidence of measurable adverse effects on performance, productivity, learning, stress, or subjective well-being.

Our point of departure is the problem of maximizing aggregate welfare when individuals differ in both infection risk and the utility they derive from being cleared. For concreteness, consider a social planner---such as a testing administrator, public-health authority, or institutional decision maker---who wants to provide safe in-person access under a limited testing budget. Each individual has a probability of being healthy and a utility from being cleared for in-person activity, and individuals may differ substantially in utility even within the same institution: a laboratory researcher whose experiments require specialized equipment, a facilities worker whose tasks must be performed on campus, and a doctoral student whose progress depends on in-person supervision and laboratory access may all stand to gain differently from being cleared. The planner chooses pools subject to a budget and a pool-size constraint. An individual receives utility if they appear in at least one pool that tests negative; individuals who are not cleared remain isolated, with utility normalized to~zero.%
\footnote{Unlike in some pooled-testing regimes, we do not perform subsequent individual or multi-stage adaptive testing following a positive pooled result, as this would be prohibitively expensive in a strongly resource-constrained setting.}

A central operational choice is whether pools may overlap. Allowing overlap can increase welfare because the same individual can appear in several tests, and a negative result in any one of those tests can clear them. Overlap is therefore especially tempting when some individuals are highly likely to be healthy, highly valuable to clear, or both. But overlap is operationally burdensome: tracking individual samples across multiple pools creates dependencies across test outcomes and is difficult to implement at scale with limited laboratory personnel and workflow infrastructure~\citep{cleary2021using}. Non-overlapping allocations, in which each individual is assigned to at most one pool, are far simpler: each person submits one sample, each negative result has a transparent interpretation, and welfare decomposes additively across pools. The natural question is therefore not merely whether overlap can help, but how much welfare is lost by insisting on the operational simplicity of non-overlap. The social planner's problem is economic as well as computational, and we address both dimensions.

\subsection{Contributions}
Our contributions span four areas: we characterize the welfare cost of restricting to non-overlapping tests, establish the computational complexity of welfare-maximizing pooled testing, develop efficient exact and approximate algorithms, and validate our framework through numerical experiments and a field pilot.

First, we quantify the welfare loss from restricting attention to non-overlapping tests. We define the overlap welfare ratio $\gain(B,J)=\opt^*(B,J)/\opt(B,J)$, comparing the optimal overlapping allocation with the optimal non-overlapping allocation for a fixed population $J$ and testing budget $B$ (with the convention $\gain(B,J)=1$ in the degenerate case $\opt(B,J)=0$, where no allocation achieves positive welfare).
For three tests, we provide an example that establishes a lower bound of $\frac{19}{16}$. Moreover, we establish a universal guarantee: for every testing budget $B$ and every population $J$, $\gain(B,J)\leq 2$. We then sharpen this worst-case result in two important regimes. If every individual satisfies $q_i \leq \frac{1}{2}$, overlap provides no welfare gain. Conversely, when tested individuals have high health probabilities and pool sizes are capped, we derive explicit bounds that can be substantially smaller than two.

Second, we establish the computational complexity of welfare-maximizing pooled testing. Evaluating a non-overlapping allocation is straightforward because welfare decomposes across pools. In contrast, evaluating a given overlapping allocation is \#P-hard for pool sizes $G\geq 3$, and the corresponding threshold problem is PP-complete. For optimization, we show that the two problems of computing an optimal non-overlapping and overlapping allocation are both NP-hard for any fixed $G  \geq 3$. In both regimes, hardness also extends to inapproximability: neither problem admits an FPTAS unless $\mathrm{P}=\mathrm{NP}$, in contrast to the single-test problem. The hardness result for non-overlapping testing is tight in the pool-size parameter: for $G \leq 2$, the non-overlapping allocation problem is polynomial-time solvable. We also show that the single-test problem is W[1]-hard with respect to the pool-size bound~$G$, ruling out fixed-parameter tractability under standard assumptions.

Third, we develop exact and approximate solution methods. For a single test, we design an FPTAS and provide a mixed-integer conic formulation. Building on these single-test methods, we design greedy algorithms with complementary guarantees. The basic \greedy{} algorithm, using an exact single-test subroutine, achieves at least a $\frac{1}{e+1}$ fraction of the optimal overlapping welfare. A modified version retains a $\frac{1}{e+1+\delta}$ guarantee with an approximate single-test subroutine in time $O(2^G\cdot \mathrm{poly}(n,B,1/\delta))$. A fully polynomial-time version achieves at least a $\frac{1}{5+\delta}$ fraction of the optimal non-overlapping welfare. We also formulate a mixed-integer linear program for the non-overlapping regime whose accuracy can be traded off against computation time, allowing it to serve as a near-exact benchmark in the numerical experiments.

Fourth, we implement the framework and evaluate it on data from a field pilot at \redacted{the Potosinian Institute of Scientific and Technological Research (IPICYT)}, a research institute in Mexico. The pilot supplies real institutional estimates of utilities and infection probabilities, which we use in our computational experiments. Across these experiments, the greedy algorithm performs extremely close to the best non-overlapping benchmark, achieving at least $99.37\%$ of optimal non-overlapping welfare in the pilot's setting. The pilot was designed as a randomized controlled trial, and the empirical results provide no evidence that the testing mechanism worsened participants' productivity, performance, learning, stress, or subjective well-being relative to unrestricted campus access.

\subsection{Application}

Our framework is motivated by, and implemented at, \redacted{the Potosinian Institute of Scientific and Technological Research (IPICYT)}, a higher education research institution in Mexico. By 2021, the institute had shifted entirely to remote work, and the administration sought a safe, resource-efficient way to restore in-person access while easing the economic and social costs of continued isolation. In September 2022, 131 individuals consented, completed the baseline survey, and entered the pilot study. Of these, 130 had the health-probability and utility data required for the testing allocation, and 122 completed the endline survey. The treatment group was invited for qPCR testing based on pooled-testing allocations computed by our mechanism and was allowed access to campus only after a negative qPCR test result. We developed a survey protocol to estimate individual utilities and worked with epidemiologists at \redacted{IPICYT} and in the state of \redacted{San Luis Potosí} to estimate infection probabilities for each member of the population; see \cref{section:population-data} for details.

The pilot study served a dual purpose: the real-world population data directly informed our numerical experiments, and the deployment validated the operational feasibility of our algorithms and supporting web platform in a live institutional setting.\footnote{The web app code is open source at \redacted{\url{https://github.com/edwinlock/csef.git}} and a demo is available at \url{https://demo.c-sef.com}.} Our randomized controlled trial finds no evidence that participants' productivity, performance, learning, stress, or subjective well-being were negatively affected relative to a counterfactual of unrestricted campus access. At the same time, the protocol restricts in-person access to individuals cleared by qPCR testing, safeguarding population health at a fraction of the cost of individual testing.\footnote{At the time of reopening, \redacted{San Luis Potos\'i} had 221,870 cumulative COVID-19 cases, of which 615 were active. KN95 masks were mandatory for everyone returning to \redacted{IPICYT}.}

\subsection{A motivating example}
The welfare gains from pooled testing depend not only on who is tested, but on how the pools are structured. Although restricting to non-overlapping tests may sacrifice some welfare, the following example shows that this tradeoff arises even in the simplest cases.

\begin{example}
\label{example:four-individuals-intro}
Suppose we have a testing budget of $B=3$, and consider a population of four individuals $\{1,2,3,4\}$ with health probabilities $q_1=q_2=q_3=\frac12$, $q_4=1$, and utilities $u_1=u_2=u_3=u_4=1$. In any non-overlapping allocation, individual $4$ can appear in at most one test. If individual $4$ is tested alone, then at most two of individuals $1,2,3$ can be tested separately, so the welfare is at most $1+\frac12+\frac12=2$. If instead individual $4$ is pooled with one of $\{1,2,3\}$, that test contributes $2\cdot \frac12=1$, and the remaining two risky individuals can contribute at most $\frac12$ each, again giving total welfare at most $2$. Thus the optimal non-overlapping welfare is $2$.

By contrast, the overlapping allocation $T^*=(\{1,4\},\{2,4\},\{3,4\})$ tests individual $4$ three times. Each of individuals $1,2,3$ is cleared with probability $\frac12$, while individual $4$ is cleared unless all three risky individuals are infected, which happens with probability $\left(\frac12\right)^3$. Hence $u(T^*) = 3\cdot \frac12 + \left(1-\left(\frac12\right)^3\right)=\frac{19}{8}$. Therefore this population achieves overlap welfare ratio $\frac{19/8}{2}=\frac{19}{16}$.
\end{example}

The gain arises because individual $4$, who is always healthy, can be included in all three pools without reducing the probability that any pool tests negative. This example shows that allowing overlaps can already raise welfare to $\frac{19}{16}$ of the non-overlapping optimum when $B=3$. In \cref{section:gain-of-overlaps}, we complement it with 
the upper bound $\gain(3) \leq 2$, and regime-dependent guarantees that show when the welfare loss from non-overlap is small.

\subsection{Related work}

Pooled testing dates back to the seminal work of \citet{dorfman1943detection}, and has since become a mature field. Much of this literature aims to ascertain the infection status of \emph{all} individuals in a population with as few tests as possible.\footnote{In practice, many resource-constrained populations face testing budgets far below the information-theoretic lower bounds needed to ascertain the full health profile of the population. Moreover, complex pooled-testing regimes can be difficult to implement at scale with limited laboratory personnel and workflow infrastructure~\citep{cleary2021using}.} It has been applied to a range of diseases, especially HIV/AIDS~\citep{tu1995informativeness,wein1996pooled,emmanuel1988pooling}, and became particularly prominent during COVID-19 once qPCR-based pooling was shown to be practical~\citep{Sanghani2021,mutesa2021pooled,nalbantoglu2020group}.

The most closely related literature studies welfare-maximizing or resource-constrained test allocation in heterogeneous populations. \citet{Lipnowski2021pooled} characterize optimal non-overlapping policies in a continuous-population model and show that optimal allocations exhibit assortative batching, with quarantine decisions allowed to depend on posterior infection probabilities (whereas in our setting only negatively tested individuals are released). \citet{Bobkova2023} study a continuous population partitioned into low- and high-risk groups in which each individual can be tested at most twice, permitting limited overlap. Both papers model heterogeneity in infection risk but not in the utility individuals derive from being cleared, and neither studies the computational problem of finding near-optimal allocations in discrete populations under explicit pool-size and budget constraints. Our paper adds utility heterogeneity, compares overlapping and non-overlapping regimes, and develops exact and approximate algorithms for the resulting discrete optimization problem.

In a complementary direction, \citet{bastani2021efficient} develop a reinforcement learning system for the efficient, data-driven allocation of individual COVID-19 tests at borders under budget constraints. \citet{bastani2022interpretable} extend this into an operational pipeline with supply-chain coordination, Bayesian prevalence estimation, delayed-feedback bandits, and interpretable, human-in-the-loop decision support. Like our paper, these works address scarce testing resources and heterogeneous populations, but they focus on individual testing and border surveillance rather than welfare-maximizing pooled testing for in-person access.

Relatedly, \citet{aprahamian2019optimal} study two-stage group testing with imperfect sensitivity and specificity, where the population is first partitioned into disjoint pools and positive pools are subsequently retested individually.\footnote{Performance is measured in terms of false positives, false negatives, and the expected number of tests, and the authors reduce the core optimization problem to a shortest-path formulation.} \citet{ely2021optimal} study budget allocation across tests of different types; \citet{brault2021group} analyze pooled screening with strong penalties for false negatives; \citet{gollier2020group} study prevalence estimation and the identification of healthy individuals; and \citet{Augenblick2020} study how pooling can extend rapid, frequent testing (see also \citealt{larremore2021test}). Pooled testing has been studied using many other mathematical and algorithmic techniques.\footnote{See, e.g.,~\citep{li1962sequential,hwang1972method,hong2022group,cohen2021multi,petersen2020practical,ghosh2020tapestry,ahn2021adaptive,nikolopoulos2020community,cuturi2020noisy,du2000combinatorial,aldridge19}.} Unlike our paper, this literature typically does not model utility heterogeneity or the welfare-maximizing allocation of pooled tests for selective in-person access. An exception is \citet{GOLDBERG2020}, which, though not framed as pooled testing, can be interpreted as studying the optimal allocation of a single pooled test in a heterogeneous population; their FPTAS for the single-test problem is a key ingredient in our algorithmic development.

\subsection{Organization}
The remainder of the paper is structured as follows. \Cref{section:model} describes our population and testing model. \Cref{section:gain-of-overlaps} quantifies the welfare loss from non-overlapping testing through examples, a universal factor-two bound, and sharper guarantees for low and high health probabilities. \Cref{section:complexity-allocations} establishes the complexity of evaluating and computing test allocations, while \cref{section:finding-allocations} develops the FPTAS, mixed-integer conic formulation, \greedy{} algorithms, and MILP formulation. \Cref{section:algorithms-in-practice} presents the numerical experiments and evidence from the pilot study at \redacted{IPICYT}, and \cref{section:discussion} concludes. \Cref{section:data-code-availability} describes the data, code, and other replication materials. The Appendix contains the full proofs and supplementary theoretical, numerical, and empirical material.

\section{Model}
\label{section:model}

Fix a population size $n$, a testing budget $B \in \mathbb{N}$, and a pool-size bound $G$. Let $[n] \coloneqq \{1,\ldots,n\}$ denote the individuals in the population. A \textit{population} $J$ is a tuple $(q_1,\ldots,q_n,u_1,\ldots,u_n)$ that assigns each individual $i \in [n]$ an independent probability $q_i$ of being healthy and a utility $u_i \geq 0$ representing the gain from being cleared for in-person activity.%
\footnote{Utility might reflect people's socio-economic status, the type of occupation, or mental health considerations. See \cref{section:population-data} for details on the utilities in our pilot.} Denote the universe of all populations by $\mathcal{J} \coloneqq [0,1]^n \times \mathbb{R}_+^n$.

A single test is a pool $t \subseteq [n]$ with $|t| \leq G$. We are primarily interested in the case $G<n$.%
\footnote{Pool sizes in pooled tests are typically limited to $G \leq 10$ due to biological constraints. Our partners in Mexico have replicated techniques from \citet{Sanghani2021} to achieve a maximal pool size of 5 using saliva samples.} For any pool $t$, let $q_t = \prod_{i \in t} q_i$ denote the probability that all individuals in $t$ are healthy, equivalently, that test $t$ is negative. A \textit{test allocation} is an ordered collection $T=(t_1,\ldots,t_B)$ of $B$ tests satisfying $|t_j| \leq G$ for all $j \in [B]$. It is \textit{non-overlapping} if $t_j \cap t_k = \emptyset$ for every $j \neq k$. Without loss of generality, we assume $B<n$; otherwise, testing every individual separately is optimal.

For a given test allocation $T$, let $P_i^T$ denote the probability that individual $i$ is included in at least one negative test. The \textit{welfare} of $T$ is the aggregate expected utility of the individuals who are cleared by the allocation, and is given by
\[
u(T)=\sum_{i \in [n]} u_i \cdot P_i^T.
\]
When the population is not fixed, we write $u(T,J)$. For a single pooled test $t$, this reduces to $u(t) \coloneqq u(\{t\}) = q_t \left(\sum_{i \in t} u_i\right)$.%
\footnote{We omit the term ``expected'' for brevity and assume that all welfares and utilities are determined in expectation.} For non-overlapping allocations, test outcomes are independent and the welfare decomposes as $u(T)=\sum_{t \in T} u(t)$. In the overlapping case, by contrast, the probabilities $P_i^T$ depend on correlations across tests and are generally more complicated.

Let $\mathcal{T}^*(B)$ denote the set of all test allocations with budget $B$, and let $\mathcal{T}(B)$ denote the subset of \textit{non-overlapping} allocations. For a fixed population $J$, define
\[
\opt^*(B,J) \coloneqq \max_{T^* \in \mathcal{T}^*(B)} u(T^*,J)
\quad\text{and}\quad
\opt(B,J) \coloneqq \max_{T \in \mathcal{T}(B)} u(T,J).
\]
Any maximizer for $\opt^*(B,J)$ is an optimal overlapping allocation, and any maximizer for $\opt(B,J)$ is an optimal non-overlapping allocation.

We make two modeling assumptions.
\paragraph{Independence of infections.}
In general, infection events in a population may be correlated. Our model is intended for a regime in which the population is otherwise in full lockdown, so workplace interactions do not themselves create additional correlation in infection risk: either individuals stay at home and do not interact, or they interact at the workplace only because they have already been cleared by a negative test.\footnote{While employees may interact outside office hours, here this was not commonly observed during lockdowns.}

\paragraph{Accuracy of tests.} A key assumption in our model is that test results are perfectly accurate. In practice, there is evidence that the clinical specificity of qPCR COVID tests (the complement of the false positive rate) might be lower than 100\%.%
\footnote{See, e.g.,~\citep{Kortela-2021,Binny-2022}.}
In our practical application we use saliva testing with small pool sizes, for which comparably precise specificity estimates are not available. We therefore work with the approximate assumption of 100\% specificity throughout. Extending the model to specificity below 100\% is beyond the scope of this paper: it would require changing the welfare expressions and, consequently, reformulating all of our exact and approximate solution methods. We leave that extension to future work.

\section{Performance of non-overlapping testing}
\label{section:gain-of-overlaps}
\label{SECTION:GAIN-OF-OVERLAPS}

Non-overlapping testing is operationally attractive: each individual submits one sample, each negative result has an unambiguous interpretation, and the welfare of an allocation decomposes additively across tests. The restriction is not obviously innocuous, however, because overlapping tests can reuse particularly valuable or particularly likely healthy individuals across several pools.

This raises the central question of the section: how much welfare can be lost by insisting on non-overlapping tests? We answer this question in three steps. We first give a two-test example, where the welfare loss is a factor of $\frac76$. We then prove the main result of the section: for any budget and any population, the loss from restricting to non-overlapping allocations is at most a factor of $2$. Finally, we record two regimes where this factor-$2$ bound is conservative: no welfare is lost when all individuals in the population have $q_i\leq\frac12$, and high health probabilities yield sharper explicit bounds under a pool-size cap.

\begin{definition}
For a fixed budget $B$ and population $J$ with $\opt(B,J)>0$, define the \emph{overlap welfare ratio} by $\gain(B,J)\coloneqq \frac{\opt^*(B,J)}{\opt(B,J)}$. If $\opt(B,J)=0$, then $u_iq_i=0$ for every individual, so every allocation (overlapping or not) has zero welfare; in this degenerate case we set $\gain(B,J)\coloneqq 1$. Let $\gain(B)\coloneqq \sup_{J}\gain(B,J)$.
\end{definition}

Thus $\gain(B,J)$ measures the welfare advantage of optimal overlapping testing over optimal non-overlapping testing on a given instance, while $\gain(B)$ is the worst-case ratio over all populations.

The smallest nontrivial case already shows why overlaps can help. This example has the same structure as the motivating example in \cref{example:four-individuals-intro}: overlap allows a high-health-probability individual to support multiple lower-health-probability individuals.

\begin{example}
\label{example:three-individuals}
Suppose $B=2$ and consider three individuals $\{1,2,3\}$ with $q_1=q_2=\frac12$, $q_3=1$, and $u_1=u_2=u_3=1$. Up to symmetry, the best non-overlapping allocation achieves welfare $\frac32$, for example with $(\{1,3\},\{2\})$. The overlapping allocation $T^*=(\{1,3\},\{2,3\})$ achieves welfare $u(T^*)=\frac12+\frac12+\left(1-\left(1-\frac12\right)^2\right)=\frac74$. Hence $\gain(2,J)=\frac{7/4}{3/2}=\frac76$.
\end{example}

\subsection{A universal factor-two bound}

\begin{theorem}
\label{theorem:universal-factor-two}
\hyperref[proof:theorem:universal-factor-two]{\normalfont[Proof in \cref{proof:theorem:universal-factor-two}]}
For every testing budget $B$, the overlap welfare gain is $\gain(B)\leq 2$.
\end{theorem}

\paragraph{Proof idea.}
Start with an optimal overlapping allocation $T^* = (t_1^*, \ldots, t_B^*)$ with a minimal number of nonempty tests. This minimality forces a \emph{split-stability} property of $T^*$: no test admits a partition into two parts whose health probabilities sum to less than one. If it did, one of the parts would already do at least as well on its own, contradicting optimality.

We use $T^*$ to build a random non-overlapping allocation $N = (N_1, \ldots, N_B)$ in which each individual is allocated (with some probability) to test $N_j$ only if $t_j$ is one of her incident tests. Specifically, an individual appearing in exactly one test $t_j$ of $T^*$ is kept in $N_j$ with probability $1/2$ and not assigned to any tests otherwise; one appearing in $d_i \geq 2$ tests $t_j$ is allocated uniformly to one of the corresponding $N_j$. These assignments are made independently across individuals. Every realization of $N$ is feasible and non-overlapping, with $N_j \subseteq t_j^*$.

The heart of the proof is a coordinatewise comparison: for every individual $i$,
\[
\mathbb{E}[P_i^N] \geq \frac{1}{2} P_i^{T^*}.
\]
If the individual $i$ appears only in one test of $T^*$, this is straightforward. If individual $i$ appears in two or more tests, we use split-stability to control the conditional clearance of $i$ inside her randomly chosen pool. Summing this inequality against the nonnegative utilities gives $\mathbb{E}[u(N, J)] \geq u(T^*, J) / 2$. Since $\mathbb{E}[u(N, J)] \leq \opt(B, J)$, we conclude $\opt^*(B, J) \leq 2 \opt(B, J)$, implying $\gain(B, J) \leq 2$.

We note that this bound is not known to be tight. \cref{example:four-individuals-intro} in the introduction gives the highest gain from overlaps we know of, which is $\frac{19}{16}$. Closing this gap is an open question.

\subsection{Regime-dependent bounds}

The universal factor-$2$ theorem gives a worst-case guarantee, but there are natural regimes in which overlap is less valuable. The first is the low-$q$ regime: if every individual in the population has health probability at most one half, then overlap cannot improve welfare at all.

\begin{proposition}
\label{proposition:no-overlap-gain-below-half}
\hyperref[proof:proposition:no-overlap-gain-below-half]{\normalfont[Proof in \cref{proof:proposition:no-overlap-gain-below-half}]}
If every individual in the population satisfies $q_i \leq \frac12$, then $\gain(B,J)=1$.
\end{proposition}

The strict case $q_i<\frac12$ is direct: we first show that every optimal test must be a singleton. All optimal allocations are therefore non-overlapping, and $\gain(B,J)=1$. The boundary case $q_i\leq\frac12$ then follows by perturbing each $q_i$ down to $(1-\varepsilon)q_i$ and letting $\varepsilon\downarrow 0$, using that $\opt^*$ and $\opt$ are continuous in the health probabilities.

The opposite regime is also important in applications. Suppose pool sizes are capped at $G$ and every tested individual has health probability at least $\rho$. Then a suitable non-overlapping subtest of each overlapping pool retains a controlled fraction of its welfare. Optimizing over the subtest size gives the following high-$q$ bound.

\begin{theorem}
\label{theorem:high-q-bound}
\hyperref[proof:theorem:high-q-bound]{\normalfont[Proof in \cref{proof:theorem:high-q-bound}]}
Fix a pool-size bound $G$ and $\rho\in(0,1]$, and define
\[
\gamma_G(\rho)\coloneqq
\min_{1\leq s\leq G}\frac{G}{s\rho^{s-1}}.
\]
Suppose every individual in population $J$ satisfies $q_i\geq\rho$. Then $\gain(B, J) \leq \gamma_G(\rho)$ for any $B$.
\end{theorem}

The proof only uses the weaker condition that, in some optimal overlapping allocation $T^*$, every nonempty test has size at most $G$ and every tested individual satisfies $q_i\geq\rho$.

\paragraph{Proof idea.}
The argument mirrors that of \cref{theorem:universal-factor-two}: we construct a random non-overlapping allocation $N$ from an optimal overlapping $T^*$, show that for every individual $i$,
\[
\mathbb{E}[P_i^N] \geq \frac{P_i^{T^*}}{\gamma_G(\rho)},
\]
and conclude that some realization achieves welfare at least $u(T^*, J) / \gamma_G(\rho)$, hence $\gain(B, J) \leq \gamma_G(\rho)$. The construction of $N$ and the per-individual bound differ.

The construction begins by collapsing $T^*$ to its \emph{first-appearance projection} $T = (t_1, \ldots, t_B)$, in which each tested individual is kept only in the earliest test of $T^*$ where it appears. The first-appearance pools $t_j$ are pairwise disjoint, with sizes $m_j \leq G$. For each $j$, choose $s_j \in [m_j]$ minimizing $m_j / (s \rho^{s-1})$, so that this minimum is $\gamma_{m_j}(\rho)$. Then, independently for each $j$, set $N_j$ to a uniformly random $s_j$-subset of $t_j$ with probability $\gamma_{m_j}(\rho) / \gamma_G(\rho)$, and to $\emptyset$ otherwise. As shown in the full proof in the appendix, $\gamma_m(\rho)$ is non-decreasing in $m$, so this probability is at most $1$. It is then straightforward that every realization of $N$ is feasible and non-overlapping.

The coordinatewise bound replaces the split-stability argument of \cref{theorem:universal-factor-two} with a direct calculation that exploits the high-$q$ assumption. Each individual $i$ belongs to a unique first-appearance pool $t_j$, so its only chance of being cleared in $N$ is via $N_j$. The probability $i \in N_j$ is $(\gamma_{m_j}(\rho) / \gamma_G(\rho)) \cdot s_j / m_j$, and conditional on this event the other $s_j - 1$ members of $N_j$ each have health probability at least $\rho$, giving conditional clearance at least $q_i \rho^{s_j - 1}$. Multiplying yields $\mathbb{E}[P_i^N] \geq q_i / \gamma_G(\rho) \geq P_i^{T^*} / \gamma_G(\rho)$, where the last step uses that $i$ must be healthy to be cleared.
\Cref{table:owr-high-q-bounds} turns the high-$q$ formula into representative values. The table reports the best of the high-$q$ estimate and the universal factor-two bound. For example, with pool-size cap $G=5$, the bound is already below $1.25$ when all tested individuals have $q_i\geq 0.95$.

\begin{table}[t]
\centering
\begin{threeparttable}
\caption{High-$q$ upper bounds on the overlap welfare ratio.}
\label{table:owr-high-q-bounds}
{\setlength{\tabcolsep}{4pt}
\begin{tabular}{@{}c c c c c c c c@{}}
\toprule
$G$ & $\rho=0.50$ & $\rho=0.60$ & $\rho=0.70$ & $\rho=0.80$ &
$\rho=0.85$ & $\rho=0.90$ & $\rho=0.95$ \\
\midrule
$2$  & $2$ & $1.67$ & $1.43$ & $1.25$ & $1.18$ & $1.11$ & $1.05$ \\
$3$  & $2$ & $2$ & $2$ & $1.56$ & $1.38$ & $1.24$ & $1.11$ \\
$4$  & $2$ & $2$ & $2$ & $1.95$ & $1.63$ & $1.37$ & $1.17$ \\
$5$  & $2$ & $2$ & $2$ & $2$ & $1.92$ & $1.52$ & $1.23$ \\
$8$  & $2$ & $2$ & $2$ & $2$ & $2$ & $2$ & $1.43$ \\
$10$ & $2$ & $2$ & $2$ & $2$ & $2$ & $2$ & $1.59$ \\
\bottomrule
\end{tabular}}
\begin{tablenotes}[flushleft]
\footnotesize
\item[] \emph{Notes.} Every test has size at most $G$, and every
tested individual satisfies $q_i\geq\rho$. Each entry reports
$\min\{2,\gamma_G(\rho)\}$.
\end{tablenotes}
\end{threeparttable}
\end{table}

\Cref{figure:owr-high-q-curves} plots the resulting upper bound for the pool-size caps $G=5$ and $G=10$ used in our numerical experiments. The dashed line is the universal factor-$2$ guarantee; the high-$q$ theorem becomes strictly sharper once $\rho$ is high enough, and the bound approaches $1$ as $\rho$ approaches $1$.

\begin{figure}[t]
\centering
\begin{subfigure}[t]{0.48\linewidth}
\centering
\resizebox{\linewidth}{!}{%
\begin{tikzpicture}
\begin{axis}[
    width=\linewidth,
    height=0.72\linewidth,
    xmin=0.70, xmax=1.00,
    ymin=0.95, ymax=2.08,
    xlabel={Lower bound $\rho$ on $q_i$},
    ylabel={Upper bound on $\gain(B,J)$},
    xtick={0.70,0.80,0.90,1.00},
    ytick={1.0,1.25,1.5,1.75,2.0},
    grid=both,
    major grid style={draw=gray!30},
    minor grid style={draw=gray!15},
    legend style={draw=none, fill=white, fill opacity=0.9,
        text opacity=1, font=\scriptsize, at={(0.04,0.04)}, anchor=south west},
]
\addplot[
    thick,
    color=blue!70!black,
    domain=0.70:1.00,
    samples=160,
]
{min(2,min(5,min(2.5/x,min(1.6666666667/(x^2),min(1.25/(x^3),1/(x^4))))))};
\addlegendentry{$\min\{2,\gamma_5(\rho)\}$}
\addplot[
    dashed,
    thick,
    color=black!65,
    domain=0.70:1.00,
    samples=2,
] {2};
\addlegendentry{Universal bound}
\addplot[
    densely dotted,
    thick,
    color=blue!70!black,
] coordinates {(0.8408964153,0.95) (0.8408964153,2)};
\node[anchor=north west, font=\scriptsize, color=blue!70!black]
    at (axis cs:0.845,1.98) {$2^{-1/4}$};
\end{axis}
\end{tikzpicture}
}
\caption{Pool-size cap $G=5$.}
\end{subfigure}\hfill
\begin{subfigure}[t]{0.48\linewidth}
\centering
\resizebox{\linewidth}{!}{%
\begin{tikzpicture}
\begin{axis}[
    width=\linewidth,
    height=0.72\linewidth,
    xmin=0.70, xmax=1.00,
    ymin=0.95, ymax=2.08,
    xlabel={Lower bound $\rho$ on $q_i$},
    ylabel={Upper bound on $\gain(B,J)$},
    ylabel style={text opacity=0},
    xtick={0.70,0.80,0.90,1.00},
    ytick={1.0,1.25,1.5,1.75,2.0},
    yticklabel style={text opacity=0},
    grid=both,
    major grid style={draw=gray!30},
    minor grid style={draw=gray!15},
    legend style={draw=none, fill=white, fill opacity=0.9,
        text opacity=1, font=\scriptsize, at={(0.04,0.04)}, anchor=south west},
]
\addplot[
    thick,
    color=teal!70!black,
    domain=0.70:1.00,
    samples=200,
]
{min(2,min(10,min(5/x,min(3.3333333333/(x^2),min(2.5/(x^3),min(2/(x^4),min(1.6666666667/(x^5),min(1.4285714286/(x^6),min(1.25/(x^7),min(1.1111111111/(x^8),1/(x^9)))))))))))};
\addlegendentry{$\min\{2,\gamma_{10}(\rho)\}$}
\addplot[
    dashed,
    thick,
    color=black!65,
    domain=0.70:1.00,
    samples=2,
] {2};
\addlegendentry{Universal bound}
\addplot[
    densely dotted,
    thick,
    color=teal!70!black,
] coordinates {(0.9258747123,0.95) (0.9258747123,2)};
\node[anchor=north west, font=\scriptsize, color=teal!70!black]
    at (axis cs:0.929,1.98) {$2^{-1/9}$};
\end{axis}
\end{tikzpicture}
}
\caption{Pool-size cap $G=10$.}
\end{subfigure}
\begingroup
\makeatletter
\let\@makecaption\@maketablecaption
\makeatother
\caption{High-$q$ overlap welfare bounds.}
\label{figure:owr-high-q-curves}
\endgroup
\par\smallskip
\begin{minipage}{\linewidth}
\footnotesize\raggedright
\emph{Notes.} Panels (a) and (b) use pool-size caps
$G=5$ and $G=10$, respectively. The solid curves plot
$\min\{2,\gamma_G(\rho)\}$ for $\rho\in[0.7,1]$; the dotted vertical lines
mark the point where the high-$q$ bound first improves on the universal
factor-$2$ guarantee.
\end{minipage}
\end{figure}

These bounds are useful for interpreting the later numerical results. Whenever an instance class satisfies $\gain(B,J)\leq \alpha$, we have $\frac{u(\greedy)}{\opt^*(B,J)} = \frac{u(\greedy)}{\opt(B,J)} \cdot \frac{\opt(B,J)}{\opt^*(B,J)} \geq \frac{u(\greedy)}{\opt(B,J)}\cdot \frac{1}{\alpha}$. Thus the overlap-ratio bounds in this section explain when near-optimal performance against the non-overlapping benchmark also implies strong performance against the overlapping benchmark. This is especially relevant for the pilot-data simulations in \cref{section:algorithms-in-practice}, where estimated health probabilities are high.

\section{The Complexity of Welfare-Maximizing Pooled Testing}
\label{section:complexity-allocations}

We now turn to the problem of computing a test allocation with $B$ tests. This section maps the complexity landscape before we develop algorithmic methods in \cref{section:finding-allocations}. A recurring theme is the sharp threshold at pool size $G = 3$: evaluating overlapping allocations and optimizing non-overlapping allocations are polynomial-time solvable for $G \leq 2$, whereas $G \geq 3$ marks the onset of computational hardness for both problems.

We begin by showing that even \emph{evaluating} the welfare of a given overlapping allocation is \#P-hard for $G \geq 3$, providing a complexity-theoretic separation between overlapping and non-overlapping testing. \Cref{section:single-test-complexity} addresses the single-test problem, establishing that finding an optimal pool is W[1]-hard with respect to~$G$; this rules out an $f(G)\operatorname{poly}(n)$-time algorithm under the widely held assumption $\mathrm{FPT}\neq\mathrm{W[1]}$. \Cref{section:multi-test-complexity} then studies multi-test allocations, showing that the non-overlapping regime is NP-complete for any fixed $G \geq 3$ and polynomial-time solvable for $G \leq 2$. The same reduction also implies that the overlapping regime is NP-hard, although we conjecture that it is strictly harder than NP.
\subsection{Evaluating the welfare of an allocation}
Evaluating the welfare of a non-overlapping allocation is straightforward: the welfare of each test is a product of health probabilities multiplied by the sum of utilities, and the total welfare is the sum over tests. In contrast, even computing the welfare of a given overlapping allocation is intractable when pool sizes are at least~$3$.

\begin{proposition}
\label{prop:evaluation-complexity}
\hyperref[proof:prop:evaluation-complexity]{\normalfont[Proof in \cref{proof:prop:evaluation-complexity}]}
Fix $G \in \mathbb{N}$.
\begin{enumerate}[label=(\roman*)]
\item If $G \leq 2$, then the welfare $u(T)$ of a given test allocation $T$ with tests of size at most $G$ can be computed in polynomial time.
\item If $G \geq 3$, then the computational problem of exactly evaluating $u(T)$ for a given (possibly overlapping) test allocation $T$ with tests of size at most $G$ is \#P-hard. That is, under standard complexity-theoretic assumptions, no polynomial-time algorithm can compute the welfare of overlapping allocations.
\end{enumerate}
\end{proposition}

The proof reduces the problem of counting independent sets in a graph, a canonical \#P-complete problem, to welfare evaluation. Intuitively, when tests overlap, the probability that an individual is cleared depends on the joint outcomes of multiple correlated tests, and summing over these dependencies is as hard as counting combinatorial structures in a graph. We also show that the closely related threshold problem, deciding whether $u(T) \geq \tau$ for a given threshold $\tau$, is PP-complete for $G \geq 3$, placing the overlapping threshold decision problem at the center of \cref{section:complexity-allocations} in the complexity class $\text{NP}^{\text{PP}}$.

This separation is evidence that overlapping testing is computationally harder than non-overlapping testing, although it does not constitute a full complexity classification.

\subsection{The complexity of finding a single test}
\label{section:single-test-complexity}

The single-test problem---computing an optimal test allocation with budget $B=1$, where the non-overlapping and overlapping regimes coincide---was shown to be NP-hard for unbounded pool sizes by \citet{GOLDBERG2020} in their work on the maximum expected value all-or-nothing subset problem. However, in practice, pool sizes are typically bounded by a constant $G < n$. In this computational setting, an exhaustive search over all possible tests checks $\sum_{k \in [G]}{\binom{n}{k}} = O(n^G)$ possible constellations. Even though the running time of this procedure is technically polynomial for a fixed pool size, practical choices such as $G=5$ or $G=10$ for saliva testing (as implemented by our partner institution), and $G=32$, $G=48$, and $G=57$ for nasal swab testing \citep{Yelin-2020,Shental-2020,Theagarajan-2020}, make exhaustive search prohibitively expensive. For example, with $n=130$ individuals and $G=5$, as in our pilot study, exhaustive search requires evaluating over $286$ million candidate tests, just to find one optimal single test.

A natural question is whether there exists an algorithm with a more amenable running time that separates the dependency on $G$ from the remaining problem parameters. In the language of parameterized complexity, we ask whether the single-test problem is \emph{fixed-parameter tractable} with respect to $G$: does there exist an algorithm with running time $f(G) \cdot \text{poly}(n)$ for some function $f$? We show that the answer is no, under standard complexity-theoretic assumptions.

\begin{theorem}
\label{theorem:w1-hardness}
\hyperref[proof:theorem:w1-hardness]{\normalfont[Proof in \cref{proof:theorem:w1-hardness}]}
The problem of computing an optimal single test is W[1]-hard when parameterized by the pool-size bound~$G$: it is unlikely that there exists an algorithm with running time $f(G) \cdot \mathrm{poly}(n)$ for any computable function $f$, unless the widely held assumption $\mathrm{W[1]} \neq \mathrm{FPT}$ fails.
\end{theorem}

In other words, the combinatorial explosion in the pool size parameter $G$ cannot be avoided: any exact algorithm must have a running time whose polynomial degree grows with $G$. The proof, given in the appendix, reduces from the $k$-\textsc{Sum} problem via an intermediate problem we call \textsc{Relaxed}-$k$-\textsc{Sum}. The reduction uses the same welfare-function construction $f(x) = xr^x$ that will underpin the NP-hardness result for multi-test allocations in \cref{section:multi-test-complexity}: the base $r$ is chosen so that $f$ is uniquely maximized at a target value~$T$, encoding the combinatorial structure of the source problem into the welfare function. The main technical challenge is that the integers in the $k$-\textsc{Sum} instance may be exponentially large, so the health probabilities in the reduction may require exponentially many bits to represent. We resolve this by rounding probabilities to nearby rational numbers with polynomial encoding length, and showing via a gap lemma that this rounding does not affect which test is optimal.

\subsection{The complexity of computing multi-test allocations}
\label{section:multi-test-complexity}

We now consider allocations with multiple tests. For a single test, the non-overlapping and overlapping regimes coincide. With multiple tests they diverge, and the complexity picture sharpens: the non-overlapping problem is NP-complete for any fixed $G \geq 3$, while the overlapping problem is NP-hard and conjectured to be harder still.

\begin{theorem}
\label{theorem:np-completeness}
\hyperref[proof:theorem:np-completeness]{\normalfont[Proof in \cref{proof:theorem:np-completeness}]}
For any fixed $G \geq 3$, the problem of computing an optimal non-overlapping test allocation is NP-complete.
\end{theorem}

The proof, given in the Appendix, reduces from \textsc{3-Partition}, a problem that is strongly NP-complete. A problem instance of \textsc{3-Partition} consists of $3m$ positive integers $a_1, \ldots, a_{3m}$ and a target value $T$. The computational task is to decide whether these integers can be divided into $m$ triples that each sum to $T$. Importantly, the problem remains hard when all input numbers are bounded by a polynomial in $m$. This strong NP-completeness contrasts with weakly NP-hard number problems, whose hardness may rely on integer values that are large relative to the number of items but compact when written in binary.

Given an instance of \textsc{3-Partition} with integers $a_1, \ldots, a_{3m}$ and target~$T$, we construct a population with utilities $u_i = a_i$ and health probabilities $q_i = r^{a_i}$, where $r$ is chosen as the midpoint between $\frac{T-1}{T}$ and $\frac{T}{T+1}$. Because we reduce from the strongly NP-complete restricted version, the integers $a_i$ and $T$ may be taken to be polynomially bounded, so the rational probabilities $q_i = r^{a_i}$ have polynomial encoding length. The welfare of a single test~$t$ then takes the form $f(a(t)) = a(t) r^{a(t)}$, where $a(t) = \sum_{i \in t} a_i$. The key property is that $f$ is uniquely maximized at $x = T$: welfare is highest when the sum of utilities in a test equals exactly~$T$. Consequently, total welfare is maximized if and only if the individuals can be partitioned into groups whose utilities each sum to $T$. This is precisely the condition for a \textsc{3-Partition} solution to exist.

Given this hardness, we turn to approximation hardness. An FPTAS is the strongest usual form of efficient approximation guarantee: for any accuracy parameter $\varepsilon>0$, it returns a solution within a factor $1-\varepsilon$ of optimal in time polynomial in both the input size and $1/\varepsilon$. The following corollary rules out such a scheme for computing optimal non-overlapping allocations for any fixed pool-size bound $G \geq 3$, unless $\mathrm{P}=\mathrm{NP}$. Appendix \ref{appendix:complexity} reviews approximation schemes in more detail.

\begin{corollary}
\label{cor:inapproximability}
\hyperref[proof:cor:inapproximability]{\normalfont[Proof in \cref{proof:cor:inapproximability}]}
For any fixed $G \geq 3$, there is no FPTAS for computing an optimal non-overlapping test allocation, unless $\mathrm{P} = \mathrm{NP}$.
\end{corollary}

This contrasts with the single-test problem, which admits an FPTAS: multi-test allocation is strictly harder, ruling out approximation schemes whose running time is polynomial in both the input size and $1/\varepsilon$.

The overlapping regime is at least as hard. Using the \textsc{3-Partition} reduction with a tighter analysis shows that finding an optimal overlapping allocation is also NP-hard for any fixed $G \geq 3$.

\begin{theorem}
\label{theorem:overlapping-np-hard}
\hyperref[proof:theorem:overlapping-np-hard]{\normalfont[Proof in \cref{proof:theorem:overlapping-np-hard}]}
For any fixed $G \geq 3$, the problem of computing an optimal overlapping test allocation is NP-hard.
\end{theorem}

\begin{corollary}
\label{cor:overlapping-inapproximability}
\hyperref[proof:cor:overlapping-inapproximability]{\normalfont[Proof in \cref{proof:cor:overlapping-inapproximability}]}
For any fixed $G \geq 3$, there is no FPTAS for computing an optimal overlapping test allocation, unless $\mathrm{P}=\mathrm{NP}$.
\end{corollary}

Importantly, we do not show that this problem lies in NP; indeed, the \#P-hardness of welfare evaluation from \cref{prop:evaluation-complexity} suggests that it does not. While we conjecture that the problem is $\mathrm{NP}^{\mathrm{PP}}$-hard, establishing this requires new techniques.

These hardness results exhibit a sharp threshold in the pool size parameter for the problems covered by the reductions above. For $G \in \{1,2\}$, single-test optimization is polynomial-time solvable, non-overlapping multi-test optimization is polynomial-time solvable via reduction to maximum-weight matching with a cardinality constraint, and any fixed overlapping allocation can be evaluated in polynomial time (see the appendix for details). Thus $G = 3$ is the exact boundary for the evaluation hardness result and for non-overlapping multi-test optimization.

In summary, exact optimization of test allocations is out of reach for $G \geq 3$, even in the non-overlapping regime. Given \cref{cor:inapproximability}, we cannot hope for an FPTAS for the non-overlapping multi-test problem; instead, we develop a constant-factor approximation algorithm.

\section{Finding optimal test allocations}
\label{section:finding-allocations}

We now turn to the problem of finding optimal test allocations. As we know that finding exact allocations is intractable, the purpose of this section is to develop algorithms for computing approximately optimal allocations. We distinguish between two problems of increasing difficulty: computing a single optimal test, and computing a full test allocation with budget~$B$. As both problems are computationally hard for pool sizes $G \geq 3$, there are two natural algorithmic strategies: one can seek an efficient algorithm that is guaranteed to find an approximate solution up to a constant, or one can formulate an optimization problem that may require more computation time but gets arbitrarily close to optimum. We pursue both avenues.

\Cref{section:single-test-allocations} addresses the single-test problem by developing two approximation methods: a fully polynomial-time approximation scheme (FPTAS) and a mixed-integer conic program (MICP). \Cref{section:approximation-algorithms} develops greedy algorithms with provable constant-factor (approximation) guarantees, while \cref{section:MILP} develops a mixed-integer linear programming formulation for the non-overlapping regime whose accuracy can be traded off against computation time. The \greedy{} algorithms repeatedly apply a single-test subroutine to construct a non-overlapping allocation. Our initial \greedy{} algorithm has access to an \textit{exact} single-test subroutine and achieves at least a $\frac{1}{e+1}$ fraction of the welfare of the optimal \emph{overlapping} test allocation for any population and budget (\cref{theorem:greedy-vs-overlapping}). As we know that exact single-test computation is intractable, we propose a modification that uses our approximate single-test subroutines to attain a guarantee of $1/(e+1+\delta)$ in time $O(2^G \cdot \mathrm{poly}(n, B, 1/\delta))$ (\cref{theorem:modified-greedy}), which is fixed parameter tractable in the pool size $G$. \cref{section:MILP} formulates the non-overlapping testing problem as a mixed-integer linear program that serves as a benchmark implementation with tunable accuracy. In our computational experiments, \greedy{} achieves at least $99.3\%$ of optimal non-overlapping welfare (\cref{table:experiment1,table:experiment2}) on realistic data. Combined with the high-$q$ overlap-ratio bound of \cref{theorem:high-q-bound}, this near-optimality against the non-overlapping benchmark translates directly into near-optimality against the overlapping benchmark whenever tested individuals have high health probabilities, as is the case in the pilot data.

\subsection{Optimally allocating a single test}
\label{section:single-test-allocations}

Consider the task of finding a single optimal test. The impossibility result from \cref{section:single-test-complexity} motivates us to explore approximation algorithms. For any $\varepsilon > 0$, we seek an algorithm that finds a test with welfare at least $(1-\varepsilon)$ times the optimal. We develop two such methods: an FPTAS adapted from \citet{GOLDBERG2020}, and a conic optimization formulation.

\paragraph{Designing an FPTAS.}
\Citet{GOLDBERG2020} introduce a type of algorithm commonly known as a fully polynomial-time approximation scheme (FPTAS) that runs in time $O \left( n^5/\varepsilon \right)$.\footnote{An FPTAS is an algorithm that achieves $(1-\varepsilon)$ optimality, where for a given $\varepsilon$, the running time of the algorithm is polynomial in the problem instance (in this case $n$) as well as $1/\varepsilon$.} This FPTAS returns a test with an almost-optimal welfare; the test returned achieves a factor of $1-\varepsilon$ of the optimum achievable welfare. Here $\varepsilon > 0$ is a positive value that can be chosen arbitrarily small. The algorithm of \citet{GOLDBERG2020} returns a test with an arbitrary pool size. In \cref{appendix:proof-FPTAS}, we show how their algorithm can be adapted to return an almost-optimal single-test allocation when a constant pool size constraint is imposed.

To provide intuition for the FPTAS, we first describe at a high level a dynamic program that exactly solves for an optimal test when the utilities are non-negative integers. For $i \in [n]$, let $P(i,C,L) = \max\{ q_S \mid S \subseteq [i] \text{, } |S| = L \text{, and } \sum_{\ell \in S} u_\ell = C \}$ denote the largest negative probability of a subset of $[i]$ of cardinality $L$ and total utility $C$. Note that $\bar{C} = \sum_{i \in [n]} u_i$ is an upper bound on the sum of utilities of an optimal test, and pooled tests are bounded by $G$. We can compute $P(i,C,L)$ for $C \in \{0,\ldots,\bar{C}\}$ and $L \in \{0,\ldots,G\}$ via dynamic programming in $O(n G \bar{C}) = O(n^2 \bar{C})$ time. The test with maximal utility is the test that maximizes $C \cdot P(n,C,L)$. Our approach is similar to the dynamic program used in \cite{GOLDBERG2020}, but differs in that we keep track of pool sizes using parameter~$L$.

This dynamic program does not run in polynomial time due to the time dependence on $\bar{C}$, which is exponential in the binary representation of the population data. Instead, our polynomial-time algorithm uses a similar dynamic program as described above, but replaces each agent's utility by $\lfloor u_i / \kappa \rfloor$.%
\footnote{This approach is commonly used to design FPTASs for knapsack problems in the computer science literature.} Larger values of $\kappa$ give rise to better running times (as the resulting problem is smaller), but result in worse approximation guarantees. Choosing the value of $\kappa$ carefully as a function of the desired approximation parameter $\varepsilon$, we achieve a polynomial running time of $O(n^5/\varepsilon)$ while retaining an approximation factor of $1-\varepsilon$.

\begin{proposition}
\label{prop:single-test-fptas}
\hyperref[proof:prop:single-test-fptas]{\normalfont[Proof in \cref{proof:prop:single-test-fptas}]}
The FPTAS presented in \cref{appendix:proof-FPTAS} finds an almost-optimal single-test allocation with pool size at most $G$ in time $O(n^5/\varepsilon)$.
\end{proposition}

\paragraph{Formulating a conic optimization problem.}
The problem of allocating a single test can also be formulated as a mixed-integer conic optimization program (MICP), and can be solved using a commercial conic solver. This implementation is used as a subroutine of \greedy{} in the web app accompanying our pilot study.

Define the indicator vector $\bm{x} \in \{0,1\}^n$ with $x_i = 1$ if individual $i$ is included in the (single) test, and ${x_i = 0}$ otherwise. Our objective is to maximize the expected utility from the test given by $\sum_{i \in [n]} u_ix_i \cdot \prod_{i \in [n]}q_i^{x_i}$. Constraint $1 \leq \sum_{i \in [n]}x_i \leq G$ imposes pool sizes between $1$ and $G$. We isolate the non-linear element of the optimization problem by maximizing the logarithm of the original objective and introducing variables $z = \sum_{i \in [n]} u_i x_i$ and $y = \log z$.\footnote{For logarithmic formulations, we first discard any individual with $u_i=0$ or $q_i=0$, not changing the optimum as including such an individual in a test cannot increase its expected utility. If no individuals remain, the optimum is zero.} Note that we can relax the equality in the only remaining non-linear constraint $y = \log z$ to $y \leq \log z$ without affecting the outcome. The resulting optimization problem is
\vspace{-5pt}
\begin{maxi!}
  {\bm{x} \in \{0,1\}^n}{y + \sum_{i \in [n]}x_i \log q_i}{\label{opt:conic-program}}{}
  \addConstraint{y}{\leq \log(z)}{}\label{constraint:log}
  \addConstraint{z}{= \sum_{i \in [n]} u_i x_i}{}
  \addConstraint{1}{\leq \sum_{i \in [n]} x_i \leq G}{}
\end{maxi!}

Moreover, Constraint \ref{constraint:log} can be formulated as the conic constraint $(z, 1, y) \in K_{\exp}$. Here $K_{\exp}$ is the exponential cone defined as $K_{\exp} =
    \{(x_1, x_2, x_3) \mid x_1 \geq x_2e^{x_3 / x_2}, x_2 > 0 \}
    \cup \{ (x_1,0,x_3) \mid x_1 \geq 0, x_3 \leq 0 \}$.
We show in our numerical experiments that the resulting MICP can be solved rapidly with the conic solver MOSEK \citep{mosek}. Example running times can be found in \cref{table:experiment1} (\cref{section:simulations}) and \cref{table:experiment3} (\cref{apx:simulations} in the appendix).

In practice, these methods allows us to compute an almost-optimal single test efficiently. The single-test problem is thus hard to solve exactly for $G \geq 3$, but admits good approximation.

\subsection{Approximation algorithms}
\label{section:approximation-algorithms}

Given the hardness results of the previous sections, we study \greedy{}, which sequentially applies the single-test subroutine from \cref{section:single-test-allocations}: at each step, the subroutine finds an (approximately) optimal test among the untested individuals, adds it to the allocation, and removes those individuals from the population. The procedure continues until the testing budget or the population is exhausted. Because previously tested individuals are never reconsidered, the output is non-overlapping. The quality of the resulting allocation depends on the quality of the single-test subroutine. For both our single-test subroutines, we can write the error as a factor $1 -\varepsilon$, where the approximation error $\varepsilon$ is $10^{-7}$ for the conic program, and can be set arbitrarily small for the FPTAS.

By construction, \greedy{} produces a non-overlapping test allocation. The next proposition nevertheless compares it with the optimal \emph{overlapping} allocation. Since overlapping allocations can achieve higher welfare than non-overlapping ones, this comparison is stronger than benchmarking only against the optimal non-overlapping allocation.

\begin{proposition}
\label{theorem:greedy-vs-overlapping}
\hyperref[proof:theorem:greedy-vs-overlapping]{\normalfont[Proof in \cref{proof:theorem:greedy-vs-overlapping}]}
For any population and budget, \greedy{} with an exact single-test oracle achieves at least $\frac{1}{e+1}$ of the welfare of the optimal (overlapping) allocation.
\end{proposition}

The proof, given in the Appendix, uses a ``residual benchmark'' technique. At each round~$j$, we track the optimal welfare achievable on the residual (untested) population with the remaining budget. Removing \greedy{}'s chosen test from the optimal allocation incurs two types of loss: (i) a \emph{truncation loss} that arises from partitioning optimal tests that share individuals with \greedy{}'s choice, which we bound by $e$ times the greedy test's welfare via a partition inequality rooted in the Schur-concavity of an auxiliary function; and (ii) a \emph{budget loss} due to the fact that the optimal allocation has one fewer test available, so we must discard its lowest-marginal test, contributing at most $1$ times the greedy test's welfare. Telescoping over all $B$ rounds yields the $\frac{1}{e+1}$ bound.

However, \cref{theorem:greedy-vs-overlapping} requires an exact single-test oracle, which is W[1]-hard by \cref{theorem:w1-hardness}. If we instead use the $\varepsilon$-approximate single-test subroutine from the FPTAS, the approximation guarantee deteriorates exponentially in $B$. To avoid this, we modify \greedy{} slightly: after the approximate subroutine returns a test, we search over all subsets of that test (at most $2^G$ subsets) and select the one with the highest welfare. This post-processing step restores a crucial ``subset-domination'' property needed for the partition inequality, ensuring that the approximation error only affects the budget-loss term and not the truncation loss. %

\begin{theorem}
\label{theorem:modified-greedy}
\hyperref[proof:theorem:modified-greedy]{\normalfont[Proof in \cref{proof:theorem:modified-greedy}]}
Fix any $\delta > 0$. Modified \greedy{} with post-processing achieves welfare at least $\frac{1}{e+1+\delta}$ of the welfare of the optimal (possibly overlapping) test allocation, in time $O(2^G \cdot \mathrm{poly}(n, B, 1/\delta))$.
\end{theorem}

In the language of parameterized complexity, this running time is fixed-parameter tractable in the pool-size parameter~$G$: \cref{theorem:w1-hardness} rules out an exact fixed-parameter tractable (FPT) algorithm in~$G$, but \cref{theorem:modified-greedy} shows that giving up exactness restores tractability with only a constant-factor loss. Therefore, this is efficient in practice: for $G = 5$ the post-processing evaluates at most $32$ subsets per round, and for $G = 10$ at most $1024$. For more information on fixed parameter tractability, we refer to Appendix~\ref{appendix:complexity}.

Complementing these guarantees, we provide a family of instances for which the approximation ratio of \greedy{} tends to $\frac{1}{e}$ as the number of individuals grows. The gap between the $\frac{1}{e+1}$ lower bound and the $\frac{1}{e}$ upper bound remains open.

If we restrict attention to comparing \greedy{} against the optimal \emph{non-overlapping} allocation only, a stronger polynomial-time guarantee is available. Recall that $\varepsilon$ is the accuracy of the single-test subroutine, returning welfare within a factor $(1-\varepsilon)$ of the optimal single-test~welfare.

\begin{theorem}
\label{theorem:greedy-approximation}
\hyperref[proof:theorem:greedy-approximation]{\normalfont[Proof in \cref{proof:theorem:greedy-approximation}]}
Fix arbitrarily small $\delta > 0$. For any population and testing budget, \greedy{} achieves at least $\frac{1}{5+\delta}$ of the welfare of the optimal non-overlapping test allocation if we set the error tolerance of the single-test oracle to $\varepsilon = \frac{\delta}{9 + 2\delta}$.
\end{theorem}
This algorithm runs in polynomial time, avoiding the $2^G$ overhead of \cref{theorem:modified-greedy}.
The proof decomposes the optimal welfare into contributions from individuals tested by \greedy{} and individuals left untested. The approximation error $\varepsilon$ of the single-test subroutine enters both parts of the argument, and choosing $\varepsilon$ sufficiently small as a function of $\delta$ yields the result.

Consider the individuals that \greedy{} does \emph{not} test but that the optimal allocation~$T^*$ does. At each step, \greedy{} selects a single test that is within a $(1-\varepsilon)$ factor of the best available test among remaining individuals. Since the individuals assigned to any optimal test are still available when \greedy{} makes its corresponding choice, \greedy{}'s chosen test must yield at least $(1-\varepsilon)$ times the welfare the optimal allocation earns from those untested individuals.

For the individuals that \greedy{} \emph{does} test, the argument shows that each test chosen by \greedy{} extracts a constant fraction of the maximum possible welfare from its members. The key insight is that \greedy{}'s (approximate) greedy selection rule prevents any test from being ``split'' into two subsets that are both unlikely to test negative: if such a split existed, \greedy{} would have preferred one of the subsets instead. This property (analogous to Lemma~3 used in the proof of Theorem~1) ensures that within each chosen pool, every individual's contribution is weighted by a health probability of at least roughly $\frac{1-\varepsilon}{2}$ from the other pool members. Consequently, \greedy{} captures at least roughly $\frac{(1-\varepsilon)^2}{4}$ of the total expected healthy-individual utility in its tested population.

The optimal allocation's welfare can be decomposed into its contributions from \greedy{}-tested and \greedy{}-untested individuals. The first bound limits the latter, and the second limits the former. Summing yields that the optimal welfare is at most $(5+\delta)$ times \greedy{}'s welfare when $\varepsilon$ is set to $\frac{\delta}{9+2\delta}$, establishing the $\frac{1}{5+\delta}$-approximation guarantee.

In \cref{appendix:homogeneous-utilities} of the appendix, we consider the special case of homogeneous utilities and a variant of our \greedy{} algorithm called \vargreedy{}, which sorts the individuals in decreasing order of their probabilities of being healthy and, in each step, adds individuals to the current test as long as the expected utility of the test increases. We show that \vargreedy{} returns a $\frac{1}{e}$-approximate non-overlapping test allocation. We also consider another special setup with clusters, whereby the population can be partitioned into a finite number of disjoint clusters consisting of individuals with identical utilities and health probabilities. We show that any $(1-\varepsilon)$ optimal test can be repeated a finite number of times to achieve $(1-\varepsilon)$-welfare in the clustered population. Details are given in \cref{appendix:greedy-cluster} of the appendix.

In practice, even the simple \greedy{} with the conic optimisation subroutine performs extremely well, achieving at least $99.37\%$ for pool size $G=5$ and $96.98\%$ for pool size $G=10$ of optimal non-overlapping welfare in our computational experiments (\cref{table:experiment1,table:experiment2}).

\subsection{A formulation via mixed-integer linear programming}
\label{section:MILP}
While \greedy{} provides strong approximation guarantees and runs efficiently, it may be desirable, particularly for small testing budgets or to evaluate the \greedy{} algorithm, to compute exact or near-exact optimal allocations. For optimal testing with overlaps, we were not able to find a method that reliably computes near-optimal allocations with non-trivial instances. In the non-overlapping regime, however, we develop a mixed-integer linear programming (MILP) formulation that allows the decision-maker to trade computation time for solution accuracy guarantees. The MILP approximately solves the non-overlapping allocation problem by replacing exponential constraints with piecewise-linear functions. The accuracy of this approximation is controlled by a tuning parameter $K$, the number of segments in the piecewise-linear function: increasing $K$ improves accuracy at the cost of introducing more integer variables and longer solve times. We provide practical additive approximation guarantees as a function of $K$. The MILP formulation is competitive when test budgets are low but becomes computationally intensive as budgets increase; in our experiments, it serves as a near-exact benchmark against which we evaluate \greedy{}.
\footnote{For the MILP program it can also be valuable to cluster identical individuals in the population. A formulation of the MILP with clusters, which may speed up computations, is provided in \cref{appendix:greedy-cluster} of the appendix.}

\paragraph{Formulating a mixed-integer linear program.}
\label{section:milp-formulation}
We can assume that the testing budget $B$ is at most the population size $n$, and so pool sizes lie between $1$ and $G$. We proceed similarly to our formulation of the conic program. For each test $j \in [B]$, we introduce an indicator vector $\bm{x}^j \in \{0,1\}^n$ with $x^j_i = 1$ if individual~$i$ is included in $j$ and $x^j_i = 0$ otherwise, and let variable $w^j$ denote its expected utility $w^j = u \cdot x^j \prod_{i \in [n]}q_i^{x^j_i}$. We impose pool sizes between $1$ and $G$ with constraints $1 \leq \sum_{i \in [n]} x^j_i \leq G$ for all $j \in [B]$, and non-overlapping testing with constraints $\sum_{j \in [B]} x^j_i \leq 1$ for all $i \in [n]$. Our objective is to maximize welfare $\sum_{j \in [B]}w^j$.
In order to isolate the non-linear elements of the optimization problem, we reformulate the problem with additional variables: variables $l^j$ denote the logarithm of $w^j$, and variables $y^j$ and $z^j$ allow us to isolate the non-linear elements of each $w^j$ into constraints \eqref{constr:exp} and \eqref{constr:log}. 
\begin{maxi!}
  {}{\sum_{j \in [B]} w^j}{\label{opt:unpacked-problem}}{}
  \addConstraint{w^j}{= \exp l^j,}{\quad \forall j \in [B]}{\label{constr:exp}}
  \addConstraint{l^j}{ = y^j + \sum_{i \in [n]} x^j_i \log q_i,}{\quad \forall j \in [B]}
  \addConstraint{y^j}{= \log z^j,}{\quad \forall j \in [B]}{\label{constr:log}}
  \addConstraint{z^j}{= u \cdot x^j,}{\quad \forall j \in [B]}
  \addConstraint{\sum_{j \in [B]} x^j_i}{\leq 1,}{\quad \forall i \in [n]}{\label{constr:non-overlapping}}
  \addConstraint{1 \leq \sum_{i \in [n]} x^j_i}{\leq G,}{\quad \forall j \in [B]}
  \addConstraint{x^j_i}{\in \{0,1\},}{\quad \forall i \in [n], \forall j \in [B]}{\label{constr:xvars}}
\end{maxi!}

In order to make the problem tractable, we assume that the utility vector~$u$ is integral and non-negative. This assumption is benign, as the problem is invariant to scaling of utilities. We describe in \cref{appendix:milp-details} of the appendix how the non-linear constraints \eqref{constr:exp} and \eqref{constr:log} can respectively be captured approximately and exactly by integer linear constraints.
The exponential constraints are approximated with piecewise-linear functions that can be formulated as a collection of mixed integer linear constraints. The accuracy of this approximation can be adjusted by tuning the number $K$ of segments of the piecewise-linear function, at the cost of introducing more (integer) variables and thus the time to solve the program. This allows us to compute (additive) approximation guarantees for the MILP as a function of $K$.

\section{Algorithm Performance and Pilot Evidence}
\label{section:algorithms-in-practice}

We evaluate our algorithms using data from a pilot study conducted in partnership with the \redacted{Potosinian Institute for Scientific and Technological Research (IPICYT)} in \redacted{San Luis Potosí}, Mexico. \Cref{section:simulations} compares \greedy{} to the MILP benchmark on the pilot population and on synthetic populations spanning a broader range of infection rates. \Cref{section:pilot} summarizes the pilot implementation, in which we deployed a version of \greedy{} in a two-group randomized controlled trial with 130 participants, a weekly testing budget of $B=30$, and a pool-size bound of $G=5$. Full details of the experimental design and population data construction are provided in \cref{section:design,section:population-data} of the appendix.

\subsection{Performance evaluation of \greedy{} and the MILP}
\label{section:simulations}

We evaluate the accuracy and running times of \greedy{} and the MILP in two experiments: one using the pilot-study population, and one using synthetic populations that vary health probabilities and utilities more broadly.%
\footnote{We verified on a set of small instances that the MILP produces welfare values identical to those of brute-force enumeration and exact search up to numerical tolerance $10^{-5}$.}

In the first experiment, we test both algorithms on population data from the pilot study with testing budgets $B$ up to $34$ and pool size constraints~$G$ of $5$ and $10$.%
\footnote{As the MILP is designed to admit integral utilities only, and the problem of computing test allocations is invariant to scaling utilities, we first scale up the utilities of all individuals in the population by a factor of $50$, and then round the resulting number to the nearest integer. Choosing a larger scaling factor increases the running time, as the number of variables in the MILP increases (cf.~\cref{appendix:milp-details} in the appendix).} The pilot-data experiment illustrates the performance of \greedy{} in a low-incidence setting. In the second experiment, to study performance across more heterogeneous risk profiles, we generate random populations of size $n=200$ with health probabilities drawn independently and uniformly at random from the interval $[0.5,1]$ and utilities drawn from a normal distribution fitted to the utilities observed in the pilot study. We run \greedy{} and the MILP for both pool sizes $G \in \{5,10\}$ and all testing budgets $B \in \{2,4,\ldots, 12\}$, recording the welfare achieved by both algorithms as well as their running times (in milliseconds).%
\footnote{The experiments were run on an AWS EC2 instance type `c6g.8xlarge' with 32 vCPUs and 64GiB memory. Gurobi 9.5.0 was used for the MILP algorithm, and MOSEK 10 for the MICP.}

For both experiments, we document the true welfare achieved by the test allocations returned by both algorithms, and not the objective values of the underlying MILP and conic optimization problems. For the MILP algorithm, we tune the parameter $K$ (number of linear segments in constraint approximations) of the formulation so that the additive approximation guarantee (cf.~\cref{appendix:milp-details}) is small ($K=25$ for the simulations on pilot data, and $K=20$ for the simulations on synthetic data).
The replication materials for these simulations are described in \cref{section:data-code-availability}.

\paragraph{Results.}
\Cref{table:experiment1} lists the welfares achieved by \greedy{} and the MILP on the pilot data for pool size $G=5$, as well as the running times for both algorithms, the approximation guarantee achieved by the MILP, and an upper bound on the ratio of optimal non-overlapping welfare to the welfare achieved by \greedy{}. This upper bound is determined by summing the welfare and the approximation guarantee of the MILP to upper bound the optimum achievable welfare, and then dividing by the welfare achieved by \greedy{}. \Cref{table:experiment2} in \cref{apx:simulations} of the appendix shows analogous results for pool size constraint $G=10$. We observe that \greedy{} achieves near-optimal welfare for budgets up to $34$ for both pool sizes. Moreover, the running time of the MILP increases substantially faster with the test budget $B$ than \greedy{}, and the latter is extremely fast (even for larger populations and testing budgets). This makes \greedy{} attractive for implementations that require quick turnaround, run on modest hardware, or avoid cloud computing because of cost or concerns about sensitive health and utility data. %

\begin{table}[tb!]
    \centering
    \begin{threeparttable}
    \caption{Performance of the MILP and \greedy{} on pilot data ($G=5$).}
    \label{table:experiment1}
    {\small
    \renewcommand{\arraystretch}{1.15}
    \begin{tabular}{@{} crrrrrr @{}}
        \toprule
        & \multicolumn{3}{c}{{MILP}} & \multicolumn{3}{c}{{\greedy{}}} \\
        \cmidrule(lr){2-4} \cmidrule(l){5-7}
        {Budget} & Welfare & Guarantee & Time & Welfare & Apx To Optimal & Time \\
        \midrule
        2 & 461.24 & 0.18 & 292 ms & 461.22 & 1.0004 & 49 ms \\
        6 & 1292.04 & 0.53 & 1.91 s & 1291.91 & 1.0005 & 48 ms \\
        10 & 2070.82 & 0.89 & 5.34 s & 2070.58 & 1.0005 & 71 ms \\
        14 & 2814.82 & 1.25 & 42.51 min & 2814.50 & 1.0006 & 400 ms \\
        18 & 3524.78 & 1.61 & 25.39 min & 3524.24 & 1.0006 & 2.72 s \\
        22 & 4189.57 & 1.97 & 9.49 min & 4188.85 & 1.0006 & 42.87 s \\
        26 & 4790.46 & 2.32 & 202.68 min & 4789.37 & 1.0007 & 12.46 min \\
        30 & 4805.24 & 2.68 & 1159.03 min & 4789.37 & 1.0039 & 12.50 min \\
        34 & 4816.37 & 3.04 & 6813.38 min & 4789.37 & 1.0063 & 12.63 min \\
        \bottomrule
    \end{tabular}
    }
    \begin{tablenotes}[flushleft]
    \footnotesize
    \item[] \emph{Notes.} %
    Welfare and computation times for the MILP and \greedy{} on the pilot population ($n=130$), with $G=5$ and $B \in \{2,6,\ldots,34\}$. Welfare values are deterministic for this population. ``Guarantee'' reports the MILP's a priori additive approximation guarantee relative to optimal non-overlapping welfare. ``Apx To Optimal'' reports an upper bound on the ratio of optimal non-overlapping welfare to \greedy{} welfare, computed as MILP welfare plus the guarantee, divided by \greedy{} welfare.
    \end{tablenotes}
    \end{threeparttable}
\end{table}

In our synthetic simulations with lower health probabilities, \greedy{} performs as well as the MILP when $G=5$, and remains highly competitive also when $G=10$. \Cref{fig:experiment3,fig:experiment4} in \cref{apx:simulations} of the appendix show the distributions of the MILP-to-\greedy{} welfare ratios for pool size constraints $G=5$ and $G=10$. For each population, the ratio is the welfare achieved by the MILP divided by the welfare achieved by \greedy{}; panel (a) depicts the resulting ratio as a black dot, and panel (b) shows the ratios as per-budget histograms. \cref{table:experiment3,table:experiment4} in \cref{apx:simulations} of the appendix list the mean welfare and running times of both algorithms, as well as the approximation guarantee of the MILP and the MILP-to-\greedy{} welfare ratio, for $G \in \{5,10\}$. %

Comparing the outcomes between different pool sizes $G \in \{5, 10\}$, we see that increasing pool sizes from 5 to 10 significantly increases mean welfare if health probabilities are very high (cf.~\cref{table:experiment1,table:experiment2}). This effect is less pronounced in the experiment with synthetic data, in which participants have lower health probabilities on average (cf.~\cref{table:experiment3,table:experiment4} in the appendix). These results suggest that the pool size limit of $5$ imposed by saliva sampling, as opposed to the typical limit of $10$ for nasopharyngeal samples, may be restrictive in some scenarios, and institutions may wish to weigh the positives and negatives of saliva and nasopharyngeal sampling carefully.

\subsection{Pilot Study}
\label{section:pilot}

We report the outcomes of the implementation of our test allocation framework at \redacted{IPICYT}. After the federally mandated reopening of educational institutions in September 2022, our treatment group followed a scheduled pooled testing protocol using \greedy{}, with on-site access conditional on a negative result; the control group returned to campus with full, unrestricted mobility. This makes the control group a strong benchmark: it represents outcomes one would expect under a complete lifting of restrictions with no COVID-19 containment measures in place. Full experimental details are provided in \cref{appendix:design} of the appendix.

\paragraph{Algorithm deployment.}
\greedy{} was implemented in a dedicated web application. To account for participants’ scheduling preferences, individuals could express preferred testing windows via a virtual token budget over consecutive two-day windows, matching the approximately 48 hours of actual access available after receipt of a negative result within the 72-hour clearance window measured from sample submission, thereby avoiding unnecessary tests on days they did not intend to be on campus. Additionally, since an invited participant may choose not to submit a saliva sample, we performed a two-stage optimization: a weekly allocation followed by a second daily round restricted to submitted samples.\footnote{Re-optimizing the daily allocation over the sub-population of submitted samples weakly dominates the corresponding restriction of the weekly allocation, so the second stage cannot decrease expected welfare.}

\paragraph{Evaluation.}
We evaluated the impact of the testing protocol along five dimensions, measured via validated survey instruments administered to all participants at the start (\textit{baseline}) and end (\textit{endline}) of the pilot trial:\footnote{131 trial participants completed the baseline survey; 122 completed the endline survey. 130 baseline participants had health-probability and utility data required for the testing allocation.} stress and subjective well-being (mental health), and self-assessed performance, productivity, and learning (work and study outcomes). The initial volunteer cohort was assigned using complete cluster randomization at the working-group level; a small set ($\approx18$) of late enrollers was assigned separately under individual randomization, implemented by treating each late enroller as a singleton cluster. The full sample combines two randomization designs.

Our model specifications account for the baseline/endline structure and minor covariate imbalance documented in \cref{tab:cov-balance}. We estimate covariate-adjusted linear models with heteroskedasticity-robust (HC1) errors in two complementary forms: (i) an ANCOVA levels model regressing each endline outcome on treatment, the corresponding baseline outcome, and a fixed set of pre-treatment covariates; and (ii) a change score model regressing $\Delta Y = Y_{\text{endline}} - Y_{\text{baseline}}$ on treatment and the same baseline covariates, excluding those that are themselves outcomes.\footnote{The full covariate list can be found in \cref{section:app:metrics-and-methods}.}

To address the mixed randomization design, we re-estimate the main models on the initial cluster-randomized cohort using CR2 cluster-robust standard errors, and also report CR2 standard errors for the full sample using an \emph{effective cluster} variable that matches the randomization unit \citep{abadie2023clustering}. Because our main hypothesis is mean equivalence rather than a positive treatment effect, we additionally report two-sample equivalence tests of the endline group means using two one-sided tests (TOST) \citep{lakens2017equivalence}. We further assess robustness to multiple testing using Holm--Bonferroni and Benjamini--Hochberg corrections, to differential attrition using the sharp bounds of \citet{lee2009bounds}, and to the ordinal nature of the four-category performance, productivity, and goal-achievement outcomes using ordered logistic regressions.

\paragraph{Results.}

Across all five outcomes, the covariate-adjusted level regressions show no statistically significant treatment effect, and the change-score models agree; per-outcome coefficients and standard errors are reported in \cref{tab:results-mentalhealth,tab:results-productivity}. The point estimates show no directional pattern across outcomes, and magnitudes remain small throughout. In the separate TOST analysis of unadjusted endline group means (Cohen's $d=0.5$), four of the five outcomes are \textit{mean} equivalent at $p<0.05$, with stress narrowly outside the threshold ($p=0.079$; \cref{tab:power-analysisTOST}).

\paragraph{Robustness.}
The differences between treatment and control remain non-significant under multiple-hypotheses-testing correction across ten tests (Holm-Bonferroni and Benjamini-Hochberg adjusted $p \geq 0.25$; \cref{tab:mt-correction}). The conclusions are robust to cluster-randomization inference: refitting the main models on the initial cluster-randomized cohort with CR2 cluster-robust standard errors at the working group (cf.~\cref{tab:lm-mentalhealth-cov-cluster,tab:lm-performance-cov-cluster}) and on the full sample with CR2 standard errors at the effective cluster\footnote{Cluster id for the primary cohort, and user id for the late enrollers.} (cf.~\cref{tab:lm-mentalhealth-cov-effective}) yields qualitatively identical estimates, and all treatment effects remain non-significant under small-sample Satterthwaite tests (\cref{tab:cr2-satterthwaite}). \citet{lee2009bounds} sharp bounds under monotone differential attrition, with bootstrap Imbens--Manski confidence intervals, include zero for all five outcomes: the null finding is not overturned by worst-case selection (\cref{tab:lee-bounds}). In particular, both groups followed virtually identical sociability trajectories: $n=40$ participants in each group maintained or increased their socializing in-person, suggesting that the protocol did not impede social interactions associated with mental-health recovery after lockdown. Throughout the trial, only one pooled test returned positive; upon re-testing, a single infected individual was identified and isolated.\footnote{As per regulations of the Mexican Ministry of Health, all individuals in the positive pool were immediately re-tested using already-submitted saliva samples; within 24 hours, non-infected participants were cleared and re-queued for the next invitation window.} Together, these results suggest that welfare-maximizing \textit{pooled} test allocation grants in-person access at no statistically detectable cost to the broader population.

\section{Discussion}
\label{section:discussion}
This work introduces a utility-based approach to pooled testing in resource-constrained environments. On the structural side, we show that the welfare loss from restricting attention to non-overlapping allocations is universally bounded by a factor of $2$. We provide examples establishing lower bounds of $\frac76$ for two tests and $\frac{19}{16}$ for three tests, and identify low-$q$ and high-$q$ regimes in which overlap yields no gain or admits sharper guarantees. These results complement the algorithmic guarantee that \greedy{} achieves at least a $\frac{1}{e+1}$ fraction of the welfare of the optimal overlapping allocation, thereby giving a general non-overlapping bound even without additional regime assumptions. We also establish the main complexity results for the problem: evaluating a given overlapping allocation is \#P-hard for $G \geq 3$, computing an optimal allocation is NP-hard in the non-overlapping regime for fixed $G \geq 3$, and computing an optimal single test is W[1]-hard with respect to~$G$. The NP-hardness extends to the overlapping regime, and in both regimes the multi-test problem admits no FPTAS unless $\mathrm{P}=\mathrm{NP}$. Despite these hardness results, our numerical experiments on real-world and synthetic data show that \greedy{} performs almost optimally and runs much faster than the MILP benchmark. %

Our field experiment complements these results with evidence of operational feasibility: the mechanism allocated scarce campus access through pooled qPCR testing in an institutional setting. Relative to unrestricted access, the randomized trial finds no statistically detectable adverse effects on five work, study, and well-being outcomes. 

There are many directions for future work. On the structural side, the exact values of $\gain(2)$ and $\gain(3)$ remain open: the current bounds are $\frac76\leq\gain(2)\leq2$ and $\frac{19}{16}\leq\gain(3)\leq2$ for $G\geq2$. More generally, the low-$q$ and high-$q$ factor-$2$ windows leave an intermediate strip whose exact behavior is not yet understood. For the approximation ratio of \greedy{}, a gap remains between our lower bound of $\frac{1}{e+1}$ and the upper bound of $\frac{1}{e}$ against the overlapping optimum. Beyond the tight approximation bound of $\frac{1}{e}$ for \vargreedy{} in the case of identical utilities, tighter guarantees may be achievable when utilities take a fixed number of values (e.g.,~for dichotomous or trichotomous populations).
On a practical level, the overall testing and re-integration policy we propose is static in nature, as we consider the one-shot setting where a testing budget is to be fully utilized by a policymaker. Testing could also be dynamic, with allocations chosen adaptively as a function of previous test results, and it is valuable to understand what potential benefits this extended functionality may bring. Additionally, policymakers potentially have access to different types of tests, each with different associated costs and performance (i.e., pool size and sensitivity), and providing optimal budget-constrained allocations in this heterogeneous test setting is a key open question. On the empirical side, future work should analyze how \greedy{}'s performance responds to population characteristics and input perturbations, and evaluate the deployed protocol in larger, longer trials across diverse epidemiological and institutional settings.

Most importantly, we hope that our insights into performance and efficacy of welfare-maximizing pooled testing can help better protect resource-constrained communities when confronted with an outbreak of an infectious disease. Our framework may find further, equally important applications in mass screening for HIV/AIDS and pooled frameworks for organ donation.

\section{Code and Data}
\label{section:data-code-availability}

The public replication package at \redacted{\url{https://github.com/michelleg06/pooled_testing/}} contains the pseudonymized participant-level baseline and endline analytic data, the R analysis code and the Julia source code, stored aggregate results and environment files for computational experiments 1--4. The enrollment and randomization rosters are not released because they contain direct identifiers and could link the pseudonymized records to named participants. The authors do not provide controlled or on-request access to these files because doing so could compromise participant confidentiality.

The replication data and code are provided solely to verify the findings reported in this paper. Any other use requires separate permission from the authors. The source code of the web application used to coordinate the field study and the experiment preregistration are listed in \cref{section:preregistration}.

\bibliographystyle{plainnat}
\bibliography{refs}

\appendix
\clearpage

\vspace*{15pt}

\begin{center}
{\Large\bfseries Appendix}
\end{center}

\vspace{15pt}

\section{An overview of computational complexity}
\label{appendix:complexity}

This appendix collects the complexity-theoretic terminology used in \cref{section:complexity-allocations}. It is intended as a quick reference for readers who have not seen these notions before; we omit formal models of computation throughout and refer to \citet{AroraBarak2009} or \citet{GareyJohnson1979} for textbook treatments.

A \emph{computational problem} is a specification mapping inputs to required outputs, such as ``given a graph $G$ and an integer $k$, decide whether $G$ contains an independent set of size at least $k$.'' Each particular pair $(G,k)$ is an \emph{instance} of the problem; the \emph{input} is the encoding of an instance as a string of bits passed to the algorithm. Throughout this appendix we illustrate the ideas using the \textsc{Independent Set} problem, which the \#P-hardness proof in \cref{prop:evaluation-complexity} reduces from. An \emph{independent set} in an undirected graph $G=(V,E)$ is a subset $S \subseteq V$ no two of whose vertices are joined by an edge.

\begin{problem}[\textsc{Independent Set}]
\label{problem:independent-set}
Given an undirected graph $G=(V,E)$ and an integer $k$, decide whether $G$ contains an independent set of size at least $k$. The optimization variant asks for an independent set of maximum size, and the counting variant $\#$\textsc{Independent Set} asks for the total number of independent sets in $G$.
\end{problem}

\subsection{Running time and input size}

The running time of an algorithm is measured by a \emph{worst-case} bound on the number of steps taken as a function of the input size. Throughout, input size means the number of bits required to encode the instance. Integers in the input are encoded in binary, so an integer $m$ contributes $O(\log m)$ bits and not $m$ bits. In \cref{problem:independent-set}, for instance, the parameter $k$ contributes only $O(\log k)$ bits to the input. Here and throughout, we use big-O notation, which suppresses constant factors and lower-order terms: $f(N)=O(g(N))$ means that $f(N)$ is bounded above by a constant multiple of $g(N)$ for all sufficiently large $N$. An algorithm whose running time is polynomial in $k$ can therefore be \emph{exponential} in the input size, whereas one whose running time is polynomial in $\log k$ is polynomial in the input size. An algorithm runs in \emph{polynomial time} if its running time is bounded by a fixed polynomial in the input size. Polynomial-time algorithms are considered tractable; algorithms whose running time grows like $2^{N}$ or worse, where $N$ denotes the input size, are considered intractable for all but the smallest instances.

For \textsc{Independent Set}, the brute-force algorithm that enumerates all $2^{|V|}$ subsets of $V$ and checks each one runs in time $2^{|V|} \cdot \mathrm{poly}(|V|)$ and is therefore exponential in $|V|$, hence in the input size. By contrast, enumerating all subsets of size at most~$k$ takes time $O(|V|^k)$, which is polynomial in $|V|$ whenever~$k$ is held fixed but blows up as $k$ grows. The same distinction appears in our setting: with $n$ denoting the population size, exhaustive search over tests of size at most $G$ runs in time $O(n^G)$, which is polynomial for each fixed $G$ but becomes prohibitive even for moderate $G$, as discussed in \cref{section:single-test-complexity}.

\subsection{Decision problems and reductions}
\label{section:complexity-reductions}

A \emph{decision problem} asks for a yes/no answer about an input; for example, ``does $G$ contain an independent set of size at least $k$?'' A \emph{search problem} asks for the object itself, and a \emph{counting problem} asks how many such objects exist. Most of the classes below are defined in terms of decision problems but extend to search and counting variants in the natural way.

A \emph{polynomial-time many-one reduction} from a decision problem $A$ to a decision problem $B$, written $A \leq_p B$, is a polynomial-time computable map $f$ from instances of $A$ to instances of $B$ such that $x$ is a yes-instance of $A$ if and only if $f(x)$ is a yes-instance of $B$. The intended reading is that $B$ is at least as hard as $A$: any polynomial-time algorithm for $B$ yields a polynomial-time algorithm for $A$ by first computing $f(x)$ and then running the algorithm for $B$. For search and optimization problems the definition is augmented by a second polynomial-time map $g$ that converts a solution of $f(x)$ back into a solution of $x$; the two maps, together with some additional structure, ensure that a polynomial-time solver for $B$ yields one for $A$. In the sections below, we define in more detail the specific reductions and structures required for the complexity classes we study.

A standard device used throughout this paper is to derive hardness of a search or optimization problem from hardness of an associated \emph{threshold decision problem}. Given a maximization problem with objective $u$, the threshold version asks whether there exists a feasible solution with $u(\cdot) \geq \tau$ for an input threshold $\tau$. Any algorithm that returns an optimal solution in polynomial time also decides the threshold problem in polynomial time, simply by comparing its output's objective value to $\tau$.

\subsection{The classes P and NP}

The class $\mathrm{P}$ consists of decision problems solvable in polynomial time. The class $\mathrm{NP}$ consists of decision problems for which a yes-instance $x$ admits a (polynomial-length) \emph{certificate} $y$, paired with a polynomial-time \emph{verifier} $V$ that decides whether $y$ proves the answer is yes (we write $V(x,y)=1$ in this case). For \textsc{Independent Set}, a candidate set $S \subseteq V$ of size at least $k$ serves as a certificate, and the verifier checks that $|S| \geq k$ and that no two vertices of $S$ are joined by an edge.

A decision problem is \emph{NP-hard} if every problem in $\mathrm{NP}$ reduces to it in polynomial time, and \emph{NP-complete} if it is both NP-hard and itself in $\mathrm{NP}$. \textsc{Independent Set} and \textsc{3-Partition} are NP-complete. %
The NP-hardness definition above uses \emph{polynomial-time many-one reductions} in the sense of \cref{section:complexity-reductions}. NP-hardness extends to non-decision problems in several equivalent ways: most generally via \emph{polynomial-time Turing reductions}, where a search, optimization, or counting problem is NP-hard if a polynomial-time algorithm for it would yield a polynomial-time algorithm for some NP-hard decision problem when used as a subroutine. The route used in this paper takes a more direct form: it passes through the \emph{threshold decision version} of the optimization problem. We first show that the threshold problem itself is NP-hard via a polynomial-time reduction from \textsc{3-Partition}, and then transfer this hardness to the optimization problem using the observation that an exact solver for the optimization problem decides the threshold problem in polynomial time. \Cref{theorem:np-completeness,theorem:overlapping-np-hard} are obtained in exactly this way; in the proof of \cref{theorem:np-completeness} the reduction is tight enough that the threshold version is itself in NP, yielding NP-completeness, whereas in the overlapping setting of \cref{theorem:overlapping-np-hard} we obtain NP-hardness without claiming membership in NP.

If any NP-hard problem had a polynomial-time algorithm, then every problem in $\mathrm{NP}$ would have one. The prevailing conjecture is $\mathrm{P}\neq\mathrm{NP}$, so NP-hardness of a problem is taken as evidence that no polynomial-time algorithm exists for it.

\subsection{Counting problems: \texorpdfstring{$\#\mathrm{P}$}{\#P} and PP}

The class $\#\mathrm{P}$ consists of counting problems of the form ``given an instance $x$ and a polynomial-time verifier $V$, how many certificates $y$ satisfy $V(x,y)=1$?''. A canonical $\#\mathrm{P}$-complete problem is $\#$\textsc{Independent Set}, which asks for the number of independent sets in a given graph. Counting is in general at least as hard as deciding existence, and it can be strictly harder.
The class $\mathrm{PP}$ is the decision analogue of $\#\mathrm{P}$: a problem is in $\mathrm{PP}$ if it can be phrased as ``given an instance $x$, a verifier $V$, and an efficiently computable threshold $\tau(x) \in (0,1)$, do at least a $\tau(x)$-fraction of all candidate certificates $y$ of a prescribed polynomial length satisfy $V(x,y)=1$?''. It is known that $\mathrm{NP} \subseteq \mathrm{PP}$, and $\mathrm{PP}$ is conjectured to be strictly larger.
The threshold variant of welfare evaluation studied in \cref{prop:evaluation-threshold-complexity} is $\mathrm{PP}$-complete, which is why the threshold version of the overlapping optimization problem lies in $\mathrm{NP}^{\mathrm{PP}}$ for fixed $G$: an $\mathrm{NP}$ machine guesses an allocation and consults a $\mathrm{PP}$ oracle to check its welfare against a threshold. For more detail on oracle machines, we refer to \cite{AroraBarak2009}.

$\#\mathrm{P}$-hardness is defined via \emph{polynomial-time Turing reductions}: a function problem is $\#\mathrm{P}$-hard if a polynomial-time algorithm for it would solve every problem in $\#\mathrm{P}$ in polynomial time when used as an oracle. $\mathrm{PP}$-hardness, on the other hand, follows the same many-one template as NP-hardness, since $\mathrm{PP}$ is a class of decision problems.

\subsection{Parameterized complexity: FPT and W[1]}

Many problems that are NP-hard in the worst case become tractable when a particular parameter of the input is small. Parameterized complexity formalizes this by treating the input as a pair $(x,k)$, where $k$ is the parameter. A problem is fixed-parameter tractable, or in $\mathrm{FPT}$, if it admits an algorithm with running time $f(k)\cdot\mathrm{poly}(|x|)$ for some computable function $f$.%
\footnote{To clarify, the parameter $k$ is a quantity expressing an intrinsic structural property of the instance $x$, and as such we allow runtime to depend on $k$ directly rather than its encoding length $|k|$.}
Importantly, the combinatorial explosion is confined to $f(k)$: for any fixed $k$, the running time is polynomial in the input size with a constant degree, in contrast to the weaker guarantee of an $|x|^{f(k)}$ algorithm.

This distinction is exemplified by two classical graph problems. \textsc{Vertex Cover} parameterized by solution size $k$ has algorithms that run in time $2^k\cdot\mathrm{poly}(|V|)$, and is therefore in $\mathrm{FPT}$. For \textsc{Independent Set} parameterized by solution size $k$, the straightforward brute force approach checks all $k$-vertex subsets and runs in time $O(|V|^k)$, whose polynomial degree grows with $k$, so the problem is not obviously fixed-parameter tractable.

The class $\mathrm{W}[1]$ captures parameterized problems, including parameterized \textsc{Independent Set}, for which no $f(k)\cdot\mathrm{poly}(|x|)$ algorithm is expected. Hardness for $\mathrm{W}[1]$ is defined via \emph{FPT (parameterized) reductions} rather than polynomial-time reductions. An FPT reduction from a problem $A$ with parameter $k$ to a problem $B$ with parameter $k'$ is a map from instances of $A$ to instances of $B$ that runs in time $f(k) \cdot \mathrm{poly}(|x|)$, preserves yes-answers, and bounds $k'$ by some computable function of $k$. Plain polynomial-time many-one reductions are too coarse, since they may blow the parameter up arbitrarily and so do not preserve fixed-parameter tractability.

A problem is $\mathrm{W}[1]$-hard if every problem in $\mathrm{W}[1]$ admits such a reduction to it. Unless $\mathrm{FPT}=\mathrm{W}[1]$, which is considered unlikely, a $\mathrm{W}[1]$-hard problem is not fixed-parameter tractable. Thus \cref{theorem:w1-hardness} shows that single-test optimization is unlikely to admit an $f(G)\cdot\mathrm{poly}(N)$ algorithm, where $G$ is the pool-size bound and $N$ is the input size.

\subsection{Approximation algorithms}

When exact polynomial-time algorithms are out of reach, one settles for approximate solutions. For a maximization problem with optimal value $\mathrm{OPT}$ and $\alpha \in (0,1]$, an \emph{$\alpha$-approximation algorithm} runs in polynomial time and returns a feasible solution of value at least $\alpha \cdot \mathrm{OPT}$. The greedy algorithm of \cref{section:approximation-algorithms} is a $\frac{1}{e+1}$-approximation in this sense.

A finer notion is an \emph{approximation scheme}, which takes $\varepsilon > 0$ as an additional input and returns a $(1-\varepsilon)$-approximate solution. A \emph{polynomial-time approximation scheme} (PTAS) runs in time polynomial in the input size for each fixed $\varepsilon$, and a \emph{fully polynomial-time approximation scheme} (FPTAS) runs in time polynomial in both the input size and $1/\varepsilon$. \textsc{Independent Set} on general graphs admits no constant-factor approximation under standard assumptions, while $0/1$-\textsc{Knapsack} admits a textbook FPTAS. The single-test problem of \cref{section:single-test-allocations} sits on the FPTAS side of this divide \citep{GOLDBERG2020}, whereas \cref{cor:inapproximability} rules out an FPTAS for the multi-test non-overlapping problem unless $\mathrm{P}=\mathrm{NP}$, which is why \cref{section:approximation-algorithms} settles for a constant-factor guarantee.

\section{Proofs for Section~\ref{section:gain-of-overlaps} (Performance of non-overlapping testing)}
\label{appendix:gain-of-overlaps-proofs}

Throughout this section, for any allocation $T$, let $P_i^T$ denote the probability that individual $i$ belongs to at least one negative test under $T$. We also write $q_S=\prod_{i\in S}q_i$.

For sets $X,Y,Z$, we write $X = Y \sqcup Z$ to mean that $X$ is the disjoint union of $Y$ and $Z$: that is, $X = Y \cup Z$ and $Y \cap Z = \emptyset$.

For a fixed population $(u,q)$, budget $B$, and pool size $G$, a feasible allocation $T^*$ is a \emph{size-minimal maximizer} if it maximizes $u(T)$ over all feasible allocations with at most $B$ tests and, among all such maximizers, minimizes the number of nonempty tests. The allocation is \emph{split-stable} if every test $t_j$ satisfies $q_A + q_C \geq 1$ for every partition $t_j=A \sqcup C$.

\begin{lemma}
\label{lemma:proof-two-split-stability}
\label{lemma:overlap-gain}
\hyperref[proof:lemma:proof-two-split-stability]{\normalfont[Proof in \cref{proof:lemma:proof-two-split-stability}]}
Any size-minimal maximizing test allocation $T^*$ is split-stable.
\end{lemma}

\begin{proof}[Proof of \cref{lemma:proof-two-split-stability}]
\phantomsection\label{proof:lemma:proof-two-split-stability}
Tests with at most one individual are trivially split-stable, since one side of any partition is empty and $q_\emptyset=1$. Otherwise, suppose for contradiction that $q_A+q_C<1$ for a partition $A\sqcup C$ of some test with at least two individuals; relabel so this test is last, i.e.\ $t_B^*=A\sqcup C$.

For an allocation $T=(t_1,\dots,t_B)$, let $P_{i,r}^T$ be the probability that $t_r$ is the first test, in the given order, that contains $i$ and is negative; then $P_i^T=\sum_r P_{i,r}^T$ and $u(T)=\sum_{r,i}u_iP_{i,r}^T$. Because individuals inside and outside $t_r$ are independent,
\[
P_{i,r}^T=q_{t_r}\,\pi_{i,r}(t_r;T),
\]
where
\[
\pi_{i,r}(D;T):=\Pr\!\left\{\text{every earlier test }t_s \ni i \text{ with } s < r \text{ has an unhealthy member in }t_s\setminus D\right\}.
\]
Informally, $\pi_{i,r}(D;T)$ is the probability that every test before $r$ containing $i$ carries an unhealthy witness outside $D$; the smaller $D$, the larger the window $t_s \setminus D$ for such a witness, so $\pi_{i,r}(\cdot;T)$ is monotone decreasing in $D$. For $D\subseteq t_B^*$, let $T^D$ denote $T^*$ with its last test replaced by $D$, and write
\[
X_D:=q_D\sum_{i\in D}u_i\,\pi_{i,B}(D;T^*),
\qquad
U_{<B}:=\sum_{r<B}\sum_i u_iP_{i,r}^{T^*}.
\]
Since the first $B-1$ tests are identical in $T^*$ and $T^D$, every earlier first-negative contribution is unchanged, so $u(T^D;u,q)=U_{<B}+X_D$. In particular, with $M_B:=q_Aq_C\sum_{i\in t_B^*}u_i\,\pi_{i,B}(t_B^*;T^*)$, we have $u(T^*;u,q)=U_{<B}+M_B$.

Minimality on the number of nonempty tests forces $M_B>0$: otherwise $T^\emptyset$ would preserve welfare while strictly reducing that count. For any $D\subseteq t_B^*$ and any earlier test $t_r$, $t_r\setminus t_B^*\subseteq t_r\setminus D$, so $\pi_{i,B}(t_B^*;T^*)\le\pi_{i,B}(D;T^*)$. Splitting the sum defining $M_B$ into $i\in A$ and $i\in C$ and applying this monotonicity gives
\[
M_B\le q_C X_A+q_A X_C.
\]
Since $M_B > 0$, at least one of $X_A,X_C$ is positive; combined with $q_A+q_C<1$,
\[
q_CX_A+q_AX_C\le(q_A+q_C)\max\{X_A,X_C\}<\max\{X_A,X_C\}.
\]
Choosing $D\in\{A,C\}$ that attains this maximum gives $u(T^D;u,q)>u(T^*;u,q)$, contradicting the maximality of $T^*$.
\end{proof}

\subsection{Proof of \texorpdfstring{\cref{theorem:universal-factor-two}}{the universal factor-two bound}}

For any allocation $T=(t_1,\dots,t_B)$, write the \textit{incidence set} of $i$ as $\sigma_T(i):=\{j\in[B]\mid i\in t_j\}$.

\begin{proof}[Proof of \cref{theorem:universal-factor-two}]
\phantomsection\label{proof:theorem:universal-factor-two}
To prove the statement, we start with a size-minimal maximizing test allocation $T^* = (t^*_1, \ldots, t^*_B)$ and construct a random non-overlapping test allocation $N$ from it, which is a discrete probability distribution over a finite number of (non-overlapping) allocations. We will show that 
\begin{equation}
\label{pivot-probability-bound}
\mathbb{E} [P_i^N] \geq \frac{1}{2} P_i^{T^*}, \quad \forall i \in [n].
\end{equation}

This claim allows us to complete the proof as follows. With a slight abuse of notation, we write $T \in N$ if $T$ is in the support of $N$. The welfare of $N$ is $u(N) = \sum_{T \in N} \Pr(T) u(T)$, implying the upper bound $u(N) \leq \opt(B)$. Secondly, writing $\mathbb{E}[P^N_i] = \sum_{T \in N} \Pr(T) P^T_i$, we see that $u(N) = \sum_{i \in [n]} u_i \mathbb{E}[P^N_i]$. Thus,
\[
u(T^*) = \sum_{i \in [n]} u_i P^{T^*}_i \leq 2 \sum_{i \in [n]} u_i \mathbb{E}[P^N_i] = 2 u(N) \leq 2 \opt(B).
\]

It now remains to prove \eqref{pivot-probability-bound}. First we construct the random allocation $N = (N_1, \ldots, N_B)$. For each individual $i$, let $d_i:=|\sigma_{T^*}(i)|$. Individuals with $d_i=0$ are never cleared by $T^*$, so we ignore them. If $d_i=1$, assign $i$ to its unique incident test $N_j$ with probability $1/2$, and leave it unassigned otherwise. If $d_i \geq 2$, assign $i$ to one of its $d_i$ incident tests uniformly. (These choices are made independently across individuals.) In every realization of $N$, the pools $N_j$ are disjoint and satisfy $N_j\subseteq t_j^*$, so the realization is feasible and non-overlapping.

Fix $i$ with $d_i\ge1$. All expectations in this proof are over the random construction of $N$. Whenever we condition on $i$ being assigned to an incident test $N_j$, write $H:=t_j^*\setminus\{i\}$, and let $p_{kj}$ be the probability that $k\in H$ is assigned to $N_j$. In this event, $N_j$ is the only test containing $i$, so $P_i^N=q_{N_j}$. Let $I_k$ be the indicator that $k$ is assigned to $N_j$, so
\[
q_{N_j}=q_i\prod_{k\in H} q_k^{I_k}.
\]
The assignment choices for distinct individuals are independent, so the random variables $q_k^{I_k}$, $k\in H$, are independent. Hence,
\[
\mathbb E \left[\prod_{k\in H}q_k^{I_k}\right]
=\prod_{k\in H}\mathbb E [q_k^{I_k}].
\]
For each $k$,
\[
\mathbb E [q_k^{I_k}]
=(1-p_{kj})+p_{kj}q_k
=1-p_{kj}(1-q_k).
\]
Thus
\begin{equation}
\label{eq:conditional-expectation}
\mathbb E\!\left[P_i^N\mid i\text{ is assigned to }N_j \right]
=q_i\prod_{k\in H}\bigl(1-p_{kj}(1-q_k)\bigr).
\end{equation}

If $d_i=1$, let $N_j$ be its unique incident test. Each factor in the product of the RHS above is at least $q_k$, so the conditional expectation is at least $q_{t_j^*}$. Since $i$ is assigned with probability $1/2$,
\[
\mathbb E [P_i^N] \geq \frac{1}{2}q_{t_j^*}
=\frac{1}{2}P_i^{T^*}.
\]

Now suppose $d_i\ge2$, and fix $j\in\sigma_{T^*}(i)$. Because the assignment choices of different individuals are independent, conditioning on $i$ being assigned to $N_j$ does not change the assignment probabilities of the individuals in $H$. For each $k \in H$, the probability $p_{kj}$ that $k$ is assigned to $N_j$ is at most $1/2$: it is $1/2$ if $k$ belongs only to $j$, and it is
$1/d_k \leq 1/2$ otherwise. Since $q_k \leq 1$, this gives
\[
    1-p_{kj}(1-q_k) \geq 1-\frac{1}{2}(1-q_k) =\frac{1+q_k}{2}.
\]
Substituting this bound into \eqref{eq:conditional-expectation} gives
\[
\mathbb E\!\left[P_i^N \mid i \text{ is assigned to } N_j \right]
\geq
q_i\prod_{k\in H}\frac{1+q_k}{2}.
\]
Finally,
\[
\prod_{k\in H}(1+q_k)
=\sum_{R\subseteq H}\prod_{k\in R}q_k
=\sum_{R\subseteq H}q_R,
\]
so 
\[
\mathbb E\!\left[P_i^N \mid i \text{ is assigned to } N_j \right] \geq q_i\,2^{-|H|}\sum_{R\subseteq H}q_R.
\]

As $T^*$ is a size-minimal maximizing test allocation by assumption, it satisfies split-stability by \cref{lemma:proof-two-split-stability}. For every $R\subseteq H$, split-stability applied to the partition $(\{i\}\cup R)\sqcup(H\setminus R)$ of $t_j^*$ gives $q_iq_R+q_{H\setminus R} \geq 1$. Summing over all $R\subseteq H$ yields
\[
(1+q_i)\sum_{R\subseteq H}q_R\ge 2^{|H|}.
\]
Therefore
\[
\mathbb E\!\left[P_i^N \mid i \text{ is assigned to } N_j \right] \geq \frac{q_i}{1+q_i} \geq \frac{q_i}{2} \geq \frac{1}{2} P^{T^*}_i.
\]
The last inequality follows from the fact that the overlapping clearance of $i$ is at most $q_i$, because $i$ must be healthy to belong to any negative test. This implies \eqref{pivot-probability-bound}.
\end{proof}

\subsection{Proof of \texorpdfstring{\cref{proposition:no-overlap-gain-below-half}}{the no-overlap-gain-below-half result}}

\begin{proof}[Proof of \cref{proposition:no-overlap-gain-below-half}]
\phantomsection\label{proof:proposition:no-overlap-gain-below-half}
Choose $T^*$ to be a size-minimal maximizing overlapping allocation. By \cref{lemma:proof-two-split-stability}, $T^*$ is split-stable. Suppose first the strict case that $q_i<1/2$ for every individual. We now prove that every test in $T^*$ is a singleton.

Suppose $t_j^*$ of $T^*$ is a test of size at least two. For any $i  \in t_j^*$, split-stability applied to $\{i\}\sqcup(t_j^*\setminus\{i\})$ gives $q_{t_j^*\setminus\{i\}}\ge 1-q_i$. But since all health probabilities are below $1/2$,
\[
q_{t_j^*\setminus\{i\}}\le \max_k q_k<1-\max_k q_k\le 1-q_i,
\]
a contradiction. It follows that all tests in $T^*$ are empty or singletons. The allocation $T^*$ is therefore non-overlapping, so $\opt^*(B,J) = \opt(B,J)$.

For the boundary case $q_i\le 1/2$, replace each health probability by $(1-\varepsilon)q_i$. The strict case gives equality of the overlapping and non-overlapping optima for every $\varepsilon>0$. Since both optima are maxima over finitely many allocations of continuous functions of the health probabilities, they are continuous in $q$. Letting $\varepsilon\downarrow0$ gives $\opt^*(B,J)=\opt(B,J)$, and hence $\gain(B,J)=1$.
\end{proof}

\subsection{Proof of \texorpdfstring{\cref{theorem:high-q-bound}}{the high-q bound}}

This proof follows the same approach as the proof of \cref{theorem:universal-factor-two}: we construct a random non-overlapping allocation $N$, show that $\mathbb{E}[P_i^N] \geq \alpha \, P_i^{T^*}$ for every individual $i$ and a constant $\alpha$, and conclude that some realization of $N$ has welfare at least $\alpha \, u(T^*)$. The proofs differ in the construction of $N$ and the constant $\alpha$. In \cref{theorem:universal-factor-two}, $\alpha = 1/2$ and $N$ is built by routing each individual to one of its incident tests, with the per-individual bound coming from split-stability of a size-minimal maximizer. In the following proof, $\alpha$ is a more complicated function depending on $G$, and $N$ is built from the first-appearance projection of $T^*$ by sampling a uniformly random $s_j$-sized subset of each pool, where $s_j$ is chosen to optimize the resulting per-individual coverage.

\begin{proof}[Proof of \cref{theorem:high-q-bound}]
\phantomsection\label{proof:theorem:high-q-bound}
Let $T^*$ be an optimal overlapping allocation and let $T=(t_1,\dots,t_B)$ be the first-appearance projection of $T^*$: each tested individual is kept only in the first test of $T^*$ in which it appears. The tests $t_j$ are pairwise disjoint and satisfy $t_j\subseteq t_j^*$. We construct a random non-overlapping allocation $N=(N_1,\dots,N_B)$ such that $\mathbb{E}[P_i^N]\ge P_i^{T^*}/\gamma_G(\rho)$ for every individual $i$. This gives $\mathbb{E}[u(N)]=\sum_i u_i\,\mathbb{E}[P_i^N]\ge u(T^*)/\gamma_G(\rho)$, so some realization $N$ has $u(N)\ge u(T^*)/\gamma_G(\rho)$, and hence $\gain(B,J)\le\gamma_G(\rho)$.

Define $\gamma_m(\rho):=\min_{s \in [m]} \frac{m}{s\rho^{s-1}}$. To see that this function is nondecreasing in $m$, note that, for each $s \in [m]$, the term $\frac{m+1}{s\rho^{s-1}}$ of $\gamma_{m+1}(\rho)$ exceeds the term $\frac{m}{s\rho^{s-1}}\ge\gamma_m(\rho)$ of $\gamma_m(\rho)$, while the new $s=m+1$ term of $\gamma_{m+1}(\rho)$ satisfies $\rho^{-m} \geq \rho^{1-m} \geq \gamma_m(\rho)$ since $\rho \leq 1$ and $\rho^{1-m}$ is the $s=m$ term of $\gamma_m(\rho)$.

We now construct $N$. For each nonempty $t_j$, let $m_j = |t_j|$ and choose $s_j \in [m_j]$ to be the argmin of $m_j / (s \rho^{s-1})$ over $s \in [m_j]$, so that $m_j / (s_j \rho^{s_j - 1}) = \gamma_{m_j}(\rho)$. Independently for each $j \in [B]$, set $N_j$ to a uniformly random $s_j$-sized subset of $t_j$ with probability $\gamma_{m_j}(\rho) / \gamma_G(\rho)$, and to $\emptyset$ otherwise; if $t_j$ is empty, set $N_j=\emptyset$. The pools $N_j$ are pairwise disjoint and satisfy $N_j\subseteq t_j^*$, so every realization of $N$ is feasible and non-overlapping.

The first-appearance pools are disjoint, each individual $i$ belongs to a unique $t_j$, and within $N$ can only be cleared via $N_j$, so $\mathbb{E}[P_i^N]=\Pr[i\in N_j]\cdot\mathbb{E}[q_{N_j}\mid i\in N_j]$. The probability that the sampled $N_j$ contains $i$ is $(\gamma_{m_j}(\rho) / \gamma_G(\rho)) \cdot s_j / m_j$, since a uniformly random $s_j$-subset of $t_j$ contains $i$ with probability $\binom{m_j - 1}{s_j - 1} / \binom{m_j}{s_j} = s_j / m_j$. Conditional on this event, $N_j$ is a uniformly random $s_j$-subset of $t_j$ containing $i$, and its other $s_j-1$ members each have health probability at least $\rho$, so $\mathbb{E}[q_{N_j}\mid i\in N_j]\ge q_i\rho^{s_j-1}$. Multiplying and using $s_j\rho^{s_j-1}/m_j=1/\gamma_{m_j}(\rho)$,
\[
\mathbb{E}[P_i^N]\ge \frac{\gamma_{m_j}(\rho)}{\gamma_G(\rho)}\cdot\frac{s_j}{m_j}\cdot q_i\rho^{s_j-1}=\frac{q_i}{\gamma_G(\rho)}\ge\frac{P_i^{T^*}}{\gamma_G(\rho)}.
\]
Here the last inequality uses $q_i\ge P_i^{T^*}$, as an individual must be healthy to be cleared.
\end{proof}

\section{Proofs for Section~\ref{section:complexity-allocations} (Complexity)}
\label{appendix:complexity-proofs}

Throughout this appendix we assume the input is finitely encoded: the health probabilities $q_i$, the utilities $u_i$, and any threshold $\tau$ are binary-encoded rationals, and the input size is the total number of bits in this encoding. While the model of \cref{section:model} permits arbitrary real $q_i$ and $u_i$, the membership and running-time claims below are stated relative to this standard rational encoding.

\subsection{Proof of \texorpdfstring{\cref{prop:evaluation-complexity}}{the welfare-evaluation complexity result}}

We first define computational problems that we will use to establish hardness of evaluating welfare of a given allocation.

\begin{problem}[Fixed-allocation evaluation]
Fix a pool-size bound $G$. The input is a population instance and a test allocation $T$ whose tests have size at most $G$. The task is to compute the welfare $u(T)$.
\end{problem}

\begin{problem}[\countis{}]
The input is a graph $H$. The task is to compute the number of independent sets of $H$, denoted $\#\mathrm{IS}(H)$. An independent set is a set of vertices containing no edge of $H$.
\end{problem}

\begin{proof}[Proof of \cref{prop:evaluation-complexity}]
\phantomsection\label{proof:prop:evaluation-complexity}
For the first claim, fix an individual $i$. If singleton test $\{i\}$ appears in $T$, then $i$ is cleared exactly when $i$ is healthy, so $P_i^T=q_i$. Otherwise, since every test has size at most $2$, every test containing $i$ is of the form $\{i,j\}$. Let
\[
N_T(i)=\{j\neq i:\{i,j\}\text{ appears in }T\}.
\]
Writing $X_k$ for the event that individual $k$ is healthy, we have
\[
P_i^T
=\Pr\!\left[X_i\wedge\bigvee_{j\in N_T(i)}X_j\right]
=q_i\Pr\!\left[\bigvee_{j\in N_T(i)}X_j\right]
=q_i\left(1-\prod_{j\in N_T(i)}(1-q_j)\right).
\]
Hence each $P_i^T$ can be computed in polynomial time, and so $u(T)=\sum_i u_iP_i^T$ can also be computed in polynomial time.

For the second claim, we reduce from \countis{}. Let $H=(V,E)$ be a graph with $V=\{1,\dots,n\}$. Introduce one special individual $s$ with $u_s=1$ and $q_s=1$. For each vertex $v\in V$, introduce an individual, again denoted $v$, with $u_v=0$ and $q_v=1/2$. For each edge $\{a,b\}\in E$, include the test $\{s,a,b\}$.

Since only $s$ has positive utility, $u(T)=P_s^T$. The individual $s$ is cleared if and only if at least one test containing $s$ is negative, i.e.
\[
P_s^T=\Pr\!\left[\bigvee_{\{a,b\}\in E}(X_a\wedge X_b)\right].
\]
Thus $P_s^T$ is the probability that the set of healthy vertices contains an edge. Since each $q_v=1/2$, every subset of $V$ occurs as the healthy set with probability $2^{-n}$. Therefore
\[
u(T)=P_s^T=1-\frac{\#\mathrm{IS}(H)}{2^n},
\]
where $\#\mathrm{IS}(H)$ is the number of independent sets of $H$. Computing $u(T)$ therefore determines $\#\mathrm{IS}(H)$, proving \#P-hardness.

The construction uses tests of size $3$, so the result holds for $G=3$. For $G>3$, the same allocation remains feasible because every test has size at most $G$.
\end{proof}

\subsection{Threshold evaluation and membership of the overlapping problem}
\label{proof:threshold-evaluation}

\begin{problem}[Threshold evaluation]
Fix a pool-size bound $G$. The input is a population instance, a possibly overlapping allocation $T$ whose tests have size at most $G$, and a rational threshold $\tau$. The task is to decide whether $u(T)\geq\tau$.
\end{problem}

\begin{problem}[\thresholdis{}]
The input is a graph $H$ and an integer $K$. The task is to decide whether $\#\mathrm{IS}(H)\leq K$.
\end{problem}

This problem is PP-complete: the usual threshold version with inequality $\#\mathrm{IS}(H)\geq K$ is PP-complete, and the direction of the inequality can be reversed because PP is closed under complement and the threshold can be shifted by one.

\begin{proposition}
\label{prop:evaluation-threshold-complexity}
\hyperref[proof:prop:evaluation-threshold-complexity]{\normalfont[Proof in \cref{proof:prop:evaluation-threshold-complexity}]}
Fix $G\geq 3$. Given a population instance, a possibly overlapping test allocation $T$ with tests of size at most $G$, and a rational threshold $\tau$, deciding whether $u(T)\geq\tau$ is PP-complete.
\end{proposition}

\begin{proof}[Proof of \cref{prop:evaluation-threshold-complexity}]
\phantomsection\label{proof:prop:evaluation-threshold-complexity}
We first prove membership in PP. Multiplying all utilities and the threshold by the same positive integer preserves the inequality $u(T)\geq\tau$. By choosing this integer to be the product of the denominators of the utilities and the threshold, we may assume without loss of generality that all $u_i$ and $\tau$ are integers. Write each health probability in lowest terms:
\[
q_j=\frac{a_j}{b_j}.
\]
Let $B_0=\prod_{j=1}^n b_j$, let $W=\max_i u_i$, and set $C=\max\{1,W\}$.

To express the expected welfare $u(T)$ as an integer count, introduce the auxiliary space $\Omega=\prod_{j=1}^n[b_j]$. Each $\omega=(\omega_1,\dots,\omega_n)\in\Omega$ encodes a health state in which individual $j$ is healthy if and only if $\omega_j\leq a_j$. This induces the product distribution $(q_1,\dots,q_n)$, and for each individual $i$, the number of outcomes in which $i$ is cleared equals $B_0P_i^T$. For each $S\subseteq[n]$, there are exactly $\prod_{j\in S}a_j\prod_{j\notin S}(b_j-a_j)$ outcomes $\omega\in\Omega$ for which $S$ is the set of healthy individuals. Hence
\[
B_0u(T)
=\sum_{\omega\in\Omega}\sum_{i=1}^n u_i\mathbf{1}[i\text{ is cleared under }T\text{ in the health state encoded by }\omega].
\]

We now show that the right-hand side is the number of accepting paths of a nondeterministic polynomial-time machine. The machine guesses fixed-length binary strings representing $\omega_1,\dots,\omega_n$, an index $i$, and an integer $c$, rejects if any represented value lies outside $\omega_j\in[b_j]$, $i\in[n]$, or $c\in[C]$, and accepts iff $c\leq u_i$ and individual $i$ is cleared in the health state encoded by $\omega$. For each fixed pair $(\omega,i)$, the accepting choices of $c$ are exactly $1,\dots,u_i$ if individual $i$ is cleared, and there are no accepting choices otherwise. Thus this pair contributes exactly
\[
u_i\mathbf{1}[i\text{ is cleared under }T\text{ in the health state encoded by }\omega]
\]
accepting paths. Summing over all choices of $\omega$ and $i$ gives precisely the displayed right-hand side. Hence $B_0u(T)$ is a \#P function. The integer $B_0\tau$ is computable in polynomial time and has polynomial bit length. By the standard characterization of PP as threshold comparison for \#P counts, since $u(T)\geq\tau$ iff $B_0u(T)\geq B_0\tau$, the threshold problem belongs to PP.

For hardness, reduce from \thresholdis{}. Let $(H,K)$ be an instance, where $H=(V,E)$ has $V=\{1,\dots,n\}$. Construct a population with one special individual $s$ satisfying $q_s=1$ and $u_s=1$, and one individual for each vertex $v\in V$ satisfying $q_v=1/2$ and $u_v=0$. Let the allocation $T$ contain the test $\{s,a,b\}$ for each edge $\{a,b\}\in E$. Then, as in the proof of \cref{prop:evaluation-complexity},
\[
u(T)=1-\frac{\#\mathrm{IS}(H)}{2^n}.
\]
Set $\tau=1-K/2^n$. Then
\[
u(T)\geq\tau
\iff
\#\mathrm{IS}(H)\leq K.
\]
Thus threshold evaluation is PP-hard.
\end{proof}

\begin{problem}[$G$-\overlap{}]
Fix a pool-size bound $G$. The input is a population instance, a budget $B$, and a rational threshold $\tau$. The task is to decide whether there exists a possibly overlapping allocation with at most $B$ tests, each of size at most $G$, whose welfare is at least $\tau$.
\end{problem}

\begin{corollary}
\label{cor:overlap-in-nppp}
\hyperref[proof:cor:overlap-in-nppp]{\normalfont[Proof in \cref{proof:cor:overlap-in-nppp}]}
For every fixed $G\geq 3$, $G$-\overlap{} belongs to $\mathrm{NP}^{\mathrm{PP}}$.
\end{corollary}

Equivalently, since PP and \#P are polynomial-time Turing equivalent, it belongs to $\mathrm{NP}^{\#\mathrm{P}}$.

\begin{proof}[Proof of \cref{cor:overlap-in-nppp}]
\phantomsection\label{proof:cor:overlap-in-nppp}
Because $G$ is fixed, a candidate allocation with $B$ tests of size at most $G$ has polynomial encoding length. We may therefore nondeterministically guess such an allocation and then invoke the PP procedure above to check whether its welfare is at least the threshold.
\end{proof}

\subsection{Shared lemmas for the reduction proofs}
\label{proof:complexity-shared-lemmas}

The single-test W[1]-hardness proof and the multi-test hardness proofs use the same one-dimensional welfare function. For $T\geq 2$, define
\[
f(x)=xr^x, \text{ with } r=\frac{2T^2-1}{2T(T+1)}.
\]

\begin{lemma}
\label{lemma:f-maximizer}
\hyperref[proof:lemma:f-maximizer]{\normalfont[Proof in \cref{proof:lemma:f-maximizer}]}
For integers $x\geq 1$, the function $f(x)=xr^x$ is uniquely maximized at $x=T$.
\end{lemma}

\begin{proof}[Proof of \cref{lemma:f-maximizer}]
\phantomsection\label{proof:lemma:f-maximizer}
We have
\[
\frac{f(x+1)}{f(x)}=r\frac{x+1}{x}.
\]
Hence $f(x+1)>f(x)$ iff $r>x/(x+1)$. Since $x/(x+1)$ is strictly increasing in $x$, and
\[
\frac{T-1}{T}<r<\frac{T}{T+1},
\]
the function $f$ increases up to $T$ and decreases afterwards. Thus $T$ is the unique maximizer.
\end{proof}

\begin{observation}
\label{observation:rT-bound}
For $T\geq 2$, we have $(T-1)/T<r<1$ and
\[
r^{T-1}>r^T>\left(1-\frac{1}{T}\right)^T\geq\frac14.
\]
\end{observation}

\begin{lemma}[Gap lemma]
\label{lemma:gap}
\hyperref[proof:lemma:gap]{\normalfont[Proof in \cref{proof:lemma:gap}]}
We have
\[
\Delta:=\min_{x\in\N,\,x\neq T}\bigl(f(T)-f(x)\bigr)\geq\frac{1}{8(T+1)}.
\]
\end{lemma}

\begin{proof}[Proof of \cref{lemma:gap}]
\phantomsection\label{proof:lemma:gap}
By \cref{lemma:f-maximizer}, the smallest gap from the maximum occurs at one of the neighbors:
\[
\Delta=\min\{f(T)-f(T-1),f(T)-f(T+1)\}.
\]
Since
\[
r-\frac{T-1}{T}=\frac{1}{2T(T+1)},\qquad
\frac{T}{T+1}-r=\frac{1}{2T(T+1)},
\]
we get
\[
f(T)-f(T-1)
=r^{T-1}\bigl(Tr-(T-1)\bigr)
=\frac{r^{T-1}}{2(T+1)}
\]
and
\[
f(T)-f(T+1)
=r^T\bigl(T-(T+1)r\bigr)
=\frac{r^T}{2T}.
\]
Using the preceding observation, both terms are at least the claimed bound after taking the minimum.
\end{proof}

\subsection{Proof of \texorpdfstring{\cref{theorem:w1-hardness}}{W[1]-hardness}}

\begin{problem}[\ksum{}]
An instance consists of nonnegative integers $a_1,\dots,a_n$, a target integer $T$, and a parameter $k$. The task is to decide whether there exists a set $S\subseteq[n]$ of cardinality exactly $k$ such that $\sum_{i\in S}a_i=T$.
\end{problem}

\begin{problem}[\rksum{}]
An instance consists of positive integers $b_1,\dots,b_n\geq 1$, a target integer $T\geq 2$, and a parameter $k\geq 1$. The task is to decide whether there exists a set $S\subseteq[n]$ of cardinality at most $k$ such that $\sum_{i\in S}b_i=T$.
\end{problem}

\begin{problem}[\singletest{}]
The input is a population instance, a pool-size bound $G$, and a rational threshold $\tau$. The task is to decide whether there exists a test $t$ of size at most $G$ with welfare at least $\tau$.
\end{problem}

We reduce from \ksum{} through \rksum{}.

\begin{lemma}
\label{lemma:rksum-hard}
\hyperref[proof:lemma:rksum-hard]{\normalfont[Proof in \cref{proof:lemma:rksum-hard}]}
\rksum{} is W[1]-hard parameterized by $k$.
\end{lemma}

\begin{proof}[Proof of \cref{lemma:rksum-hard}]
\phantomsection\label{proof:lemma:rksum-hard}
We reduce from \ksum{}, which is W[1]-hard parameterized by $k$. Let $(a_1,\dots,a_n,T,k)$ be an instance of \ksum{}; assume $k\geq 1$. Define
\[
L=2+\sum_{i=1}^n a_i,\qquad b_i=a_i+L,\qquad T'=T+kL.
\]
All $b_i$ are positive and $T'\geq 2$. Suppose $S$ is a solution to the constructed \rksum{} instance. Then $|S|\leq k$ and
\[
T+kL=\sum_{i\in S}b_i=\sum_{i\in S}a_i+|S|L<( |S|+1)L,
\]
since $\sum_{i\in S}a_i\leq\sum_{i=1}^n a_i<L$. Hence $k<|S|+1$, so $|S|\geq k$. Together with $|S|\leq k$, this gives $|S|=k$. For sets of cardinality $k$,
\[
\sum_{i\in S}b_i=T'
\iff
\sum_{i\in S}a_i=T.
\]
The reduction is polynomial time and preserves the parameter.
\end{proof}

Let $a_1,\dots,a_n,T,k$ be an instance of \rksum{}. We may assume that $a_i\leq T$ for all $i$, since any larger item cannot belong to a solution. The exact reduction would use utilities $u_i=a_i$, health probabilities $q_i=r^{a_i}$, threshold $Tr^T$, and pool-size bound $G=k$. Then a test $t$ has welfare $u(t)=a(t)r^{a(t)}=f(a(t))$, where $a(t)=\sum_{i\in t}a_i$. The following rounding argument shows that this construction can be implemented with rational probabilities of polynomial encoding length without changing the answer.
Set
\[
\lambda=\frac{1}{8(T+1)},\qquad \eta=\frac{1}{128kT(T+1)}.
\]
Let $p_i=r^{a_i}$. For each $i$, choose a dyadic rational $\tilde q_i$ satisfying $|\tilde q_i-p_i|\leq\eta/4$, and let $\widetilde u(t)=a(t)\prod_{i\in t}\tilde q_i$ denote welfare under the rounded probabilities.

\begin{lemma}
\label{lemma:stability}
\hyperref[proof:lemma:stability]{\normalfont[Proof in \cref{proof:lemma:stability}]}
For every test $t$ with $|t|\leq k$, we have $|\widetilde u(t)-u(t)|\leq\frac{\lambda}{4}$.
\end{lemma}

\begin{proof}[Proof of \cref{lemma:stability}]
\phantomsection\label{proof:lemma:stability}
By the observation above, $r^T\geq 1/4$. Since $a_i\leq T$, each $p_i=r^{a_i}\geq r^T\geq 1/4$. Write $\tilde q_i=p_i(1+\varepsilon_i)$. Then $|\varepsilon_i|\leq\eta$. For $|t|\leq k$,
\[
(1-\eta)^k\leq\prod_{i\in t}(1+\varepsilon_i)\leq(1+\eta)^k.
\]
Using $(1+\eta)^k\leq 1+2k\eta$, we obtain
\[
\left|\prod_{i\in t}\tilde q_i-\prod_{i\in t}p_i\right|
\leq 2k\eta\prod_{i\in t}p_i.
\]
Multiplying by $a(t)$ gives
\[
|\widetilde u(t)-u(t)|\leq 2k\eta u(t)\leq 2k\eta T=\frac{1}{64(T+1)}\leq\frac{\lambda}{4},
\]
where $u(t)\leq f(T)=Tr^T\leq T$.
\end{proof}

Let $g$ be a dyadic rational satisfying $|g-f(T)|\leq\lambda/8$, and set
\[
\tau=g-\frac{\lambda}{2}.
\]
In the following lemma, we use the rounded instance with utilities $u_i=a_i$, health probabilities $\tilde q_i$, pool-size bound $G=k$, and threshold $\tau$.

\begin{lemma}
\label{lemma:correctness-rounded}
\hyperref[proof:lemma:correctness-rounded]{\normalfont[Proof in \cref{proof:lemma:correctness-rounded}]}
The rounded instance with utilities $u_i=a_i$, health probabilities $\tilde q_i$, pool-size bound $G=k$, and threshold $\tau$ is a yes-instance if and only if the original \rksum{} instance is a yes-instance.
\end{lemma}

\begin{proof}[Proof of \cref{lemma:correctness-rounded}]
\phantomsection\label{proof:lemma:correctness-rounded}
If there is a test $t$ with $|t|\leq k$ and $a(t)=T$, then $u(t)=f(T)$, and by \cref{lemma:stability},
\[
\widetilde u(t)\geq f(T)-\frac{\lambda}{4}\geq g-\frac{3\lambda}{8}>g-\frac{\lambda}{2}=\tau.
\]
Conversely, if $a(t)\neq T$, then by \cref{lemma:gap},
\[
u(t)\leq f(T)-\Delta\leq f(T)-\lambda.
\]
Using \cref{lemma:stability} and the definition of $g$,
\[
\widetilde u(t)\leq f(T)-\lambda+\frac{\lambda}{4}
\leq g+\frac{\lambda}{8}-\frac{3\lambda}{4}
=g-\frac{5\lambda}{8}<\tau.
\]
Thus $\widetilde u(t)\geq\tau$ holds exactly when $a(t)=T$.
\end{proof}

The required approximation error is $\eta/4$. Let $M=\left\lceil \log_2(16T/\eta)\right\rceil=O(\log k+\log T)$. Round $r$ down to an $M$-bit dyadic rational, and compute each $\tilde q_i$ by repeated squaring, rounding every intermediate product down to $M$ fractional bits. All intermediate values lie in $[0,1]$. If $\widehat p_a$ denotes the resulting approximation to $r^a$, then induction on the exponentiation expression gives $0\leq r^a-\widehat p_a\leq(2a-1)2^{-M}$: the initial rounding of $r$ contributes at most $a2^{-M}$, and the at most $a-1$ rounded multiplications contribute at most a further $(a-1)2^{-M}$. Since $a_i\leq T$, this gives $|r^{a_i}-\tilde q_i|\leq(2T-1)2^{-M}<\eta/8<\eta/4$. Thus the rounded probabilities have polynomial encoding length and can be computed in time polynomial in the input length. Applying the same procedure with exponent $T$ and setting $g=T\widehat p_T$ gives $|g-f(T)|\leq T\eta/8\leq\lambda/8$.

\begin{proof}[Proof of \cref{theorem:w1-hardness}]
\phantomsection\label{proof:theorem:w1-hardness}
By \cref{lemma:rksum-hard,lemma:correctness-rounded}, there is an \textrm{FPT}-reduction from \rksum{} to \singletest{}. Since \rksum{} is W[1]-hard parameterized by $k$, \singletest{} is W[1]-hard parameterized by the pool-size bound $G$.
\end{proof}

\subsection{Proof of \texorpdfstring{\cref{theorem:np-completeness}}{NP-completeness}}

\begin{problem}[\threepart{}]
An instance consists of positive integers $a_1,\dots,a_{3m}$ and $T$ such that $\sum_{i=1}^{3m}a_i=mT$ and $a_i\in(T/4,T/2)$ for every $i$. The task is to decide whether the integers can be partitioned into $m$ triples, each summing to $T$. We use the strongly NP-complete version, in which the integers (not just their bit representations) are polynomially bounded in $m$.
\end{problem}

\begin{problem}[$G$-\nonoverlap{}]
Fix a pool-size bound $G$. The input is a population instance, a budget $B$, and a rational threshold $\tau$. The task is to decide whether there exists a non-overlapping allocation with at most $B$ tests, each of size at most $G$, whose welfare is at least $\tau$.
\end{problem}

\begin{lemma}
\label{lemma:welfare-TrT}
\hyperref[proof:lemma:welfare-TrT]{\normalfont[Proof in \cref{proof:lemma:welfare-TrT}]}
In the instance constructed below, every test $t$ has welfare at most $Tr^T$. It achieves welfare $Tr^T$ if and only if $a(t)=T$. In this case, $|t|=3$.
\end{lemma}

\begin{proof}[Proof of \cref{lemma:welfare-TrT}]
\phantomsection\label{proof:lemma:welfare-TrT}
The welfare of a test $t$ is $f(a(t))$, where $f(x)=xr^x$. By \cref{lemma:f-maximizer}, $f$ is uniquely maximized at $x=T$, so $u(t)=Tr^T$ if and only if $a(t)=T$. Since each \threepart{} integer satisfies $a_i\in(T/4,T/2)$, any set summing to $T$ must have exactly three elements.
\end{proof}

\begin{proof}[Proof of \cref{theorem:np-completeness}]
\phantomsection\label{proof:theorem:np-completeness}
Membership in NP is immediate because the welfare of a non-overlapping allocation can be computed by summing the welfare of its tests.

For hardness, reduce from \threepart{}. Given an instance $a_1,\dots,a_{3m},T$, construct a population with $u_i=a_i$, budget $B=m$, and
\[
r=\frac{1}{2}\left(\frac{T-1}{T}+\frac{T}{T+1}\right)
=\frac{2T^2-1}{2T(T+1)},\qquad q_i=r^{a_i}.
\]
Set the threshold to $\tau=mTr^T$. The construction has polynomial encoding length because \threepart{} is strongly NP-complete, so the integers $a_i$ and $T$ are polynomially bounded.

If the \threepart{} instance is a yes-instance, its triples define $m$ non-overlapping tests, each with $a(t)=T$. By \cref{lemma:welfare-TrT}, the allocation has welfare $mTr^T=\tau$.

Conversely, suppose there is a non-overlapping allocation with welfare at least $\tau$. Each test has welfare at most $Tr^T$ by \cref{lemma:welfare-TrT}. Since there are at most $m$ tests, achieving total welfare at least $mTr^T$ requires exactly $m$ tests, each with welfare $Tr^T$. Again by \cref{lemma:welfare-TrT}, each test has exactly three individuals and its utilities sum to $T$. The tests therefore form a valid \threepart{}.
\end{proof}

\subsection{Proof of \texorpdfstring{\cref{cor:inapproximability}}{the inapproximability corollary}}

\begin{proof}[Proof of \cref{cor:inapproximability}]
\phantomsection\label{proof:cor:inapproximability}
Let $A=mTr^T=mf(T)$. The proof of \cref{theorem:np-completeness} shows that yes-instances of \threepart{} correspond exactly to testing instances whose optimum is $A$.

Now consider a no-instance of \threepart{}. Any feasible allocation uses at most $m$ tests. If it uses fewer than $m$ tests, its welfare is at most $(m-1)f(T)=A-f(T)\leq A-\Delta$, where $\Delta<f(T)$ because $\Delta\leq f(T)-f(T+1)$ and $f(T+1)>0$. If it uses exactly $m$ tests, then not all tests can satisfy $a(t_i)=T$, since otherwise the allocation would give a valid \threepart{}. Hence at least one test has welfare at most
\[
\max_{x\in\N,\,x\neq T}f(x)=f(T)-\Delta,
\]
where, by \cref{lemma:gap},
\[
\Delta\geq\frac{1}{8(T+1)}.
\]
Therefore every no-instance has optimum at most $A-\Delta$.

Since $r<1$, we have $f(T)=Tr^T\leq T$, so $A\leq mT$. Hence
\[
\frac{\Delta}{A}\geq\frac{1}{8mT(T+1)},
\]
which is inverse-polynomial because $T$ is polynomially bounded in the strongly NP-complete version of \threepart{}.

Suppose there were an FPTAS for the non-overlapping problem. Run it on the constructed instance with
\[
\varepsilon=\frac{1}{16mT(T+1)}.
\]
This value is inverse-polynomial, so the running time remains polynomial. Since the welfare of a non-overlapping allocation can be computed exactly in polynomial time, we can evaluate the returned allocation. In yes-instances, the returned welfare is at least $(1-\varepsilon)A>A-\Delta/2$. In no-instances, every feasible allocation has welfare at most $A-\Delta<A-\Delta/2$. The algorithm would therefore decide \threepart{} in polynomial time, implying $\mathrm{P}=\mathrm{NP}$.
\end{proof}

\subsection{Proof of \texorpdfstring{\cref{theorem:overlapping-np-hard}}{NP-hardness with overlaps}}

This subsection concerns $G$-\overlap{}, defined in \cref{proof:threshold-evaluation}.

\begin{lemma}
\label{lemma:union-bound}
\hyperref[proof:lemma:union-bound]{\normalfont[Proof in \cref{proof:lemma:union-bound}]}
For any individual $i$ and any test allocation $T=(t_1,\dots,t_B)$,
\[
u(T)\leq\sum_{j=1}^B u(t_j).
\]
\end{lemma}

\begin{proof}[Proof of \cref{lemma:union-bound}]
\phantomsection\label{proof:lemma:union-bound}
Individual $i$ is cleared only if some test containing $i$ is negative. By Boole's inequality,
\[
P_i^T
=\Pr\!\left[\bigcup_{j:\,i\in t_j}\{t_j\text{ is negative}\}\right]
\leq\sum_{j:\,i\in t_j}q_{t_j}.
\]
Multiplying by $u_i$, summing over $i$, and interchanging the order of summation gives
\[
u(T)\leq\sum_{i=1}^n u_i\sum_{j:\,i\in t_j}q_{t_j}
=\sum_{j=1}^Bq_{t_j}\sum_{i\in t_j}u_i
=\sum_{j=1}^Bu(t_j).
\]
\end{proof}

\begin{lemma}
\label{lemma:overlap-bound}
\hyperref[proof:lemma:overlap-bound]{\normalfont[Proof in \cref{proof:lemma:overlap-bound}]}
Consider the testing instance constructed from a \threepart{} instance in the proof of \cref{theorem:np-completeness}. For any possibly overlapping allocation $T=(t_1,\dots,t_m)$ with $m$ tests of size at most $G$,
\[
u(T)\leq\sum_{j=1}^m f(a(t_j))\leq mTr^T.
\]
Moreover, if $u(T)=mTr^T$, then $T$ is non-overlapping and constitutes a valid \threepart{}.
\end{lemma}

\begin{proof}[Proof of \cref{lemma:overlap-bound}]
\phantomsection\label{proof:lemma:overlap-bound}
By \cref{lemma:union-bound},
\[
u(T)\leq\sum_{j=1}^m u(t_j)
=\sum_{j=1}^m a(t_j)r^{a(t_j)}
=\sum_{j=1}^m f(a(t_j)).
\]
By \cref{lemma:f-maximizer}, each term is at most $f(T)=Tr^T$, proving the first claim.

Now suppose $u(T)=mTr^T$. Then every inequality above is an equality. In particular, $f(a(t_j))=Tr^T$ for all $j$, so $a(t_j)=T$ and $|t_j|=3$ by \cref{lemma:welfare-TrT}. Furthermore, Boole's inequality in \cref{lemma:union-bound} must be tight for every individual. If some individual appeared in two tests, then the corresponding two negative-test events would each have probability $r^T\in(0,1)$ and would have positive-probability intersection, making Boole's inequality strict. Hence every individual appears in at most one test.

Since there are $m$ tests, each of size $3$, the total number of individual-test memberships is $3m$, equal to the population size. Thus every individual appears in exactly one test, and the tests partition the population into triples summing to $T$.
\end{proof}

\begin{proof}[Proof of \cref{theorem:overlapping-np-hard}]
\phantomsection\label{proof:theorem:overlapping-np-hard}
Use the same reduction from \threepart{} as in the proof of \cref{theorem:np-completeness}: $u_i=a_i$, $q_i=r^{a_i}$, $B=m$, and $\tau=mTr^T$.

If the \threepart{} instance is a yes-instance, the corresponding non-overlapping allocation achieves welfare $mTr^T$, so it is also a feasible overlapping allocation with welfare at least $\tau$.

Conversely, if there is an overlapping allocation $T$ with $u(T)\geq mTr^T$, then \cref{lemma:overlap-bound} implies $u(T)=mTr^T$ and that $T$ is a non-overlapping partition into triples summing to $T$. This certifies a yes-instance of \threepart{}.
\end{proof}

\begin{proof}[Proof of \cref{cor:overlapping-inapproximability}]
\phantomsection\label{proof:cor:overlapping-inapproximability}
Let $A=mTr^T=mf(T)$, and let $\Delta$ be as in \cref{lemma:gap}. We first show that every possibly overlapping allocation that does not constitute a valid \threepart{} has welfare at most $A-\Delta$.

Consider an allocation $\mathcal T$ with at most $m$ nonempty tests. If it has fewer than $m$ nonempty tests, then \cref{lemma:union-bound,lemma:f-maximizer} imply
\[
u(\mathcal T)\leq (m-1)f(T)=A-f(T)\leq A-\Delta.
\]
Suppose instead that it has $m$ nonempty tests. If $a(t_j)\neq T$ for some test $t_j$, then \cref{lemma:union-bound,lemma:gap} give
\[
u(\mathcal T)\leq\sum_{j=1}^m f(a(t_j))\leq (m-1)f(T)+f(T)-\Delta=A-\Delta.
\]

It remains to consider the case in which $a(t_j)=T$ for every $j$. By \cref{lemma:welfare-TrT}, every test then contains exactly three individuals. If the tests are non-overlapping, they form a valid \threepart{}. Otherwise, choose distinct tests $t_j,t_k$ and an individual $i\in t_j\cap t_k$. Let $E_\ell$ denote the event that test $t_\ell$ is negative. The slack in Boole's inequality for individual $i$ is at least
\[
\sum_{\ell:\,i\in t_\ell}\Pr(E_\ell)
-\Pr\!\left(\bigcup_{\ell:\,i\in t_\ell}E_\ell\right)
\geq \Pr(E_j\cap E_k).
\]
Since $\sum_{j=1}^m u(t_j)=A$, it follows that
\[
A-u(\mathcal T)
\geq u_i\Pr(E_j\cap E_k)
=a_i r^{a(t_j\cup t_k)}
>\frac{T}{4}r^{2T}.
\]
The restrictions in \threepart{} imply $T\geq3$. Moreover, the proof of \cref{lemma:gap} gives $\Delta\leq r^T/(2T)$, while $r^T>1/4$. Therefore
\[
\frac{(T/4)r^{2T}}{r^T/(2T)}
=\frac{T^2r^T}{2}
>\frac{T^2}{8}
\geq\frac98,
\]
so $A-u(\mathcal T)>\Delta$. This proves the claimed gap.

Suppose now that an FPTAS for the overlapping problem exists. Run it on the constructed instance with
\[
\varepsilon=\frac{1}{16mT(T+1)}.
\]
Because we use the strongly NP-complete version of \threepart{}, this choice preserves polynomial running time. On a yes-instance, \cref{lemma:overlap-bound} gives $\opt^*=A$. Since $A<mT$ and $\Delta\geq1/(8(T+1))$, the allocation $\widehat{\mathcal T}$ returned by the FPTAS satisfies
\[
u(\widehat{\mathcal T})\geq(1-\varepsilon)A>A-\frac{\Delta}{2}.
\]
By the gap proved above, $\widehat{\mathcal T}$ must itself be a non-overlapping partition into $m$ triples summing to $T$. This property can be checked directly in polynomial time. On a no-instance, no returned allocation can have this property. The FPTAS would therefore decide \threepart{} in polynomial time, implying $\mathrm{P}=\mathrm{NP}$.
\end{proof}

\subsection{\texorpdfstring{Polynomial-time tractability for $G=1,2$}{Polynomial-time tractability for G = 1, 2}}
\label{proof:low-G-tractability}

For $G=1$, the optimal allocation is obtained by sorting individuals by singleton welfare $q_iu_i$ and choosing the best $B$ of them.

For $G=2$, reduce to a maximum-weight matching problem with a cardinality constraint. Build a graph with one vertex for each individual and an edge $(i,j)$ of weight $u(\{i,j\})$ for each pair of individuals. To allow singleton tests, add a leaf vertex $\ell_i$ adjacent only to $i$, with edge weight $u(\{i\})=q_iu_i$. A matching with at most $B$ edges corresponds exactly to a non-overlapping allocation with at most $B$ tests of size at most $2$, and the matching weight is the allocation welfare. Conversely, every such allocation gives a matching of the same weight. Since maximum-weight matching with a cardinality constraint is solvable in polynomial time, the decision and optimization versions of the non-overlapping problem are polynomial-time solvable for $G=2$.

For overlapping allocations with $G=2$, fix an individual $i$. If $\{i\}$ appears, then $P_i^T=q_i$. Otherwise, every test containing $i$ is a pair $\{i,j\}$, so
\[
P_i^T=q_i\left(1-\prod_{j\in N_T(i)}(1-q_j)\right),
\]
as in the proof of \cref{prop:evaluation-complexity}. Thus any fixed overlapping allocation with tests of size at most $2$ can also be evaluated in polynomial time. This proves the evaluation claim for $G\leq 2$; the matching argument above proves optimization tractability for the non-overlapping problem.

\section{Proofs for Section~\ref{section:finding-allocations} (Finding Near-Optimal Test Allocations)}
\label{appendix:finding-allocations-proofs}

Throughout this appendix, for a population $R$ and integer $B \geq 0$, let
\[
\opt_B(R)=\max\{u(T): T \text{ is a feasible allocation in }R\text{ with at most } B \text{ tests}\}.
\]

\subsection{Proof of \texorpdfstring{\cref{prop:single-test-fptas}}{the single-test FPTAS}}
\label{appendix:proof-FPTAS}

We describe a fully polynomial-time approximation scheme (FPTAS) for optimally allocating a single test with pool size $G < n$. This FPTAS is obtained by modifying the FPTAS of \citet{GOLDBERG2020}, which finds an approximately optimal single test of arbitrary size. We use similar notation as \citet{GOLDBERG2020} where possible.

For $i\in\{0,\ldots,n\}$, $C\in\{0,\ldots,\bar{C}\}$, and $L\in\{0,\ldots,G\}$, let $P(i,C,L)=\max\bigl(\{q_S\mid S\subseteq[i],\ |S|=L,\ \sum_{\ell\in S}u_\ell=C\}\cup\{0\}\bigr)$, where $q_\emptyset=1$. Analogous to Eq.~(6) of \citep{GOLDBERG2020}, these values satisfy the recurrence
\begin{equation} \label{eq:FPTAS}
P(i,C,L)=
\begin{cases}
1 & \text{if } (i,C,L)=(0,0,0),\\
0 & \text{if } i=0 \text{ and } (C,L)\neq(0,0),\\
\max\{P(i-1,C,L),q_iP(i-1,C-u_i,L-1)\} & \text{if } i\geq1,\ L\geq1,\text{ and }u_i\leq C,\\
P(i-1,C,L) & \text{if } i\geq1\text{ and }(L=0\text{ or }u_i>C).
\end{cases}
\end{equation}
For every feasible state $(i,C,L)$, let $t(i,C,L)$ denote a subset of $[i]$ attaining $P(i,C,L)$, with total utility $C$ and cardinality $L$.

Let $\bar{C}$ be an upper bound on the sum of utilities of an optimal test ($\bar{C} = \sum_{i \in [n]} u_i$ suffices). It is straightforward that $t(n,C^*,L^*)$ is welfare-maximizing and has pool size at most $G$ whenever $(C^*,L^*)\in\argmax_{C\in\{0,\ldots,\bar{C}\},\,L\in\{0,\ldots,G\}} C\cdot P(n,C,L)$, where the maximum is restricted to feasible states.
The running time of this dynamic program is given by $O(n G \bar{C}) \leq O(n^2 \bar{C})$, since $G\leq n$. This is pseudo-polynomial in the input size as $\bar{C}$ is potentially exponential in the size of its representation.

In what follows, we scale and round the utilities so that the dynamic program runs in polynomial time. Fix $\varepsilon\in(0,1)$. For every individual $j$ with $u_j>0$, let $R_j=\{i\in[n]\setminus\{j\}:u_i\leq u_j\}$ and set $\kappa_j=\varepsilon u_j/n$. For each $i\in R_j$, define the scaled utility $\widehat u_i^{\,j}=\floor{u_i/\kappa_j}$. We apply \cref{eq:FPTAS} to the individuals in $R_j$, using the scaled utilities and the cardinality bound $G-1$. Denote the value of its final state $(C,L)$ by $\widehat P_j(C,L)$, and for every feasible state let $\widehat t_j(C,L)\subseteq R_j$ denote a corresponding set attaining this value. Finally, let $\overline C_j=\sum_{i\in R_j}\widehat u_i^{\,j}$ and define
\[
\widehat z_j(C,L)=(u_j+\kappa_j C)q_j\widehat P_j(C,L).
\]
The algorithm chooses a feasible state maximizing $\widehat z_j(C,L)$, adds $j$ to its reconstructed set, and returns the candidate with highest actual welfare.

\begin{algorithm}[t]
\caption{The FPTAS for computing a $(1-\varepsilon)$-approximate single-test allocation.}
\label{alg:FPTAS}
\begin{algorithmic}[1]
\State $z^*\gets0;\ t\gets\emptyset$
\For{$j\in[n]$ with $u_j>0$}
    \State $\kappa_j\gets\varepsilon u_j/n$ and $R_j\gets\{i\in[n]\setminus\{j\}:u_i\leq u_j\}$
    \State Compute $\widehat P_j(C,L)$ and corresponding sets $\widehat t_j(C,L)$ for the feasible states with $C\in\{0,\ldots,\overline C_j\}$ and $L\in\{0,\ldots,G-1\}$
    \State Choose $(\widehat C_j,\widehat L_j)\in\argmax_{(C,L)}\widehat z_j(C,L)$ over these feasible states
    \State $t_j\gets\widehat t_j(\widehat C_j,\widehat L_j)\cup\{j\}$
    \If{$u(t_j)>z^*$}
        \State $z^*\gets u(t_j);\ t\gets t_j$
    \EndIf
\EndFor
\State \Return $t$
\end{algorithmic}
\end{algorithm}

The following lemma gives the approximation guarantee for the candidate associated with the largest-utility individual in a comparison test.

\begin{lemma}
\label{lemma:FPTAS}
\hyperref[proof:lemma:FPTAS]{\normalfont[Proof in \cref{proof:lemma:FPTAS}]}
Let $t^*$ be any test with positive welfare, and choose $j\in t^*$ such that $u_j=\max_{i\in t^*}u_i$. Then the candidate $t_j$ constructed by \cref{alg:FPTAS} satisfies $u(t_j)\geq(1-\varepsilon)u(t^*)$.
\end{lemma}

\begin{proof}[Proof of \cref{lemma:FPTAS}]
\phantomsection\label{proof:lemma:FPTAS}
Let $U(S)=\sum_{i\in S}u_i$ and set $S^*=t^*\setminus\{j\}$. Since $j$ has maximum utility in $t^*$, we have $S^*\subseteq R_j$, and since $|t^*|\leq G$, the set $S^*$ uses at most $G-1$ positions. This also covers the case in which $t^*$ is a singleton, when $S^*=\emptyset$. Let $\widehat C^*=\sum_{i\in S^*}\widehat u_i^{\,j}$ and $L^*=|S^*|$. The state $(\widehat C^*,L^*)$ is feasible, so the definition of $\widehat P_j$ gives $\widehat P_j(\widehat C^*,L^*)\geq q_{S^*}$. Moreover,
\[
u_j+\kappa_j\widehat C^*\geq U(t^*)-|S^*|\kappa_j\geq U(t^*)-n\kappa_j=U(t^*)-\varepsilon u_j\geq(1-\varepsilon)U(t^*).
\]
By the choice of $(\widehat C_j,\widehat L_j)$, it follows that
\[
\begin{aligned}
\widehat z_j(\widehat C_j,\widehat L_j)
&\geq \widehat z_j(\widehat C^*,L^*)\\
&\geq (u_j+\kappa_j\widehat C^*)q_jq_{S^*}\\
&\geq (1-\varepsilon)U(t^*)q_{t^*}
=(1-\varepsilon)u(t^*).
\end{aligned}
\]
Finally, the actual utility sum of the reconstructed candidate is at least its scaled utility estimate, so
\[
\begin{aligned}
u(t_j)
&=U(t_j)q_j\widehat P_j(\widehat C_j,\widehat L_j)\\
&\geq (u_j+\kappa_j\widehat C_j)q_j\widehat P_j(\widehat C_j,\widehat L_j)
=\widehat z_j(\widehat C_j,\widehat L_j).
\end{aligned}
\]
Combining the two inequalities proves the claim.
\end{proof}

\begin{proof}[Proof of \cref{prop:single-test-fptas}]
\phantomsection\label{proof:prop:single-test-fptas}
Let $t^*$ be an optimal test. If $u(t^*)=0$, the empty test returned by the initialization is optimal. Otherwise, choose $j\in t^*$ with maximum utility. By \cref{lemma:FPTAS}, the candidate $t_j$ has welfare at least $(1-\varepsilon)u(t^*)$, and \cref{alg:FPTAS} returns a candidate with welfare at least $u(t_j)$.

It remains to bound the running time. For a fixed $j$ and every $i\in R_j$,
\[
\widehat u_i^{\,j}\leq\frac{u_i}{\kappa_j}\leq\frac{u_j}{\kappa_j}=\frac{n}{\varepsilon},
\]
so $\overline C_j\leq n^2/\varepsilon$. The dynamic program for a fixed $j$ therefore runs in time $O(nG\overline C_j)=O(n^4/\varepsilon)$, using $G\leq n$. Repeating it for at most $n$ choices of $j$ gives total running time $O(n^5/\varepsilon)$.
\end{proof}

\subsection{Proofs of \texorpdfstring{\cref{theorem:greedy-vs-overlapping,theorem:modified-greedy}}{the Greedy approximation guarantees}}
\label{proof:greedy-vs-overlapping}

We first prove a partition inequality for single tests. For this, we need the following technical lemma.

\begin{lemma}
\label{lemma:partition-analytic}
\hyperref[proof:lemma:partition-analytic]{\normalfont[Proof in \cref{proof:lemma:partition-analytic}]}
For $m \geq 2$, define
\[
G_m(x) \coloneqq \frac{1-\sum_{r=1}^m x_r^2}{\prod_{r=1}^m (1-x_r)}
\]
on the region $\mathcal{D} = \{x \in [0,1)^m : \sum_r x_r \leq 1\}$. Then
\[
G_m(x) \leq \left(\frac{m}{m-1}\right)^{m-1}
\qquad \text{for all } x \in \mathcal{D}.
\]
\end{lemma}

\begin{proof}[Proof of \cref{lemma:partition-analytic}]
\phantomsection\label{proof:lemma:partition-analytic}
We proceed in three steps. First, write $N=1-\sum_r x_r^2$ and $D=\prod_r(1-x_r)$. A direct computation gives
\[
\frac{\partial G_m}{\partial x_i}=\frac{N-2x_i(1-x_i)}{D(1-x_i)}.
\]
Since $\sum_{r\neq i}x_r\leq1-x_i$ and all coordinates are nonnegative, $\sum_{r\neq i}x_r^2\leq(1-x_i)^2$, so the numerator is nonnegative. Hence $G_m$ is coordinatewise nondecreasing on $\mathcal{D}$, and it suffices to maximize it on the simplex $\Delta=\{x\in[0,1)^m:\sum_r x_r=1\}$.

Second, we show that $G_m$ is Schur-concave on $\Delta$. Define
\[
H(x)=-\log G_m(x)=-\log\!\left(1-\sum_r x_r^2\right)+\sum_r\log(1-x_r).
\]
Writing again $N=1-\sum_r x_r^2$, for $i\neq j$ we have
\[
\frac{\partial H}{\partial x_i}-\frac{\partial H}{\partial x_j}
=(x_i-x_j)\left(\frac{2}{N}-\frac{1}{(1-x_i)(1-x_j)}\right).
\]
Using $\sum_r x_r=1$, a short calculation gives
\[
2(1-x_i)(1-x_j)-N
=\left(\sum_{r\neq i,j}x_r\right)^2+\sum_{r\neq i,j}x_r^2\geq0.
\]
Thus $H$ is Schur-convex, so $G_m$ is Schur-concave. It is therefore maximized at the uniform vector $\bar x=(1/m,\dots,1/m)$.

Finally,
\[
G_m(\bar x)
=\frac{1-m\cdot m^{-2}}{(1-1/m)^m}
=\left(\frac{m}{m-1}\right)^{m-1}.
\]

\end{proof}

\begin{proposition}
\label{prop:partition}
\hyperref[proof:prop:partition]{\normalfont[Proof in \cref{proof:prop:partition}]}
Let $t=P_1\sqcup\cdots\sqcup P_m$ be a partition of a feasible test $t$ into $m\geq1$ pieces satisfying
\[
u\!\left(\bigcup_{r\in A}P_r\right)\leq u(t)
\qquad\text{for every }A\subseteq\{1,\dots,m\}.
\]
Then
\[
\sum_{r=1}^m u(P_r)\leq e\,u(t).
\]
\end{proposition}

\begin{proof}[Proof of \cref{prop:partition}]
\phantomsection\label{proof:prop:partition}
The case $m=1$ is immediate. For $m\geq2$, it suffices to show the sharper bound
\[
\sum_{r=1}^m u(P_r)\leq u(t)\left(\frac{m}{m-1}\right)^{m-1}.
\]
If $u(t)=0$, then all pieces have zero utility and the claim is trivial. Otherwise rescale utilities so that $u(t)=1$.

Define $\alpha_r=q_{P_r}$, $w_r=\sum_{i\in P_r}u_i$, $x_r=1-\alpha_r$, and $W=\sum_r w_r$. Since $u(t)=1$, we have $W\prod_r\alpha_r=1$. Applying the hypothesis to the union of all pieces except $P_r$ gives $w_r\geq x_rW$ for each $r$, and hence $\sum_r x_r\leq1$. Therefore
\[
\sum_{r=1}^m u(P_r)
=W-\sum_r x_rw_r
\leq W\left(1-\sum_r x_r^2\right)
=\frac{1-\sum_r x_r^2}{\prod_r(1-x_r)}.
\]
The result follows from \cref{lemma:partition-analytic} and $\left(\frac{m}{m-1}\right)^{m-1}\leq e$.
\end{proof}

Let $R_j$ be the residual population at the start of round $j$ of exact \greedy{}, and let
\[
F_j=\opt_{B-j+1}(R_j).
\]
Thus $F_1$ is the optimal overlapping welfare with budget $B$, and $F_{B+1}=0$.

\begin{proposition}
\label{proposition:one-step-loss}
\hyperref[proof:proposition:one-step-loss]{\normalfont[Proof in \cref{proof:proposition:one-step-loss}]}
For every round $j\in[B]$ of exact \greedy{},
\[
F_j-F_{j+1}\leq(e+1)u(t_j).
\]
\end{proposition}

\begin{proof}[Proof of \cref{proposition:one-step-loss}]
\phantomsection\label{proof:proposition:one-step-loss}
Fix a round $j$ and set $L=B-j+1$. Let $T^*=(s_1,\dots,s_m)$ be an optimal allocation on $R_j$ using $m\leq L$ tests, so $u(T^*)=F_j$. Let $T'=(s'_1,\dots,s'_m)$ be the allocation obtained by truncating each test, $s'_\ell=s_\ell\setminus t_j$.

For individuals $i\notin t_j$, truncation can only increase clearance probability: every negative original test remains negative after deleting members. Individuals in $t_j$ are removed from every test in $T'$. Hence
\[
u(T^*)-u(T')\leq \sum_{i\in t_j}u_iP_i^{T^*}\leq\sum_{i\in t_j}u_iq_i=\sum_{i\in t_j}u(\{i\}).
\]
The singletons in $t_j$ partition $t_j$, and every union of them is a feasible test on $R_j$ with welfare at most $u(t_j)$ by exact greedy optimality. By \cref{prop:partition},
\[
u(T^*)-u(T')\leq e\,u(t_j).
\]

If $m<L$, then $T'$ already uses at most $L-1$ tests on $R_{j+1}$, so set $T''=T'$. If $m=L$, remove from $T'$ a test $s'_\ell$ whose removal causes the smallest welfare loss, and call the resulting allocation $T''$. For any removed test,
\[
u(T')-u(T'\setminus\{s'_\ell\})
\leq u(s'_\ell).
\]
Since $s'_\ell$ is a feasible single test on $R_j$, exact greedy optimality gives $u(s'_\ell)\leq u(t_j)$. Thus $u(T'')\geq u(T')-u(t_j)$ in all cases. As $T''$ is feasible on $R_{j+1}$ with at most $L-1$ tests,
\[
F_{j+1}\geq u(T'')\geq F_j-(e+1)u(t_j).
\]

\end{proof}

\begin{proof}[Proof of \cref{theorem:greedy-vs-overlapping}]
\phantomsection\label{proof:theorem:greedy-vs-overlapping}
Summing \cref{proposition:one-step-loss} over $j=1,\dots,B$ gives
\[
F_1-F_{B+1}\leq(e+1)\sum_{j=1}^B u(t_j).
\]
Since $F_1$ is the optimal overlapping welfare, $F_{B+1}=0$, and the greedy tests are non-overlapping, $\sum_j u(t_j)=u(T)$ for the allocation $T$ returned by \greedy{}. Therefore
\[
u(T)\geq\frac{1}{e+1}F_1,
\]
as claimed.
\end{proof}

For the modified guarantee, assume a single-test oracle that returns a test with welfare at least $(1-\varepsilon)\opt_1(R_j)$ in residual population $R_j$. The modified algorithm post-processes the oracle output by choosing the best subset of that output.

\begin{algorithm}[ht]
\caption{Modified $\varepsilon$-\greedy{}}\label{alg:modified-greedy}
\begin{algorithmic}[1]
\Require Population $R_1$, budget $B$, pool size $G$, approximation parameter $\varepsilon\in(0,1)$
\Ensure Non-overlapping allocation $T=(t_1,\ldots,t_B)$
\For{$j=1,2,\ldots,B$}
    \State Use the approximate oracle to obtain $r_j\subseteq R_j$ with $u(r_j)\geq(1-\varepsilon)\opt_1(R_j)$
    \State $t_j\gets\arg\max\{u(t):t\subseteq r_j\}$
    \State $R_{j+1}\gets R_j\setminus t_j$
\EndFor
\State \Return $T=(t_1,\ldots,t_B)$
\end{algorithmic}
\end{algorithm}

\begin{lemma}
\label{lem:subsetcorrected-properties}
\hyperref[proof:lem:subsetcorrected-properties]{\normalfont[Proof in \cref{proof:lem:subsetcorrected-properties}]}
For every round $j$, the chosen test $t_j$ satisfies $u(t_j)\geq(1-\varepsilon)\opt_1(R_j)$ and, for every subset $t\subseteq t_j$, $u(t)\leq u(t_j)$.
\end{lemma}

\begin{proof}[Proof of \cref{lem:subsetcorrected-properties}]
\phantomsection\label{proof:lem:subsetcorrected-properties}
Since $t_j$ is chosen as the best subset of $r_j$, $u(t_j)\geq u(r_j)\geq(1-\varepsilon)\opt_1(R_j)$. If $t\subseteq t_j$, then $t\subseteq r_j$, so the same subset maximization gives $u(t)\leq u(t_j)$.
\end{proof}

\begin{proof}[Proof of \cref{theorem:modified-greedy}]
\phantomsection\label{proof:theorem:modified-greedy}
Run modified $\varepsilon$-\greedy{} with $\varepsilon=\delta/(1+\delta)$, so $1/(1-\varepsilon)=1+\delta$.

The proof is identical to that of \cref{theorem:greedy-vs-overlapping}, with \cref{lem:subsetcorrected-properties} replacing exact optimality. For the truncation loss, the subset-domination property ensures that every union of pieces inside $t_j$ has welfare at most $u(t_j)$, so \cref{prop:partition} gives loss at most $e\,u(t_j)$. For the loss from discarding one test, the approximate optimality property gives
\[
\opt_1(R_j)\leq \frac{u(t_j)}{1-\varepsilon}=(1+\delta)u(t_j).
\]
Thus
\[
F_j-F_{j+1}\leq(e+1+\delta)u(t_j).
\]
Telescoping over all rounds gives
\[
u(T)\geq \frac{1}{e+1+\delta}\opt_B(R_1).
\]
The subset search costs $O(2^G)$ per round. With the single-test FPTAS as oracle, the total running time is $O(2^G\cdot\mathrm{poly}(n,B,1/\delta))$.
\end{proof}

\subsection{Proof of \texorpdfstring{\cref{theorem:greedy-approximation}}{the non-overlapping Greedy approximation guarantee}}

For any test allocation $T$, recall that $I^T_{i,j}$ is the indicator that takes value $1$ if individual $i$ is included in test $j$ of allocation $T$.

\begin{proof}[Proof of \cref{theorem:greedy-approximation}]
\phantomsection\label{proof:theorem:greedy-approximation}
Set the error tolerance of the single-test allocation problem to $\varepsilon = \frac{\delta}{9 + 2 \delta}$. Let $T = \{t_1, \ldots, t_B\}$ be the test allocation returned by \greedy{} using this single-test allocation algorithm, and let $T^* = \{t^*_1, \ldots, t^*_m\}$ be an optimal non-overlapping test allocation. Without loss of generality, we may assume that $m=B$ by padding $T^*$ with empty tests. Relabel the individuals so that $N'=\{1,\ldots,n'\}$ is the set of individuals included in $T$. We can write the welfares of $T$ and $T^*$ as
\begin{align*}
    u(T^*)= \sum_{j\in [B]} u(t^*_j)=  \sum_{j\in [B]} q_{t^*_j} \cdot  \left( \sum_{i\in N'}
     I^{T^*}_{i,j} \cdot u_i \right)
     +\sum_{j\in [B]} q_{t^*_j} \cdot  \left( \sum_{i\in [n]\setminus N'}
     I^{T^*}_{i,j} \cdot u_i \right),
\end{align*}
and
\begin{align*}
    u(T)= \sum_{j\in [B]} u(t_j)=
    \sum_{j\in [B]}q_{t_j}  \cdot \left(\sum_{i\in N'} I^T_{i,j} \cdot  u_i\right)=\sum_{j\in [B]}q_{t_j}  \cdot \left(\sum_{i\in t_j}  u_i\right).
\end{align*}

Now, let $T' = (t'_1, \ldots, t'_B)$ be the test allocation created from $T^*$ by removing any individuals in~$N'$, so $t'_j= t^*_j \setminus N'$ for every $j \in [B]$. Note that since every $t'_j$ consists of individuals that are not included in any test in $T$, all the individuals in $t'_j$ are available at the $j$-th iteration of \greedy{}. Thus for each $j \in [B]$, we have $u(t_j)\geq (1-\varepsilon) u(t'_j)$, as otherwise \greedy{} would have chosen $t'_j$ instead of $t_j$ at the $j$-th iteration. Thus, we get that
 \begin{align*}
    u(T)=\sum_{j\in [B]} u(t_j) \geq (1-\varepsilon) \sum_{j\in [B]} u(t'_j) = (1-\varepsilon) \cdot u(T').
 \end{align*}
Note also that
\begin{align*}
    u(T')=  \sum_{j \in [B]}  q_{t'_j}  \cdot  \left( \sum_{i \in [n]\setminus N'} I^{T'}_{i,j} \cdot u_i \right)
    &= \sum_{j \in [B]}  q_{t'_j} \cdot \left( \sum_{i \in [n]\setminus N'} I^{T^*}_{i,j}   \cdot u_i \right)\nonumber
    \geq  \sum_{j \in [B]}  q_{t^*_j} \cdot \left( \sum_{i \in [n]\setminus N'} I^{T^*}_{i,j}   \cdot u_i \right)
\end{align*}
where the second equality follows from the fact that $I^{T'}_{i,j}=I^{T^*}_{i,j}$ for any $i \in [n]\setminus N'$  and $j\in [B]$, and  the last inequality follows from the fact that   $q_{t'_j}\geq q_{t^*_j}$  since $t'_j \subseteq t^*_j$ for any $j \in [B]$. Thus,
\begin{align*}
    u(T)\geq  (1-\varepsilon)  \cdot \sum_{j \in [B]}  q_{t^*_j} \cdot \left( \sum_{i \in [n]\setminus N'} I^{T^*}_{i,j}   \cdot u_i \right).
\end{align*}

From all the above we have
\begin{align}
    \frac{u(T^*)}{u(T)}
     &\leq \frac{\sum_{j\in [B]} q_{t^*_j} \cdot  \left( \sum_{i\in t^*_j \cap N'} u_i \right)
      }{   u(T)} +\frac{1}{1-\varepsilon}\nonumber  \\
    &\leq \frac{\sum_{j\in [B]}  \left( \sum_{i\in t^*_j \cap N'} q_{i} \cdot  u_i \right)
      }{   u(T)} +\frac{1}{1-\varepsilon}\nonumber  \\
       &= \frac{\sum_{i \in N'}   q_{i} \cdot  u_i
      }{   u(T)} +\frac{1}{1-\varepsilon}. \label{ineq:greedy}
\end{align}

In what follows, we will show that each test $t_j \in T$ obtains at least a $\frac{(1-\varepsilon)^2}{4}$ ratio of the maximal possible utility to be gained from individuals in $t_j$. First, there cannot exist $S \subseteq t_j$ such that $q_S + q_{t_j \setminus S} < 1 - \varepsilon$. If such an $S$ existed, approximate optimality would imply both $u(t_j) \geq (1-\varepsilon)u(S)$ and $u(t_j) \geq (1-\varepsilon)u(t_j\setminus S)$, while the identity
\[
u(t_j)=q_{t_j\setminus S}u(S)+q_Su(t_j\setminus S)
\]
would imply $u(t_j)<(1-\varepsilon)\max\{u(S),u(t_j\setminus S)\}$, a contradiction.

Now let us consider the case where there exists $i' \in t_j$ such that $q_{i'}<(1-\varepsilon)/2$. Approximate optimality gives
\[
(1-\varepsilon)u(t_j\setminus\{i'\})\leq u(t_j),\qquad
(1-\varepsilon)u(\{i'\})\leq u(t_j),
\]
and the preceding paragraph gives $q_{t_j\setminus\{i'\}}\geq(1-\varepsilon)/2$. Thus
\[
u(t_j)\geq\frac{(1-\varepsilon)^2}{4}\sum_{i\in t_j}q_iu_i.
\]
In the remaining case, every $\ell\in t_j$ has $q_\ell\geq(1-\varepsilon)/2$. One similarly shows that $q_{t_j\setminus\{i\}}\geq(1-\varepsilon)^2/4$ for every $i\in t_j$, and therefore
\[
u(t_j)=\sum_{i\in t_j}q_{t_j\setminus\{i\}}q_iu_i
\geq\frac{(1-\varepsilon)^2}{4}\sum_{i\in t_j}q_iu_i.
\]
Summing over greedy tests gives
\[
u(T)\geq\frac{(1-\varepsilon)^2}{4}\sum_{i\in N'}q_iu_i.
\]
Together with \cref{ineq:greedy},
\[
\frac{u(T^*)}{u(T)}
    \leq \frac{4}{(1-\varepsilon)^2} + \frac{1}{1 - \varepsilon}
    < 5 + \frac{9\varepsilon}{1-2\varepsilon}
    = 5+\delta,
\]
by the choice $\varepsilon=\delta/(9+2\delta)$. Thus the theorem follows.
\end{proof}

\subsection{Details on the MILP formulation}
\label{appendix:milp-details}

We describe how to approximate the exponential constraints and reformulate the logarithmic constraints from the mixed-integer program in \cref{section:milp-formulation}.

\paragraph{Handling the logarithmic constraints.}
We can replace \eqref{constr:log} with integer linear constraints as follows. Fix some test $j \in [B]$. Note that $z^j$ takes integral values in the range $[L, U]$, where $L = \min_i u_i$ and $U = G\max_i u_i$. We introduce an indicator vector $\gamma^j \in \{0,1\}^{[L,U]}$ indexed by $k \in [L,U]$ with constraints $\sum_{k \in [L,U]} \gamma^j_k = 1$ and $\sum_{k \in [L,U]} k \cdot \gamma^j_k = z^j$ to encode which value $z$ holds, and ensure $y^j = \log(z^j)$ with the constraint $y^j = \sum_{k \in [L,U]} \log (k) \cdot \gamma^j_k$.

\paragraph{Approximating the exponential constraints.}
We now describe how to approximate \eqref{constr:exp} from above by a piecewise-linear function $f$ using integer linear constraints. Fix some test $j \in [B]$. Note first that we can relax the equality in \eqref{constr:exp} to $w^j \leq \exp(l^j)$ without affecting the outcome. The variable $l^j$ takes values between $A = \min_i(\log u_i) + G \min_i (\log q_i)$ and $C = \log(G \max_i u_i) + \max_i (\log q_i)$. We approximate $\exp$ from above by a piecewise-linear function $f : [A,C] \to \mathbb{R}$ with $K$ linear segments. Partitioning $[A,C]$ into $K$ parts $[c_k, c_{k+1}]$, $k \in [K]$, we define the $k$-th line segment as the linear function $f_k(x) = a_k x + b_k$ on domain $[c_k, c_{k+1}]$ with slope $a_k = \frac{\exp c_{k+1} - \exp c_k}{c_{k+1} - c_k}$ and residual $b_k = \exp c_{k+1} - a_k c_{k+1}$.
We introduce indicator vectors $\delta^j \in \{0,1\}^K$ to encode in which part $[c_k, c_{k+1}]$ the value of $l^j$ lies, as well as the vector $v^j \in \mathbb{R}^K$ whose $k$-th entry agrees with $l^j$ if $l^j$ lies in the $k$-th part, and is $0$ otherwise. This is guaranteed by constraints $\sum_{k \in [K]} \delta^j_k = 1$, $l^j = \sum_{k \in [K]}v^j_k$ and $c_k \cdot \delta^j_k \leq v^j_k \leq c_{k+1} \cdot \delta^j_k$ for all $k \in [K]$. Finally, we require that $w^j \leq f_k(l^j)$ for the $k$-th part $[c_k, c_{k+1}]$ that $l^j$ lies in. This is expressed by constraint $w^j \leq \sum_{k \in [K]} a_k v^j_k + b_k \cdot \delta^j_k$.

\paragraph{Bounding the approximation error.}
Recall that the piecewise-function $f$ with $K$ segments approximates $\exp$ on domain $[A,C]$ from above with error $\varepsilon$. Let $\sigma(x) = \sum_{j \in [B]}\exp(l^j)$ and $\sigma'(x) = \sum_{j \in [B]}f(l^j)$ respectively denote the corresponding objective values of the convex program \eqref{opt:unpacked-problem} and the MILP described above for testing $x$. Let $x^*$ denote an optimal non-overlapping allocation, so $x^*$ maximizes $\sigma$, and $x'$ be an optimal solution for the MILP. Clearly, $x^*$ and $x'$ are both feasible for both programs and satisfy $\sigma(x') \leq \sigma(x^*)$ as well as $\sigma'(x^*) \leq \sigma'(x')$. By construction of $f$, we have $\sigma(x) \leq \sigma'(x)$ and $\sigma(x) \geq \sigma'(x) - \varepsilon B$, which implies $\sigma(x^*) \leq \sigma'(x^*) \leq \sigma'(x') \leq \sigma(x') + \varepsilon B$. Hence, $0 \leq \sigma(x^*) - \sigma(x') \leq \varepsilon B$. This allows us to compute a bound on the additive gap between the welfare achieved by the optimal solution of our MILP and the optimal non-overlapping testing.

\section{Additional Theoretical Results for Section~\ref{section:finding-allocations}}

\subsection{Homogeneous Utilities}
\label{appendix:homogeneous-utilities}
In this section, we consider the special case in which all the individuals have the same utility, so $u_i=u_{i'}$ for all $i,i'\in [n]$. Without loss of generality, assume that $u_i=1$ for any $i \in [n]$ and $q_{i} \geq q_{i+1}$ for any $i \in [n-1]$.
The following example demonstrates that \greedy{} may produce suboptimal allocations (though necessarily bounded in their suboptimality via \cref{theorem:greedy-approximation}). In the example, all individuals have the same utility and health probability; the health probability is chosen such that \greedy{} will exhaust the entire population with a single test, rendering the remaining $B-1$ tests useless. In the limit, as the population size goes to infinity, the example shows that \greedy{} can perform a factor of $e$ worse than optimal overlapping testing.

\begin{example}
\label{example:greedy-performance}
Consider a population of $n$ individuals with homogeneous utilities $u=1$ and probabilities $q = \frac{n-1}{n}$, a testing budget of $n$, and $G \geq n$. Clearly, the optimal testing strategy is to test every person individually, which yields an expected welfare of $nq = n-1$. The \greedy{} algorithm tests the entire population with its first test, as $n q^n \geq kq^k$ for all $k \leq n$. Hence \greedy{} achieves an expected welfare of $nq^n=n(\frac{n-1}{n})^n$, implying an approximation ratio of $\left(\frac{n-1}{n}\right)^{n-1}\to\frac{1}{e}$ as $n\to\infty$.
\end{example}

We design a variant of \greedy{} which sorts the individuals in decreasing order with respect to the probability of being healthy and, in each step, adds individuals to the current test as long as the expected utility of the test increases. In \cref{appendix:greedy-algorithm}, we show that this \vargreedy{} algorithm returns a $\frac{1}{e}$-approximate non-overlapping test allocation. When applying \vargreedy{} to \cref{example:greedy-performance}, all individuals are also pooled into a single test, hence this shows that the approximation ratio achieved by \vargreedy{} is tight.

\begin{proposition}
\label{prop:homogeneous-utilities-vargreedy}
\hyperref[proof:prop:homogeneous-utilities-vargreedy]{\normalfont[Proof in \cref{proof:prop:homogeneous-utilities-vargreedy}]}
If all individuals have the same utility, \vargreedy{} returns a $\frac{1}{e}$-approximate non-overlapping test allocation.
\end{proposition}

Suppose that $T$ is a test allocation returned by \vargreedy{} which tests individuals $[n'] \subseteq [n]$. The proof of approximate optimality of $T$ rests on showing that if an individual $i$ is included in test $t_j$ of size $|t_j| = k$, then $q_i \geq (k-1)/k$. This implies that $q_{t_j} \geq q_i \left( \frac{k-1}{k} \right)^{k-1} \geq q_i \frac{1}{e}$. It follows that $u(T) \geq \frac{1}{e} \sum_{i \in [n']} q_i \cdot u_i$. On the other hand, by \cref{lem:ident-utilities,lem:ident-util-greedy-1}, there exists an optimal non-overlapping test allocation $T^*$ that tests a prefix $[n'']\subseteq[n']$. Hence, $u(T^*)\leq\sum_{i\in[n'']}q_i u_i\leq\sum_{i\in[n']}q_i u_i$. %

\subsubsection{Structural Lemmas for Optimal Test Allocations}

In the context of homogeneous utilities, we say that a test allocation $T = (t_1,\dots,t_B)$ is \textit{proper} if the following hold:
\begin{itemize}
    \item $|t_1| \geq |t_2| \geq \dots \geq |t_B|$;
    \item for all $j,j' \in [B]$ such that $j < j'$, if $i \in t_j$ and $i' \in t_{j'}$, then $i < i'$.
\end{itemize}

The following crucial lemma shows that there exist proper optimal test allocations, similar in nature to the main structural results of \citet{aprahamian2019optimal} and \citet{Lipnowski2021pooled}.

\begin{lemma}
\label{lem:ident-utilities}
\hyperref[proof:lem:ident-utilities]{\normalfont[Proof in \cref{proof:lem:ident-utilities}]}
Let $T^* = (t^*_1,\dots,t^*_B)$ be an optimal non-overlapping test allocation such that $|t^*_j|\geq |t^*_{j+1}|$ for any  $j\in [B-1]$. Then, there exists a proper optimal non-overlapping test allocation $T'$ such that for each $j\in [B]$,  $|t'_j| =|t^*_{j}|$.
\end{lemma}
\begin{proof}[Proof of \cref{lem:ident-utilities}]
\phantomsection\label{proof:lem:ident-utilities}
First we show that without loss of generality, we can assume that $T^*$ tests precisely $[k] \subseteq [n]$ for some $k \leq n$. Suppose that this is not the case, and there exist $i,i' \in [n]$ such that $i < i'$  where $i$ is not tested in $T^*$ yet $i'$ is tested in $t_j$. It is straightforward to see that the test allocation which replaces $i'$ with $i$ in $t_j$ obtains at least as much utility as $T^*$ and is hence optimal by assumption.

We prove the lemma by induction on the number of tests. We start from the case where $B=2$. Let  $T^*$ be an optimal non-overlapping test allocation with $|t^*_1|\geq |t^*_2|$. Assume that $t^*_1= S_1 \cup S'_1$, where $S_1 \subset \{1,\ldots, |t^*_1|\} $,  and $S'_1 \subset \{|t^*_1|+1, \ldots, k\}$ and  $t^*_2= S_2 \cup S'_2$, where $S_2 \subset \{1,\ldots, |t^*_1|\} $ and $S'_2 \subset \{|t^*_1|+1, \ldots, k\}$. Since $t^*_1\cup t^*_2=[k]$, we get that $|S'_1|=|S_2|$ and since $|t^*_1|\geq |t^*_2|$, we get that  $|S_1|\geq |S'_2|$. Now, consider the test allocation $T$ such that $t_1=S_1 \cup S_2$ and $t_2=S'_1 \cup S'_2$. Notice that  $t_1\cup t_2=[k]$, $|t_1|=|t^*_1|$ and $|t_2|=|t^*_2|$. Then, we have
\begin{align*}
    u(T^*)= q_{S_1} \cdot  q_{S'_1} \cdot |t^*_1| +
      q_{S_2}\cdot  q_{S'_2} \cdot  |t^*_2|
\end{align*}
and
\begin{align*}
    u(T)=  q_{S_1}\cdot q_{S_2} \cdot |t^*_1| + q_{S'_1}\cdot  q_{S'_2}  \cdot  |t^*_2|,
\end{align*}
and hence,
\begin{align}
\label{ineq:lem-ident-utilities}
 u(T)-u(T^*) =
    \left(  q_{S_2}- q_{S'_1}\right) \cdot  \left( q_{S_1} \cdot |t^*_1|
 -  q_{S'_2} \cdot  |t^*_2| \right).
\end{align}
Due to optimality of $T^*$, we have that for any  $\hat{S}_1\subseteq S_1 $
\begin{align*}
   q_{S_1}\cdot  q_{S'_1} \cdot |t^*_1| \geq
  q_{ \hat{S}_1} \cdot   q_{ S'_1} \cdot (|\hat{S}_1|+|S'_1|)
\end{align*}
as otherwise if $T'$ is a test allocation with $t'_1=\hat{S}_1 \cup S'_1$ and $t'_2=t^*_2$, then it would hold that $u(T')> u(T^*)$ which is a contradiction to the optimality of $T^*$. Now, choose an arbitrary $\hat{S_1} \subseteq S_1$ such that $|\hat{S}_1|=|S'_2|$. We know that this is feasible since $|S_1|\geq |S'_2|$. We know that $q_{S_1}\cdot q_{S'_1} \cdot |t^*_1| \geq q_{\hat{S}_1}\cdot q_{S'_1} \cdot |\hat{S}_1 \cup S'_1| = q_{ \hat{S}_1} \cdot  q_{S'_1} \cdot |t^*_2|$, where we also used the fact that $|\hat{S}_1|=|S'_2|$. Finally, by construction $q_{ \hat{S}_1} \geq q_{S'_2}$, hence it follows that $q_{S_1}\cdot q_{S'_1} \cdot |t^*_1| \geq q_{ S'_2} \cdot  q_{S'_1} \cdot |t^*_2|$. We thus obtain
\begin{align*}
   q_{S_1}\cdot  |t^*_1|  -   q_{S'_2}  \cdot |t^*_2| \geq 0.
\end{align*}
Now from \cref{ineq:lem-ident-utilities}, we have that $u(T) \geq u(T^*)$ since $ \left(  q_{S_2}- q_{S'_1}\right)\geq 0$ as for each  $i \in S_2$ and each $i' \in S'_1$ it holds that $q_i \geq q_{i'}$ and $|S'_1|=|S_2|$. Thus, we conclude that if $T^*$ is optimal, so is $T$, which is in turn a proper test allocation as desired.

Now, suppose that the claim holds for $B-1$. We will show that it holds for $B$. Let  $T^* = (t^*_1,\dots,t^*_B)$ be an optimal non-overlapping test allocation with $|t^*_j|\geq |t^*_{j+1}|$ for any $j\in [B-1]$. It must be the case that the allocation $(t^*_1,\dots,t^*_{B-1})$ is optimal for the population given by $\cup_{j = 1}^{B-1} t^*_j$, hence we can apply our inductive assumption to obtain a proper optimal test allocation $T' = (t'_1,\dots,t'_{B-1})$ for $\cup_{j = 1}^{B-1} t^*_j$ such that $|t'_j| = |t^*_j|$. Let $T^0 = (t'_1,\dots,t'_{B-1},t^*_B)$.

We create a proper test allocation for the tested population $[k]$ in $B-1$ rounds. For each $\ell\in[B-1]$, we begin with an optimal test allocation $T^{\ell-1}=(t^{\ell-1}_1,\dots,t^{\ell-1}_B)$ such that $|t^{\ell-1}_j| = |t^*_j|$ for all $j \in [B]$ and $(t^{\ell-1}_1,\dots,t^{\ell-1}_{\ell-1})$ is an optimal proper test allocation for $\cup_{j = 1}^{\ell-1} t^{\ell-1}_j = \{1,\dots, \sum_{j=1}^{\ell-1} |t^*_j|\}$.
We consider the allocation $(t^{\ell-1}_{\ell},t^{\ell-1}_B)$ which must be optimal for the population given by $t^{\ell-1}_{\ell} \cup t^{\ell-1}_B$. We can apply the lemma for the case of $B = 2$ to obtain an optimal proper test allocation for $t^{\ell-1}_{\ell} \cup t^{\ell-1}_B$ which is given by $(t^{\ell}_{\ell},t^{\ell}_B)$ such that $|t^{\ell-1}_{\ell}| = |t^{\ell}_{\ell}|$ and $|t^{\ell-1}_B| = |t^{\ell}_B|$. For $j \neq \ell,B$, we let $t^\ell_j = t^{\ell-1}_{j}$.
It follows that $T^{\ell}$ is such that $|t^\ell_j| = |t^*_j|$ for all $j \in [B]$ and $(t^{\ell}_1,\dots,t^{\ell}_{\ell})$ is an optimal proper test allocation for $\cup_{j = 1}^{\ell} t^{\ell}_j = \{1,\dots, \sum_{j=1}^{\ell} |t^*_j|\}$. Given the construction, it follows that $T^{B-1}$ is a proper optimal test allocation such that $|t^{B-1}_j|=|t^*_j|$ for all $j\in[B]$, as desired.
\end{proof}

\begin{observation}
 Using \cref{lem:ident-utilities}, we can find an optimal non-overlapping test allocation as follows: for all $k\in [n]$ and $k_1 \geq k_2 \geq \ldots \geq k_B$  with $\sum_{\ell \in [B]}k_\ell=k$, we calculate the welfare of the test allocation, $T = (t_1,\dots,t_B)$, such that
 \begin{align*}
     t_j= \Big \{ \sum_{\ell \in [j-1]}k_{\ell} +1, \ldots  \sum_{\ell \in [j-1]}k_{\ell} + k_j \Big \}.
 \end{align*}
The allocation which returns the highest welfare amongst the $O(n^{B+1}/B!)$ choices of $k$ and $k_1,\dots,k_B$ values must necessarily be optimal.
\end{observation}

\subsubsection{The \vargreedy{} Algorithm}
\label{appendix:greedy-algorithm}
Here, we show that when the utilities are identical,  we can find an  $\frac{1}{e}$-approximate test allocation with respect to the optimal non-overlapping test allocation, for any value $B$. This result is tight, as shown in \cref{example:greedy-performance}. Once more, we assume a population, $[n]$, such that $u_i = 1$ for all $i$ and $q_1 \geq q_2 \geq \dots \geq q_n$. Specifically, we consider a variation of \greedy{} introduced in~\cref{section:single-test-allocations} which we call \vargreedy{}. The algorithm proceeds in $B$ rounds. In the $j$-th round it computes a test $t_j$ composed of individuals not yet tested in previous rounds as follows: it sequentially adds the untested individual of highest health probability to $t_j$ whenever doing so weakly increases the welfare of $t_j$, and stops when the next addition would strictly reduce welfare or when $t_j$ reaches size $G$. Note that \vargreedy{} always returns a proper test allocation, $T$, which tests $[n'] \subseteq [n]$ within the population.
We start with the following lemma.

\begin{lemma}
\label{lem:ident-util-greedy-1}
\hyperref[proof:lem:ident-util-greedy-1]{\normalfont[Proof in \cref{proof:lem:ident-util-greedy-1}]}
Suppose that $T^* = (t^*_1,\dots,t^*_B)$ is a proper optimal non-overlapping test allocation and that $T$ is a proper test allocation computed by \vargreedy{}. If $T^*$ tests $[n'] \subseteq [n]$ within the population and $T$ tests $[n''] \subseteq [n]$ within the population, then $n' \leq n''$.
\end{lemma}
\begin{proof}[Proof of \cref{lem:ident-util-greedy-1}]
\phantomsection\label{proof:lem:ident-util-greedy-1}
Without loss of generality, pad both $T^*$ and $T$ with empty tests so that each allocation has exactly $B$ tests, placing the empty tests last. Since the allocations are proper, we may write $t^*_j=\{i^*_{j-1}+1,\ldots,i^*_j\}$ and $t_j=\{i_{j-1}+1,\ldots,i_j\}$, where $0=i^*_0\leq i^*_1\leq\cdots\leq i^*_B=n'$ and $0=i_0\leq i_1\leq\cdots\leq i_B=n''$, and a test is interpreted as empty when its two consecutive boundaries are equal. We show that $i^*_j\leq i_j$ for each $j\in[B]$. Suppose for contradiction that $j$ is the first index such that $i^*_j>i_j$. Then $i^*_{j-1}\leq i_{j-1}$, and hence $|t^*_j|>|t_j|$. Moreover, $t_j$ is nonempty: under the tie-breaking convention above, \vargreedy{} starts another nonempty test whenever an untested individual remains, so an empty $t_j$ would imply $i_j=n$, contradicting $i^*_j>i_j$. If $|t_j|=G$, then $|t^*_j|>|t_j|=G$ contradicts the feasibility of $t^*_j$. We may therefore assume that $1\leq|t_j|<G$. %
Given that \vargreedy{} does not pool $i_{j}+1$ in  $t_j$, we have that
\begin{align*}
    q_{i_{j-1}+1} \cdot \ldots \cdot q_{i_{j}} \cdot |t_j| &> q_{i_{j-1}+1} \cdot \ldots \cdot q_{i_{j}} \cdot q_{i_{j}+1} \cdot  \left( |t_j|+1 \right)\\
    \Rightarrow \frac{|t_j|}{|t_j|+1} &>  q_{i_{j}+1}.
\end{align*}
as otherwise, from the definition of \vargreedy{}, $i_j+1$ would have been included in $t_j$.

We also note that,
\begin{align*}
q_{i^*_{j}} <  \frac{|t_j|}{|t_j|+1} \leq   \frac{|t^*_j|-1}{|t^*_j|}.
\end{align*}
The first inequality holds because $i^*_j \geq i_j + 1$ and health probabilities are in decreasing order, so $q_{i^*_j} \leq q_{i_j + 1} < \frac{|t_j|}{|t_j|+1}$. For the second inequality, $|t^*_j|>|t_j|$ implies $|t^*_j|-1\geq|t_j|$, and the function $f(x)=\frac{x-1}{x}$ is increasing in $x$. Thus, we have
\begin{align*}
   & q_{t^*_j \setminus \{i^*_j\}}  \cdot  ( |t^*_j|-1 )  >   q_{t^*_j \setminus \{i^*_j\}} \cdot q_{i^*_j} \cdot   |t^*_j|.
\end{align*}
This means that $u(t^*_j \setminus \{i^*_j \})> u(t^*_j)$. Hence, if $T'$ is the test allocation with $t'_{j'}=t^*_{j'}$ for each $j'\neq j$ and $t'_j=t^*_j \setminus \{i^*_j\}$, then $u(T')>u(T^*)$, contradicting the optimality of $T^*$. We conclude that $i^*_j\leq i_j$ for each $j\in[B]$. In particular, $n'=i^*_B\leq i_B=n''$, and the statement follows.
\end{proof}

Now, we are ready to show that \vargreedy{} returns a $\frac{1}{e}$-approximate test allocation for any population. This guarantee is tight, as shown in \cref{example:greedy-performance}.
\begin{theorem}
\label{theorem:homogeneous-utilities-vargreedy}
\hyperref[proof:theorem:homogeneous-utilities-vargreedy]{\normalfont[Proof in \cref{proof:theorem:homogeneous-utilities-vargreedy}]}
\vargreedy{} returns a $\frac{1}{e}$-approximate test allocation.
\end{theorem}
\begin{proof}[Proof of \cref{theorem:homogeneous-utilities-vargreedy}]
\phantomsection\label{proof:prop:homogeneous-utilities-vargreedy}\label{proof:theorem:homogeneous-utilities-vargreedy}
Let $T$ be the test allocation that is returned by \vargreedy{} which tests the first $n'\leq n$ individuals. We start by showing that for each $i$, $P_i^T\geq q_i \cdot \frac{1}{e}$. Consider a test $t_j$ in $T$ of size $k$ and let $i'$ be the final individual included in $t_j$.  It must be the case that $q_{i'}\geq (k-1)/k$, for otherwise
\begin{align*}
    q_{i'} \prod_{i'' \in t_j\setminus\{i'\}} q_{i''} \cdot  k <  \prod_{i'' \in t_j\setminus\{i'\}} q_{i''} \cdot (k-1)
\end{align*}
which is a contradiction. Since $i'$ was assumed to be the final individual added to $t_j$ in \vargreedy{}, it follows that for all $i \in t_j$, $q_i \geq q_{i'} \geq (k-1)/k$. With this in hand, we get that
\begin{align}
   P^T_i= q_{t_j}=q_i \cdot \prod_{i' \in t_j\setminus\{i\}} q_{i'}\geq q_i \cdot \left(\frac{k-1}{k}\right)^{k-1} \geq q_i \cdot\frac{1}{e}. \label{eq:app-greedy}
\end{align}

By \cref{lem:ident-utilities,lem:ident-util-greedy-1}, there exists a proper optimal non-overlapping test allocation $T^*$ that tests the first $n''$ individuals, where $n''\leq n'$. Then, %
\begin{align*}
    \begin{aligned}\frac{u(T^*)}{u(T)}&=\frac{\sum_{i\in[n'']}P_i^{T^*}u_i}{\sum_{i\in[n']}P_i^Tu_i}\\&\leq\frac{\sum_{i\in[n'']}q_i u_i}{\sum_{i\in[n']}q_i\frac{1}{e}u_i}\\&\leq\frac{\sum_{i\in[n']}q_i u_i}{\sum_{i\in[n']}q_i\frac{1}{e}u_i}=e.\end{aligned} %
\end{align*}
The first inequality uses $P_i^{T^*}\leq q_i$ and \cref{eq:app-greedy}, while the second uses $n''\leq n'$. %
\end{proof}

\subsection{Optimality of \greedy{} for Clustered Populations}
\label{appendix:greedy-cluster}

We consider the scenario in which individuals can only exhibit utilities and probabilities from a finite set of values. Specifically, we assume that the population at hand can be partitioned into $m$ disjoint clusters, $C_1,\dots C_m$, where $C_\ell$, has $n_\ell$ individuals with identical utility, $u_\ell$, and probability of infection $p_\ell$ (probability of health $q_\ell$). Since individuals are indistinguishable within a cluster, we can identify a test $t$ with the number of individuals included from each cluster. We let $t(\ell)$ denote the number of individuals from cluster $C_\ell$ included in $t$. We also let $C = [m]$ be the set of cluster indices. Suppose that $t^*$ is a single test that achieves optimal utility and that $t$ is a near-optimal test with $u(t) \geq (1-\varepsilon) u(t^*)$. We can show that if the population at hand is such that $t$ can be repeated $B$ times, then not only does \greedy{} return this allocation, but it also obtains $(1-\varepsilon)$ welfare of what is obtained by an optimal overlapping test allocation.

In order to incorporate clustering into the MILP, we now let the index $i$ refer to a cluster (instead of an individual), and allow variables $x^j_i$ to take arbitrary non-negative integral values (instead of binary values in \eqref{constr:xvars}); these values represent the number of individuals from cluster $i$ that are included in test~$j$. Additionally, we relax the non-overlapping test constraint in \eqref{constr:non-overlapping} to $\sum_{j \in [B]} x^j_i \leq n_i$. As an aside, it is not difficult to show that if cluster populations are much larger than the testing budget at hand, then non-overlapping tests are optimal.
We use the notation and linearization from \cref{appendix:milp-details}. In particular, we assume that the cluster utilities are positive integers and that $q_i>0$ for every $i\in[m]$, set $L=\min_{i\in[m]}u_i$ and $U=G\max_{i\in[m]}u_i$, let $[L,U]$ denote the integers from $L$ to $U$, and, for a fixed $K\in\mathbb{N}$, use the $K$-segment upper piecewise-linear approximation of the exponential function specified there. The logarithmic constraints below then encode $y^j=\log z^j$ exactly on the positive integer range $[L,U]$. We now state the full MILP with clustering below. Note that constraints \eqref{constr:cluster-exp-start}--\eqref{constr:cluster-exp-end} capture the exponential constraint \eqref{constr:exp}, while \eqref{constr:cluster-log-start}--\eqref{constr:cluster-log-end} capture the logarithmic constraint \eqref{constr:log}.

\begin{maxi!}
  {}{\sum_{j \in [B]} w^j}{\label{opt:clustered-MILP}}{}
  \addConstraint{w^j}{\leq \sum_{k \in [K]}a_k v^j_k + b_k \cdot \delta^j_k}{\quad \forall j \in [B]}{\label{constr:cluster-exp-start}}
  \addConstraint{\sum_{k \in [K]} \delta^j_k}{=1}{\quad \forall j \in [B]}
  \addConstraint{\sum_{k \in [K]} v^j_k}{=l^j}{\quad \forall j \in [B]}
  \addConstraint{c_k \cdot \delta^j_k}{\leq v^j_k}{\quad \forall j \in [B], k \in [K]}
  \addConstraint{c_{k+1} \cdot \delta^j_k}{\geq v^j_k}{\quad \forall j \in [B], k \in [K]} {\label{constr:cluster-exp-end}}
  \addConstraint{l^j}{ = y^j + \sum_{i \in [C]} x^j_i \log q_i}{\quad \forall j \in [B]}
  \addConstraint{1}{ = \sum_{k \in [L,U]} \gamma^j_k}{\quad \forall j \in [B]}{\label{constr:cluster-log-start}}
  \addConstraint{z^j}{= \sum_{k \in [L,U]} k \cdot \gamma^j_k}{\quad \forall j \in [B]}
  \addConstraint{y^j}{= \sum_{k \in [L,U]} \log (k) \cdot \gamma^j_k}{\quad \forall j \in [B]}{\label{constr:cluster-log-end}}
  \addConstraint{z^j}{= u \cdot x^j,}{\quad \forall j \in [B]}
  \addConstraint{\sum_{j \in [B]} x^j_i}{\leq n_i,}{\quad \forall i \in [C]}
  \addConstraint{\sum_{i \in [C]} x^j_i}{\geq 1,}{\quad \forall j \in [B]}
  \addConstraint{\sum_{i \in [C]} x^j_i}{\leq G,}{\quad \forall j \in [B]}
  \addConstraint{x^j_i}{\in \mathbb{N}_0,}{\quad \forall j \in [B], i \in [C]}
  \addConstraint{v^j_k}{\in \mathbb{R},}{\quad \forall i \in [C], k \in [K]}
  \addConstraint{\delta^j_k}{\in \{0,1\},}{\quad \forall i \in [C], k \in [K]}
  \addConstraint{\gamma^j_k}{\in \{0,1\},}{\quad \forall i \in [C], k \in [L,U]}
\end{maxi!}

\begin{proposition}
\label{prop:greedy-cluster-optimal}
\hyperref[proof:prop:greedy-cluster-optimal]{\normalfont[Proof in \cref{proof:prop:greedy-cluster-optimal}]}
Suppose that $t$ is a $(1-\varepsilon)$-optimal test and $B \cdot t(\ell) \leq n_\ell$ holds for each $\ell \in [m]$. Then \greedy{} returns an approximately optimal allocation that applies $B$ tests with the same composition as~$t$. This allocation obtains $(1-\varepsilon)$ welfare of what is obtained in an optimal overlapping test allocation.
\end{proposition}

The intuition behind this result is simple. Suppose that $T^* = (t^*_1,\dots,t^*_B)$ is an optimal overlapping test allocation and that $t^*$ is an optimal single pooled test for the population. Writing welfare as the sum of marginal utility obtained from each test when $T^*$ is applied sequentially, we can show that each marginal utility is bounded by $u(t^*)$, and so $u(T^*) \leq B \cdot u(t^*)$. On the other hand, $u(t) \geq (1-\varepsilon) u(t^*)$ and the fact that $B \cdot t(\ell) \leq n_\ell$ holds for each $\ell \in [m]$ means that $t$ can be applied $B$ times in the population. This is what  \greedy{} does, returning an allocation $T$ with welfare $u(T) \geq (1- \varepsilon) u(T^*)$.

\begin{proof}[Proof of \cref{prop:greedy-cluster-optimal}]
\phantomsection\label{proof:prop:greedy-cluster-optimal}
Suppose that $T^* = (t^*_1,\dots,t^*_B)$ is an optimal overlapping test allocation and that $t^*$ is an optimal single group test for the population. Using the notation for first-negative probabilities from \cref{appendix:gain-of-overlaps-proofs}, we can write welfare as the sum of marginal contributions from the tests in $T^*$, %
$$
u(T^*) =  \sum_{j \in [B]} \sum_{i \in t^*_j} u_i \cdot P^{T^*}_{i,j},
$$
where $\sum_{i \in t^*_j} u_i \cdot P^{T^*}_{i,j}$ is the marginal utility of $t^*_j$. Since $P^{T^*}_{i,j} \leq q_{t^*_j}$, it follows that the marginal utility of $t^*_j$ is at most $q_{t^*_j} \sum_{i \in t^*_j} u_i = u(t^*_j) \leq u(t^*)$. It follows that $u(T^*) \leq B \cdot u(t^*)$. On the other hand, $u(t) \geq (1-\varepsilon) u(t^*)$. The fact that $B \cdot t(\ell) \leq n_\ell$ holds for each $\ell \in [m]$ means that $t$ can be applied $B$ times in the population, which is what \greedy{} does, returning allocation $T$. It follows that $u(T) = B \cdot u(t) \geq B (1- \varepsilon) u(t^*) \geq(1- \varepsilon) u(T^*)$.
\end{proof}

\section{Additional Simulation Results for Section~\ref{section:simulations}}
\label{apx:simulations}

Here we show figures and summary tables for our experiments comparing the MILP and \greedy{} on the pilot study data with pool size constraint $G=10$ in \cref{table:experiment2}, and on synthetic populations of size $n=200$ and pool size constraints $G \in \{5,10\}$ in \cref{fig:experiment3,fig:experiment4} and \cref{table:experiment3,table:experiment4}. For details on the experiments, we refer to \cref{section:simulations}.

\begin{table}[tb!]
    \centering
    \begin{threeparttable}
    \caption{Performance of the MILP and \greedy{} on pilot data ($G=10$).}
    \label{table:experiment2}
    {\small
    \renewcommand{\arraystretch}{1.15}
    \begin{tabular}{@{} crrrrrr @{}}
        \toprule
        & \multicolumn{3}{c}{{MILP}} & \multicolumn{3}{c}{{\greedy{}}}\\
        \cmidrule(lr){2-4} \cmidrule(l){5-7}
        {Budget} & Welfare & Guarantee & Time & Welfare & Apx To Optimal & Time \\
        \midrule
        2 & 866.58 & 0.47 & 470 ms & 866.47 & 1.0007 & 52 ms \\
        6 & 2391.49 & 1.42 & 3.50 s & 2390.99 & 1.0008 & 50 ms \\
        10 & 3775.35 & 2.36 & 29.70 min & 3774.20 & 1.0009 & 87 ms \\
        14 & 4696.30 & 3.31 & 321.55 min & 4678.52 & 1.0045 & 279 ms \\
        18 & 4741.30 & 4.25 & 1046.40 min & 4678.52 & 1.0143 & 672 ms \\
        22 & 4770.00 & 5.19 & 1473.71 min & 4678.52 & 1.0207 & 1.09 s \\
        26 & 4790.29 & 6.13 & 2232.60 min & 4678.52 & 1.0252 & 289 ms \\
        30 & 4805.25 & 7.08 & 4749.83 min & 4678.52 & 1.0286 & 1.22 s \\
        34 & 4816.15 & 8.03 & 9083.70 min & 4678.52 & 1.0311 & 1.18 s \\
        \bottomrule
    \end{tabular}
    }
    \begin{tablenotes}[flushleft]
    \footnotesize
    \item[] \emph{Notes.} Summary showing welfare and computation time for the MILP and \greedy{} on the pilot data with a population of $n=130$ and pool size constraint $G=10$, with testing budgets $B \in \{2, 6, \ldots, 34\}$. Welfare figures are deterministic values computed on the pilot population. The column ``Guarantee'' reports the a priori additive approximation guarantee of MILP relative to optimal non-overlapping welfare. The column ``Apx To Optimal'' reports the upper bound on the ratio between optimal non-overlapping welfare and \greedy{} welfare, computed as MILP welfare plus the guarantee, divided by \greedy{} welfare.
    \end{tablenotes}
    \end{threeparttable}
\end{table}

\begin{figure}[!htbp]
    \centering
    \begingroup
    \makeatletter
    \let\@makecaption\@maketablecaption
    \makeatother
    \caption{MILP-to-\greedy{} welfare ratios on synthetic populations ($G=5$).} %
    \label{fig:experiment3}
    \endgroup
    \vspace{0.45em}

    \begin{subfigure}[t]{0.44\textwidth}%
        \setlength{\columnwidth}{\linewidth}%
        \centering
        \resizebox{\linewidth}{!}{\begin{tikzpicture}
\begin{axis}[
    width=\columnwidth, height={9.5cm},
    axis y line=left, axis x line=bottom, separate axis lines,
    axis line style={gray!55, line width={0.4pt}, -},
    ymajorgrids, grid style={gray!20, line width={0.4pt}},
    xmin={1.4}, xmax={12.6}, xtick={2,4,6,8,10,12},
    ymin={99.95617049344375}, ymax={100.28128562415401},
    enlarge y limits={0.02}, ytick={100,100.1,100.2},
    xlabel={Test budget $B$},
    ylabel={MILP-to-\greedy{} welfare ratio (\%)},
    tick style={gray!55, line width={0.4pt}},
    tick label style={font={\footnotesize}},
    label style={font={\small}},
]
    \addplot[
        densely dashed, gray!70, line width={0.5pt},
        domain={1.4:12.6}, samples={2}, forget plot
    ] {100};
    \addplot[only marks, mark={*}, mark size={1.6pt}, mark options={fill={black!55}, draw={black!55}}]
        coordinates {
            (1.9662385069685298,100.0)
            (1.7074050292666647,100.00427097462847)
            (1.7473560894831774,100.0)
            (1.8019023041240805,100.0)
            (2.0354639754016683,99.99834788251472)
            (1.7595228111728294,100.0)
            (2.2655425875141644,100.08255774146025)
            (2.277935411829673,100.0)
            (1.9927155061649213,100.00000000000003)
            (2.136699287309308,100.0)
            (2.171470450485397,100.0)
            (1.970272482806394,100.00311309351859)
            (1.8810544992219749,100.04409786638743)
            (1.878545344082833,100.03048643916917)
            (2.1077612008600792,100.0)
            (1.9228441256229107,100.00000000000003)
            (2.1525025605899417,99.99656884336567)
            (2.1544509141926746,100.00796575229451)
            (2.0932191375405043,100.0)
            (2.1425452009443036,100.05917631950778)
        }
        ;
    \addplot[only marks, mark={*}, mark size={1.6pt}, mark options={fill={black!55}, draw={black!55}}]
        coordinates {
            (4.235079586391446,100.00204073044159)
            (4.225253822986561,99.98599623347147)
            (3.9268267555406693,99.9805704132381)
            (3.792249209894968,100.0166673149179)
            (4.107761511952393,100.11054263011361)
            (3.712380887179642,100.0487923398375)
            (4.062771602262767,100.10334901929217)
            (3.780394497228001,100.07388827893544)
            (4.236004640233888,99.96505105474333)
            (4.082278482931492,100.04463414467766)
            (4.219279986825443,100.002803349033)
            (3.8272825510019093,99.9995805603067)
            (4.234896281005868,99.9899819483017)
            (3.902002349162165,100.03027380408282)
            (3.811664321439719,99.97663616396267)
            (4.104481879458952,100.0263019189416)
            (4.153694119462772,99.97320301395617)
            (4.17310813235673,99.95707576839418)
            (3.9671705988715633,100.07867270117892)
            (4.191051005021272,100.28128562415401)
        }
        ;
    \addplot[only marks, mark={*}, mark size={1.6pt}, mark options={fill={black!55}, draw={black!55}}]
        coordinates {
            (6.281785063541519,99.98487580975046)
            (6.070298171287476,99.9854038732363)
            (6.185259060999817,100.0148645905728)
            (5.814217112242761,100.15378658142211)
            (5.853663570501656,100.06910045783629)
            (6.098749528166346,100.1089556270602)
            (5.922832482727272,100.11662071721952)
            (5.810631705431543,100.07342969031383)
            (6.069594013893313,99.96314077654989)
            (5.980389470579464,100.02331681721392)
            (5.781389779986485,100.0218318195909)
            (5.807727523855293,100.0339098152554)
            (6.038717171285973,99.98011270690954)
            (5.739730263424352,100.04220721656677)
            (6.28317704961323,99.9840983273004)
            (6.283067264641481,100.01996405299477)
            (5.736820097831078,100.02880278084092)
            (5.762734378640101,99.9937432156545)
            (5.975156965618212,100.0747444088173)
            (6.182312293153076,100.12454707859686)
        }
        ;
    \addplot[only marks, mark={*}, mark size={1.6pt}, mark options={fill={black!55}, draw={black!55}}]
        coordinates {
            (8.197993238688506,99.98274556680872)
            (8.203923262197076,99.97419756637544)
            (7.740007861746144,100.00406272236253)
            (8.011888241432766,100.16690832873043)
            (8.096616206480558,100.04805440334141)
            (7.901221876696428,100.09914595603763)
            (7.935917797842178,100.09490074866582)
            (7.809563978385675,100.09222241624143)
            (8.112242180568616,100.09712714712163)
            (8.12702868745406,100.06085937988023)
            (8.054015655075565,100.00198040494648)
            (7.908267564641101,100.0538416102173)
            (8.264300784962554,99.99982680344502)
            (8.000243029865791,100.0350528665932)
            (7.786662999772761,99.98914285736988)
            (7.9606460017668645,100.01197095261978)
            (7.901556649611579,100.02789577365849)
            (8.127512325817715,99.99347737644203)
            (8.146913480909358,100.08020997704867)
            (7.913838267304669,100.14171303735186)
        }
        ;
    \addplot[only marks, mark={*}, mark size={1.6pt}, mark options={fill={black!55}, draw={black!55}}]
        coordinates {
            (9.874870241054282,99.98441994550625)
            (9.92019684143767,99.95617049344375)
            (10.191808459797372,100.00587423332712)
            (10.19265293310803,100.11863499167598)
            (10.059167816391053,100.05501681188657)
            (10.081118230522309,100.08263657844967)
            (9.86173229538698,100.09687462251708)
            (9.73374671906472,100.06938856916054)
            (10.09552762642002,100.11032142409545)
            (10.08726443354131,100.06462198469923)
            (9.825215149254916,100.00059305734933)
            (10.083974726986414,100.03828008307846)
            (9.846408479951526,99.99425939221722)
            (10.005518941160759,100.03116465382706)
            (10.161605936868593,99.98316258953984)
            (10.228080337954788,100.01906270860887)
            (9.754073873722312,100.00195438755503)
            (10.25306062345904,99.99619952989957)
            (9.868625879238792,100.06779919123316)
            (10.067460302670453,100.08951537507302)
        }
        ;
    \addplot[only marks, mark={*}, mark size={1.6pt}, mark options={fill={black!55}, draw={black!55}}]
        coordinates {
            (11.865139715939904,99.98109429228738)
            (12.031648047374722,99.95966120510008)
            (12.111412933587971,99.991355096918)
            (12.12559107050711,100.10268516670067)
            (12.040838202691528,100.06085273022312)
            (12.254272188317215,100.07523193241991)
            (11.91956371752208,100.0963348307679)
            (12.298490137128036,100.06495251054571)
            (11.73280811491612,100.09882027540813)
            (11.751575294516693,100.05438157907672)
            (12.002819914228217,100.02151861975253)
            (12.269882291949076,100.02093161774621)
            (12.065187715770127,100.00090683893414)
            (11.876595994876737,100.0276931469876)
            (11.714072917445073,99.97077258946007)
            (12.034666443514833,100.01526500174374)
            (12.205974942660431,99.9956138136388)
            (11.910251804004998,99.99992283040662)
            (11.927087266422845,100.09759281317278)
            (11.952591472765778,100.08095811950379)
        }
        ;
\end{axis}
\end{tikzpicture}}%
        \caption{}
    \end{subfigure}\hfill
    \begin{subfigure}[t]{0.44\textwidth}%
        \setlength{\columnwidth}{\linewidth}%
        \centering
        \resizebox{\linewidth}{!}{\begin{tikzpicture}
\begin{groupplot}[
    group style={group size={1 by 6}, vertical sep={9pt}, x descriptions at=edge bottom},
    width=\columnwidth, height={2.9cm},
    axis y line=left, axis x line=bottom, separate axis lines,
    axis line style={gray!55, line width={0.4pt}, -},
    ymajorgrids, grid style={gray!20, line width={0.4pt}},
    xmin={99.95617049344375}, xmax={100.28128562415401}, enlarge x limits={0.02},
    xtick={100,100.1,100.2},
    ymin={0}, ymax={15}, ytick={0,5,10,15},
    xlabel={MILP-to-\greedy{} welfare ratio (\%)},
    tick style={gray!55, line width={0.4pt}},
    tick label style={font={\footnotesize}},
    label style={font={\small}},
]
    \nextgroupplot
    \node[anchor=north east, font={\footnotesize}] at (rel axis cs:0.985,0.92) {$B = 2$};
    \addplot[ybar interval, fill={rgb,255:red,100;green,149;blue,237}, draw={gray!60}, line width={0.3pt}]
        table[row sep={\\}]
        {
            x  y  \\
            99.95617049344375  0.0  \\
            99.97242624997926  0.0  \\
            99.98868200651478  15.0  \\
            100.0049377630503  1.0  \\
            100.02119351958581  1.0  \\
            100.03744927612132  1.0  \\
            100.05370503265682  1.0  \\
            100.06996078919234  1.0  \\
            100.08621654572785  0.0  \\
            100.10247230226337  0.0  \\
            100.11872805879888  0.0  \\
            100.1349838153344  0.0  \\
            100.15123957186991  0.0  \\
            100.16749532840542  0.0  \\
            100.18375108494094  0.0  \\
            100.20000684147644  0.0  \\
            100.21626259801195  0.0  \\
            100.23251835454747  0.0  \\
            100.24877411108298  0.0  \\
            100.2650298676185  0.0  \\
            100.28128562415401  0.0  \\
        }
        ;
    \draw[densely dashed, gray!70, line width={0.5pt}] ({axis cs:100.0,0}|-{rel axis cs:0,1}) -- ({axis cs:100.0,0}|-{rel axis cs:0,0});
    \nextgroupplot
    \node[anchor=north east, font={\footnotesize}] at (rel axis cs:0.985,0.92) {$B = 4$};
    \addplot[ybar interval, fill={rgb,255:red,100;green,149;blue,237}, draw={gray!60}, line width={0.3pt}]
        table[row sep={\\}]
        {
            x  y  \\
            99.95617049344375  2.0  \\
            99.97242624997926  4.0  \\
            99.98868200651478  4.0  \\
            100.0049377630503  1.0  \\
            100.02119351958581  2.0  \\
            100.03744927612132  2.0  \\
            100.05370503265682  0.0  \\
            100.06996078919234  2.0  \\
            100.08621654572785  0.0  \\
            100.10247230226337  2.0  \\
            100.11872805879888  0.0  \\
            100.1349838153344  0.0  \\
            100.15123957186991  0.0  \\
            100.16749532840542  0.0  \\
            100.18375108494094  0.0  \\
            100.20000684147644  0.0  \\
            100.21626259801195  0.0  \\
            100.23251835454747  0.0  \\
            100.24877411108298  0.0  \\
            100.2650298676185  1.0  \\
            100.28128562415401  1.0  \\
        }
        ;
    \draw[densely dashed, gray!70, line width={0.5pt}] ({axis cs:100.0,0}|-{rel axis cs:0,1}) -- ({axis cs:100.0,0}|-{rel axis cs:0,0});
    \nextgroupplot
    \node[anchor=north east, font={\footnotesize}] at (rel axis cs:0.985,0.92) {$B = 6$};
    \addplot[ybar interval, fill={rgb,255:red,100;green,149;blue,237}, draw={gray!60}, line width={0.3pt}]
        table[row sep={\\}]
        {
            x  y  \\
            99.95617049344375  1.0  \\
            99.97242624997926  4.0  \\
            99.98868200651478  1.0  \\
            100.0049377630503  2.0  \\
            100.02119351958581  4.0  \\
            100.03744927612132  1.0  \\
            100.05370503265682  1.0  \\
            100.06996078919234  2.0  \\
            100.08621654572785  0.0  \\
            100.10247230226337  2.0  \\
            100.11872805879888  1.0  \\
            100.1349838153344  0.0  \\
            100.15123957186991  1.0  \\
            100.16749532840542  0.0  \\
            100.18375108494094  0.0  \\
            100.20000684147644  0.0  \\
            100.21626259801195  0.0  \\
            100.23251835454747  0.0  \\
            100.24877411108298  0.0  \\
            100.2650298676185  0.0  \\
            100.28128562415401  0.0  \\
        }
        ;
    \draw[densely dashed, gray!70, line width={0.5pt}] ({axis cs:100.0,0}|-{rel axis cs:0,1}) -- ({axis cs:100.0,0}|-{rel axis cs:0,0});
    \nextgroupplot
    \node[anchor=north east, font={\footnotesize}] at (rel axis cs:0.985,0.92) {$B = 8$};
    \addplot[ybar interval, fill={rgb,255:red,100;green,149;blue,237}, draw={gray!60}, line width={0.3pt}]
        table[row sep={\\}]
        {
            x  y  \\
            99.95617049344375  0.0  \\
            99.97242624997926  2.0  \\
            99.98868200651478  5.0  \\
            100.0049377630503  1.0  \\
            100.02119351958581  2.0  \\
            100.03744927612132  1.0  \\
            100.05370503265682  2.0  \\
            100.06996078919234  1.0  \\
            100.08621654572785  4.0  \\
            100.10247230226337  0.0  \\
            100.11872805879888  0.0  \\
            100.1349838153344  1.0  \\
            100.15123957186991  1.0  \\
            100.16749532840542  0.0  \\
            100.18375108494094  0.0  \\
            100.20000684147644  0.0  \\
            100.21626259801195  0.0  \\
            100.23251835454747  0.0  \\
            100.24877411108298  0.0  \\
            100.2650298676185  0.0  \\
            100.28128562415401  0.0  \\
        }
        ;
    \draw[densely dashed, gray!70, line width={0.5pt}] ({axis cs:100.0,0}|-{rel axis cs:0,1}) -- ({axis cs:100.0,0}|-{rel axis cs:0,0});
    \nextgroupplot
    \node[anchor=north east, font={\footnotesize}] at (rel axis cs:0.985,0.92) {$B = 10$};
    \addplot[ybar interval, fill={rgb,255:red,100;green,149;blue,237}, draw={gray!60}, line width={0.3pt}]
        table[row sep={\\}]
        {
            x  y  \\
            99.95617049344375  1.0  \\
            99.97242624997926  2.0  \\
            99.98868200651478  4.0  \\
            100.0049377630503  2.0  \\
            100.02119351958581  1.0  \\
            100.03744927612132  1.0  \\
            100.05370503265682  4.0  \\
            100.06996078919234  1.0  \\
            100.08621654572785  2.0  \\
            100.10247230226337  2.0  \\
            100.11872805879888  0.0  \\
            100.1349838153344  0.0  \\
            100.15123957186991  0.0  \\
            100.16749532840542  0.0  \\
            100.18375108494094  0.0  \\
            100.20000684147644  0.0  \\
            100.21626259801195  0.0  \\
            100.23251835454747  0.0  \\
            100.24877411108298  0.0  \\
            100.2650298676185  0.0  \\
            100.28128562415401  0.0  \\
        }
        ;
    \draw[densely dashed, gray!70, line width={0.5pt}] ({axis cs:100.0,0}|-{rel axis cs:0,1}) -- ({axis cs:100.0,0}|-{rel axis cs:0,0});
    \nextgroupplot
    \node[anchor=north east, font={\footnotesize}] at (rel axis cs:0.985,0.92) {$B = 12$};
    \addplot[ybar interval, fill={rgb,255:red,100;green,149;blue,237}, draw={gray!60}, line width={0.3pt}]
        table[row sep={\\}]
        {
            x  y  \\
            99.95617049344375  2.0  \\
            99.97242624997926  1.0  \\
            99.98868200651478  4.0  \\
            100.0049377630503  2.0  \\
            100.02119351958581  2.0  \\
            100.03744927612132  0.0  \\
            100.05370503265682  3.0  \\
            100.06996078919234  2.0  \\
            100.08621654572785  3.0  \\
            100.10247230226337  1.0  \\
            100.11872805879888  0.0  \\
            100.1349838153344  0.0  \\
            100.15123957186991  0.0  \\
            100.16749532840542  0.0  \\
            100.18375108494094  0.0  \\
            100.20000684147644  0.0  \\
            100.21626259801195  0.0  \\
            100.23251835454747  0.0  \\
            100.24877411108298  0.0  \\
            100.2650298676185  0.0  \\
            100.28128562415401  0.0  \\
        }
        ;
    \draw[densely dashed, gray!70, line width={0.5pt}] ({axis cs:100.0,0}|-{rel axis cs:0,1}) -- ({axis cs:100.0,0}|-{rel axis cs:0,0});
\end{groupplot}
\node[rotate=90, anchor=south, font={\small}]
    at ([xshift={-24pt}]$(group c1r1.north west)!0.5!(group c1r6.south west)$)
    {Populations};
\end{tikzpicture}}%
        \caption{}
    \end{subfigure}

    \par\vspace{1em}
\begin{minipage}{\textwidth}
    \footnotesize
    \raggedright
    \emph{Notes.} Outcomes of \greedy{} and the MILP on synthetic data
    with populations of size $n=200$, pool size bound $G=5$, and testing budgets $B \in \{2, 4, 6, 8, 10, 12\}$. Panel (a) shows the MILP-to-\greedy{} welfare ratios for each budget, expressed as percentages, with one point per randomly generated population. Panel (b) shows the same ratios as per-budget histograms, with a dashed line at $100\%$ (parity). Each point / histogram entry corresponds to one of the 20 randomly generated populations. %
\end{minipage}
\end{figure}

\begin{figure}[!htbp]
    \centering
    \begingroup
    \makeatletter
    \let\@makecaption\@maketablecaption
    \makeatother
    \caption{MILP-to-\greedy{} welfare ratios on synthetic populations ($G=10$).} %
    \label{fig:experiment4}
    \endgroup
    \vspace{0.45em}

    \begin{subfigure}[t]{0.44\textwidth}%
        \setlength{\columnwidth}{\linewidth}%
        \centering
        \resizebox{\linewidth}{!}{\begin{tikzpicture}
\begin{axis}[
    width=\columnwidth, height={9.5cm},
    axis y line=left, axis x line=bottom, separate axis lines,
    axis line style={gray!55, line width={0.4pt}, -},
    ymajorgrids, grid style={gray!20, line width={0.4pt}},
    xmin={1.4}, xmax={12.6}, xtick={2,4,6,8,10,12},
    ymin={99.99365726241717}, ymax={106.36602290101844},
    enlarge y limits={0.02}, ytick={100,102,104,106},
    xlabel={Test budget $B$},
    ylabel={MILP-to-\greedy{} welfare ratio (\%)},
    tick style={gray!55, line width={0.4pt}},
    tick label style={font={\footnotesize}},
    label style={font={\small}},
]
    \addplot[
        densely dashed, gray!70, line width={0.5pt},
        domain={1.4:12.6}, samples={2}, forget plot
    ] {100};
    \addplot[only marks, mark={*}, mark size={1.6pt}, mark options={fill={black!55}, draw={black!55}}]
        coordinates {
            (1.9662385069685298,99.99999999999997)
            (1.7074050292666647,100.0)
            (1.7473560894831774,100.00000000000003)
            (1.8019023041240805,100.0)
            (2.0354639754016683,100.00582302967959)
            (1.7595228111728294,99.99365726241717)
            (2.2655425875141644,99.99999999999997)
            (2.277935411829673,100.0)
            (1.9927155061649213,100.0)
            (2.136699287309308,99.99999999999997)
            (2.171470450485397,100.05406422469673)
            (1.970272482806394,100.0)
            (1.8810544992219749,100.0)
            (1.878545344082833,100.0)
            (2.1077612008600792,99.99999999999997)
            (1.9228441256229107,100.01688180754313)
            (2.1525025605899417,100.00000000000003)
            (2.1544509141926746,99.99999999999997)
            (2.0932191375405043,99.99999999999997)
            (2.1425452009443036,100.07365356910734)
        }
        ;
    \addplot[only marks, mark={*}, mark size={1.6pt}, mark options={fill={black!55}, draw={black!55}}]
        coordinates {
            (4.235079586391446,100.14004171069455)
            (4.225253822986561,100.01008328975962)
            (3.9268267555406693,100.03512989352707)
            (3.792249209894968,99.99999999999999)
            (4.107761511952393,100.00361592779097)
            (3.712380887179642,100.00697853283226)
            (4.062771602262767,100.05138960511883)
            (3.780394497228001,100.00761398248184)
            (4.236004640233888,100.03465088049894)
            (4.082278482931492,100.0)
            (4.219279986825443,100.03856878469568)
            (3.8272825510019093,100.01071543679782)
            (4.234896281005868,100.12742168639276)
            (3.902002349162165,100.06056756921286)
            (3.811664321439719,99.99999999999999)
            (4.104481879458952,100.019806574998)
            (4.153694119462772,100.04353224105427)
            (4.17310813235673,100.00038909139084)
            (3.9671705988715633,100.02136670962211)
            (4.191051005021272,100.04664039104973)
        }
        ;
    \addplot[only marks, mark={*}, mark size={1.6pt}, mark options={fill={black!55}, draw={black!55}}]
        coordinates {
            (6.281785063541519,101.26507179905586)
            (6.070298171287476,101.09062890459984)
            (6.185259060999817,100.83019938900195)
            (5.814217112242761,100.41603440422485)
            (5.853663570501656,100.44095643259475)
            (6.098749528166346,100.44405513619787)
            (5.922832482727272,100.38587946673245)
            (5.810631705431543,100.37543578192887)
            (6.069594013893313,101.12279059909459)
            (5.980389470579464,100.688506763911)
            (5.781389779986485,100.36458721681687)
            (5.807727523855293,100.73482170279186)
            (6.038717171285973,100.42231199068972)
            (5.739730263424352,100.26165311126482)
            (6.28317704961323,100.4083586819992)
            (6.283067264641481,101.11879772094146)
            (5.736820097831078,100.55524678154566)
            (5.762734378640101,100.25207586937634)
            (5.975156965618212,100.06906233669974)
            (6.182312293153076,100.64658350723235)
        }
        ;
    \addplot[only marks, mark={*}, mark size={1.6pt}, mark options={fill={black!55}, draw={black!55}}]
        coordinates {
            (8.197993238688506,103.39252462729267)
            (8.203923262197076,102.80766508006738)
            (7.740007861746144,102.06628066936891)
            (8.011888241432766,101.163989094994)
            (8.096616206480558,101.6725583110772)
            (7.901221876696428,101.2909194521016)
            (7.935917797842178,100.92396874600979)
            (7.809563978385675,101.18204393147616)
            (8.112242180568616,102.42519147435587)
            (8.12702868745406,102.07544775429602)
            (8.054015655075565,101.92764034928746)
            (7.908267564641101,101.45659523300151)
            (8.264300784962554,101.15653848675234)
            (8.000243029865791,101.38866847530157)
            (7.786662999772761,101.00764974004868)
            (7.9606460017668645,102.13757070340716)
            (7.901556649611579,101.69468380962061)
            (8.127512325817715,101.27202975717931)
            (8.146913480909358,100.96105314442264)
            (7.913838267304669,101.57338011071553)
        }
        ;
    \addplot[only marks, mark={*}, mark size={1.6pt}, mark options={fill={black!55}, draw={black!55}}]
        coordinates {
            (9.874870241054282,104.58743768975395)
            (9.92019684143767,104.50777231211812)
            (10.191808459797372,103.05150046766371)
            (10.19265293310803,102.34445277952256)
            (10.059167816391053,102.76372194828332)
            (10.081118230522309,102.03837344628633)
            (9.86173229538698,101.87371305418822)
            (9.73374671906472,102.21420277469954)
            (10.09552762642002,103.88065947571148)
            (10.08726443354131,104.46431065507704)
            (9.825215149254916,103.4819240723599)
            (10.083974726986414,102.44066933193604)
            (9.846408479951526,102.30636854509201)
            (10.005518941160759,102.14711858331937)
            (10.161605936868593,102.28897780361173)
            (10.228080337954788,102.6681018481828)
            (9.754073873722312,102.68394913620418)
            (10.25306062345904,102.55943341596736)
            (9.868625879238792,102.62017393792628)
            (10.067460302670453,102.57495071921025)
        }
        ;
    \addplot[only marks, mark={*}, mark size={1.6pt}, mark options={fill={black!55}, draw={black!55}}]
        coordinates {
            (11.865139715939904,105.15228084738963)
            (12.031648047374722,105.61164363052842)
            (12.111412933587971,103.94360374497838)
            (12.12559107050711,103.52287792437424)
            (12.040838202691528,103.53372611247167)
            (12.254272188317215,103.04942160770949)
            (11.91956371752208,103.0980227464401)
            (12.298490137128036,103.14738813396347)
            (11.73280811491612,104.44039671844081)
            (11.751575294516693,106.36602290101844)
            (12.002819914228217,104.55330039014812)
            (12.269882291949076,103.5340670473985)
            (12.065187715770127,103.36180427620675)
            (11.876595994876737,103.85304297787057)
            (11.714072917445073,103.80547816184591)
            (12.034666443514833,103.63748596071144)
            (12.205974942660431,103.86395464616973)
            (11.910251804004998,103.55710355703916)
            (11.927087266422845,103.772761052027)
            (11.952591472765778,103.53429457451806)
        }
        ;
\end{axis}
\end{tikzpicture}}%
        \caption{}
    \end{subfigure}\hfill
    \begin{subfigure}[t]{0.44\textwidth}%
        \setlength{\columnwidth}{\linewidth}%
        \centering
        \resizebox{\linewidth}{!}{\begin{tikzpicture}
\begin{groupplot}[
    group style={group size={1 by 6}, vertical sep={9pt}, x descriptions at=edge bottom},
    width=\columnwidth, height={2.9cm},
    axis y line=left, axis x line=bottom, separate axis lines,
    axis line style={gray!55, line width={0.4pt}, -},
    ymajorgrids, grid style={gray!20, line width={0.4pt}},
    xmin={99.99365726241717}, xmax={106.36602290101844}, enlarge x limits={0.02},
    ymin={0}, ymax={20}, ytick={0,10,20},
    xlabel={MILP-to-\greedy{} welfare ratio (\%)},
    tick style={gray!55, line width={0.4pt}},
    tick label style={font={\footnotesize}},
    label style={font={\small}},
]
    \nextgroupplot
    \node[anchor=north east, font={\footnotesize}] at (rel axis cs:0.985,0.92) {$B = 2$};
    \addplot[ybar interval, fill={rgb,255:red,100;green,149;blue,237}, draw={gray!60}, line width={0.3pt}]
        table[row sep={\\}]
        {
            x  y  \\
            99.99365726241717  20.0  \\
            100.31227554434723  0.0  \\
            100.6308938262773  0.0  \\
            100.94951210820736  0.0  \\
            101.26813039013743  0.0  \\
            101.58674867206749  0.0  \\
            101.90536695399756  0.0  \\
            102.22398523592761  0.0  \\
            102.54260351785769  0.0  \\
            102.86122179978774  0.0  \\
            103.1798400817178  0.0  \\
            103.49845836364787  0.0  \\
            103.81707664557793  0.0  \\
            104.135694927508  0.0  \\
            104.45431320943806  0.0  \\
            104.77293149136813  0.0  \\
            105.09154977329818  0.0  \\
            105.41016805522825  0.0  \\
            105.72878633715831  0.0  \\
            106.04740461908838  0.0  \\
            106.36602290101844  0.0  \\
        }
        ;
    \nextgroupplot
    \node[anchor=north east, font={\footnotesize}] at (rel axis cs:0.985,0.92) {$B = 4$};
    \addplot[ybar interval, fill={rgb,255:red,100;green,149;blue,237}, draw={gray!60}, line width={0.3pt}]
        table[row sep={\\}]
        {
            x  y  \\
            99.99365726241717  20.0  \\
            100.31227554434723  0.0  \\
            100.6308938262773  0.0  \\
            100.94951210820736  0.0  \\
            101.26813039013743  0.0  \\
            101.58674867206749  0.0  \\
            101.90536695399756  0.0  \\
            102.22398523592761  0.0  \\
            102.54260351785769  0.0  \\
            102.86122179978774  0.0  \\
            103.1798400817178  0.0  \\
            103.49845836364787  0.0  \\
            103.81707664557793  0.0  \\
            104.135694927508  0.0  \\
            104.45431320943806  0.0  \\
            104.77293149136813  0.0  \\
            105.09154977329818  0.0  \\
            105.41016805522825  0.0  \\
            105.72878633715831  0.0  \\
            106.04740461908838  0.0  \\
            106.36602290101844  0.0  \\
        }
        ;
    \nextgroupplot
    \node[anchor=north east, font={\footnotesize}] at (rel axis cs:0.985,0.92) {$B = 6$};
    \addplot[ybar interval, fill={rgb,255:red,100;green,149;blue,237}, draw={gray!60}, line width={0.3pt}]
        table[row sep={\\}]
        {
            x  y  \\
            99.99365726241717  3.0  \\
            100.31227554434723  9.0  \\
            100.6308938262773  4.0  \\
            100.94951210820736  4.0  \\
            101.26813039013743  0.0  \\
            101.58674867206749  0.0  \\
            101.90536695399756  0.0  \\
            102.22398523592761  0.0  \\
            102.54260351785769  0.0  \\
            102.86122179978774  0.0  \\
            103.1798400817178  0.0  \\
            103.49845836364787  0.0  \\
            103.81707664557793  0.0  \\
            104.135694927508  0.0  \\
            104.45431320943806  0.0  \\
            104.77293149136813  0.0  \\
            105.09154977329818  0.0  \\
            105.41016805522825  0.0  \\
            105.72878633715831  0.0  \\
            106.04740461908838  0.0  \\
            106.36602290101844  0.0  \\
        }
        ;
    \nextgroupplot
    \node[anchor=north east, font={\footnotesize}] at (rel axis cs:0.985,0.92) {$B = 8$};
    \addplot[ybar interval, fill={rgb,255:red,100;green,149;blue,237}, draw={gray!60}, line width={0.3pt}]
        table[row sep={\\}]
        {
            x  y  \\
            99.99365726241717  0.0  \\
            100.31227554434723  0.0  \\
            100.6308938262773  1.0  \\
            100.94951210820736  5.0  \\
            101.26813039013743  5.0  \\
            101.58674867206749  2.0  \\
            101.90536695399756  4.0  \\
            102.22398523592761  1.0  \\
            102.54260351785769  1.0  \\
            102.86122179978774  0.0  \\
            103.1798400817178  1.0  \\
            103.49845836364787  0.0  \\
            103.81707664557793  0.0  \\
            104.135694927508  0.0  \\
            104.45431320943806  0.0  \\
            104.77293149136813  0.0  \\
            105.09154977329818  0.0  \\
            105.41016805522825  0.0  \\
            105.72878633715831  0.0  \\
            106.04740461908838  0.0  \\
            106.36602290101844  0.0  \\
        }
        ;
    \nextgroupplot
    \node[anchor=north east, font={\footnotesize}] at (rel axis cs:0.985,0.92) {$B = 10$};
    \addplot[ybar interval, fill={rgb,255:red,100;green,149;blue,237}, draw={gray!60}, line width={0.3pt}]
        table[row sep={\\}]
        {
            x  y  \\
            99.99365726241717  0.0  \\
            100.31227554434723  0.0  \\
            100.6308938262773  0.0  \\
            100.94951210820736  0.0  \\
            101.26813039013743  0.0  \\
            101.58674867206749  1.0  \\
            101.90536695399756  3.0  \\
            102.22398523592761  4.0  \\
            102.54260351785769  6.0  \\
            102.86122179978774  1.0  \\
            103.1798400817178  1.0  \\
            103.49845836364787  0.0  \\
            103.81707664557793  1.0  \\
            104.135694927508  0.0  \\
            104.45431320943806  3.0  \\
            104.77293149136813  0.0  \\
            105.09154977329818  0.0  \\
            105.41016805522825  0.0  \\
            105.72878633715831  0.0  \\
            106.04740461908838  0.0  \\
            106.36602290101844  0.0  \\
        }
        ;
    \nextgroupplot
    \node[anchor=north east, font={\footnotesize}] at (rel axis cs:0.985,0.92) {$B = 12$};
    \addplot[ybar interval, fill={rgb,255:red,100;green,149;blue,237}, draw={gray!60}, line width={0.3pt}]
        table[row sep={\\}]
        {
            x  y  \\
            99.99365726241717  0.0  \\
            100.31227554434723  0.0  \\
            100.6308938262773  0.0  \\
            100.94951210820736  0.0  \\
            101.26813039013743  0.0  \\
            101.58674867206749  0.0  \\
            101.90536695399756  0.0  \\
            102.22398523592761  0.0  \\
            102.54260351785769  0.0  \\
            102.86122179978774  3.0  \\
            103.1798400817178  1.0  \\
            103.49845836364787  8.0  \\
            103.81707664557793  3.0  \\
            104.135694927508  1.0  \\
            104.45431320943806  1.0  \\
            104.77293149136813  0.0  \\
            105.09154977329818  1.0  \\
            105.41016805522825  1.0  \\
            105.72878633715831  0.0  \\
            106.04740461908838  1.0  \\
            106.36602290101844  1.0  \\
        }
        ;
\end{groupplot}
\node[rotate=90, anchor=south, font={\small}]
    at ([xshift={-24pt}]$(group c1r1.north west)!0.5!(group c1r6.south west)$)
    {Populations};
\end{tikzpicture}}%
        \caption{}
    \end{subfigure}

    \par\vspace{1em}
    \begin{minipage}{\textwidth}
        \footnotesize
        \raggedright
        \emph{Notes.} Outcomes of \greedy{} and the MILP on synthetic data
        with populations of size $n=200$, pool size bound $G=10$, and testing budgets $B \in \{2, 4, 6, 8, 10, 12\}$. Panel (a) shows the MILP-to-\greedy{} welfare ratios for each budget, expressed as percentages, with one point per randomly generated population. Panel (b) shows the same ratios as per-budget histograms. Each point / histogram entry corresponds to one of the 20 randomly generated populations. %
    \end{minipage}
\end{figure}

\begin{table}[tb!]
    \centering
    \begin{threeparttable}
    \caption{Synthetic-population performance of the MILP and \greedy{} ($G=5$).}
    \label{table:experiment3}
    {\small
    \renewcommand{\arraystretch}{1.15}
    \begin{tabular}{@{} crrrrrr @{}}
        \toprule
        & \multicolumn{3}{c}{{MILP}} & \multicolumn{3}{c}{{\greedy{}}} \\
        \cmidrule(lr){2-4} \cmidrule(l){5-7}
        {Budget} & Welfare & Guarantee & Time & Welfare & Apx To Optimal & Time \\
        \midrule
        2 & 366.90 & 0.56 & 285 ms & 366.85 & 1.0016 & 35 ms \\
        4 & 675.51 & 1.11 & 838 ms & 675.28 & 1.0020 & 68 ms \\
        6 & 947.86 & 1.67 & 2.58 s & 947.48 & 1.0022 & 98 ms \\
        8 & 1188.80 & 2.23 & 9.93 s & 1188.24 & 1.0024 & 134 ms \\
        10 & 1399.83 & 2.78 & 3.37 min & 1399.30 & 1.0024 & 176 ms \\
        12 & 1584.99 & 3.34 & 15.82 min & 1584.43 & 1.0025 & 226 ms \\
        \bottomrule
    \end{tabular}
    }
    \begin{tablenotes}[flushleft]
        \footnotesize
    \item[] \emph{Notes.} Summary showing welfare and computation time for the MILP and \greedy{} on synthetic data with populations of size $n=200$ and pool size constraint $G=5$, with testing budgets $B \in \{2, 4, 6, 8, 10, 12\}$. Welfare and times are averaged over 20 randomly generated populations. The column ``Guarantee'' reports the a priori additive approximation guarantee of MILP relative to optimal non-overlapping welfare. The column ``Apx To Optimal'' reports the upper bound on the ratio between optimal non-overlapping welfare and \greedy{} welfare, computed as MILP welfare plus the guarantee, divided by \greedy{} welfare.
    \end{tablenotes}
    \end{threeparttable}
\end{table}

\begin{table}[tb!]
    \centering
    \begin{threeparttable}
    \caption{Synthetic-population performance of the MILP and \greedy{} ($G=10$).}
    \label{table:experiment4}
    {\small
    \renewcommand{\arraystretch}{1.15}
    \begin{tabular}{@{} crrrrrr @{}}
        \toprule
        & \multicolumn{3}{c}{{MILP}} & \multicolumn{3}{c}{{\greedy{}}}\\
        \cmidrule(lr){2-4} \cmidrule(l){5-7}
        {Budget} & Welfare & Guarantee & Time & Welfare & Apx To Optimal & Time \\
        \midrule
        2 & 587.20 & 1.24 & 390 ms & 587.16 & 1.0022 & 34 ms \\
        4 & 941.65 & 2.48 & 2.09 s & 941.34 & 1.0030 & 84 ms \\
        6 & 1177.26 & 3.71 & 9.30 s & 1170.46 & 1.0090 & 172 ms \\
        8 & 1368.17 & 4.95 & 33.18 s & 1345.93 & 1.0202 & 266 ms \\
        10 & 1533.75 & 6.19 & 1.26 min & 1491.34 & 1.0326 & 380 ms \\
        12 & 1680.95 & 7.43 & 5.83 min & 1617.25 & 1.0440 & 567 ms \\
        \bottomrule
    \end{tabular}
    }
    \begin{tablenotes}[flushleft]
    \footnotesize
    \item[] \emph{Notes.} Summary showing welfare and computation time for the MILP and \greedy{} on synthetic data with populations of size $n=200$ and pool size constraint $G=10$, with testing budgets $B \in \{2, 4, 6, 8, 10, 12\}$. Welfare and times are averaged over 20 randomly generated populations. The column ``Guarantee'' reports the a priori additive approximation guarantee of MILP relative to optimal non-overlapping welfare. The column ``Apx To Optimal'' reports the upper bound on the ratio between optimal non-overlapping welfare and \greedy{} welfare, computed as MILP welfare plus the guarantee, divided by \greedy{} welfare.
    \end{tablenotes}
    \end{threeparttable}
\end{table}

\clearpage

\section{Pilot Materials for Section~\ref{section:pilot}}
\label{appendix:RCT}

\subsection{Experimental design}
\label{section:design}
\label{appendix:design}
We implemented our utility-based non-overlapping pooled testing regime based on \greedy{} in a two-group randomized controlled trial at \redacted{IPICYT} in September 2022. The study design and pre-analysis plan were preregistered before implementation.\footnote{\redacted{Finster, Simon, et al. 2022. ``Fighting the Learning Loss: Evaluating C-SEF for University Students and Staff.''} AEA RCT Registry, May 23. \redacted{\url{https://doi.org/10.1257/rct.9466-2.0}}.} Of 131 individuals who consented, completed the baseline survey, and entered the trial, 130 had the health-probability and utility data required for the testing allocation, and 122 completed the endline survey; sample composition and balance are reported in \cref{section:app:describe-data} and \cref{tab:cov-balance}.

The trial compares two experimental conditions defined against the \redacted{IPICYT} reopening that took place shortly before the pilot started: the treatment group followed our safety protocol, including scheduled pooled testing with on-site work conditional on a negative result, while the control group followed no protocol and returned to on-site activities. The control is therefore a strong, possibly the best attainable, benchmark in terms of productivity and mental health, as it captures the outcomes one would expect without any COVID-19 restrictions or testing requirements. These benefits came without the testing-based reduction in infection risk provided by the treatment protocol.\footnote{A negative pooled qPCR result reduces but does not eliminate the possibility that an admitted participant is infectious, because false-negative results remain possible.}

Because the protocol imposes additional logistical burden on the treatment arm relative to the unrestricted control, our core hypothesis is one of \textit{mean equivalence} rather than of a positive treatment effect: we ask whether the welfare-maximizing pooled testing protocol can be deployed without statistically detectable harm. For statistical inference and robustness, we use two complementary covariate-adjusted linear specifications, an ANCOVA level model and a covariate-adjusted change score (delta) model, both with heteroskedasticity-robust (HC1) standard errors. These are complemented by equivalence tests implemented as two one-sided tests (TOST) \citep{lakens2017equivalence}; Holm-Bonferroni and Benjamini-Hochberg corrections across the ten primary tests; Lee sharp bounds on the effect of treatment assignment under monotone attrition \citep{lee2009bounds}; and cluster-robust (CR2) standard errors at the working group on the initial cluster-randomized cohort and at an effective cluster variable on the full sample, following the design-based framework of \citet{abadie2023clustering}. Full specifications and the covariate set appear in \cref{section:app:metrics-and-methods}, with robustness checks in \cref{sec:robustnessanalysis}.

All elements of the pilot study, including consent, a baseline and an endline survey, email invitations for testing, data processing, and computing test allocations, were coordinated in a web app developed specifically for the trial. A demo version is available at \url{https://demo.c-sef.com}.

At the beginning of each week, we used \greedy{} to compute a test allocation for each day among treatment-group participants, subject to a weekly budget of 30 tests, and invited the selected individuals to submit saliva samples at the \redacted{LANBAMA} facility, \redacted{IPICYT}'s in-house medical laboratory. Because invited participants could choose not to submit a sample, we reran \greedy{} daily on the set of submitted samples to determine the pools that were actually tested. A negative result allowed treatment participants to access university facilities until 72 hours after sample submission. Because results generally became available approximately 24 hours after submission, this provided about 48 hours of actual on-site access; participants otherwise worked remotely.

\subsection{Determining population data: utilities and health probabilities}
\label{section:population-data}
The crucial input for our algorithms are the individuals' utilities and health probabilities.
Our utility measure captures four dimensions of participants' need for in-person access: socioeconomic disadvantage, dependence on institutional and collaborative resources, psychosocial need, and recent work or study performance. First, the need for in-person work or study depends on the nature of an individual's activities. For example, an experimentalist may require access to a laboratory more frequently than a theoretician requires access to an office. We measured this dimension using questions about computer and internet use, access to digital resources, communication with colleagues, and teamwork. The questions were designed, ordered, and worded to make it difficult for participants to infer how particular responses would affect their priority for testing.

Second, closures of educational and working environments can disproportionately affect socioeconomically disadvantaged individuals.\footnote{See, e.g.,~\citep{azevedo2021simulating,gorgen2020report,goudeau2021lockdown,hossain2021unequal}. These studies of heterogeneous effects on vulnerable populations primarily concern students from K--12 through higher education, but they also document related effects among teaching staff and in academic environments more generally.} We measured this dimension using the number of dependants, household size, and two measures of self-perceived socio-economic status. In Mexico, socio-economic disadvantage is correlated with several protected characteristics.\footnote{Low income is correlated with membership in an ethnic minority, older age, and being female \citep{ordonez2018discrimination}.}

Third, pandemic-induced remote learning and work can adversely affect mental health,\footnote{See, e.g.,~\citep{asanov2021remote,Bertoni2022}.} and Mexican students and employees experienced mental-health difficulties associated with COVID-19 institutional closures.\footnote{See, e.g.,~\citep{limon2020social,martinez2021psychological}.} We measured psychosocial need using sociability, perceived stress, and life satisfaction. Finally, participants reporting lower recent performance, productivity, or goal achievement may have had a greater need for institutional access. We therefore included these measures as a fourth component. The corresponding survey questions and point assignments are reproduced in \cref{section:app:survey}.

For participant $i$, let $u_i^{se}$ denote the socio-economic component, $u_i^{dr}$ the digital-resources and collaboration component, $u_i^{psy}$ the psychosocial component, and $u_i^{ppa}$ the performance, productivity, and achievement component. The implemented allocation assigned equal weight to these four components:
\[
u_i
  = \frac{1}{4}\left(
      u_i^{se}
      + u_i^{dr}
      + u_i^{psy}
      + u_i^{ppa}
    \right).
\]
Thus, in the notation $u_i=\sum_k w^k u_i^k$, the weights were
\[
w^{se}=w^{dr}=w^{psy}=w^{ppa}=\frac14,
\]
and summed to one.

Each component was the mean of its available included item scores. Specifically, the socio-economic component used survey questions 2.1, 2.2, 2.6, and 2.7; the digital-resources and collaboration component used questions 2.3--2.5 and 3.1--3.3; the psychosocial component used questions 4.1, 4.3, and 4.4; and the performance, productivity, and achievement component used questions 5.1 and 5.3--5.5. Question 5.2 (learning), question 4.2 (fear of COVID-19), and question 4.5 (satisfaction with institutional safety measures) did not enter the utility. For the Perceived Stress Scale in question 4.3, the four item responses were first averaged, after reverse-coding the two positively worded items, to form a single stress score, which then entered the psychosocial component as one item.

Formally, let $\mathcal Z^k$ be the set of included items for component $k$ and let $P_{i,z}^k$ denote participant $i$'s point score for item $z$. If $\mathcal Z_i^k\subseteq\mathcal Z^k$ is the set of included items answered by participant $i$, the component score was
\[
u_i^k=\frac{1}{|\mathcal Z_i^k|}
       \sum_{z\in\mathcal Z_i^k}P_{i,z}^k.
\]
The web application assigned a component a value of zero if no included item in that component was answered.\footnote{No participant had an empty socio-economic, digital-resources, or psychosocial component. One participant had an empty performance, productivity, and achievement component and was assigned zero for that component, as specified by the implementation.}

Health probabilities were estimated for age and gender categories using Bayesian updates of local public-health data. We computed the probability of infection conditional on membership in one of six groups: $\{\text{male},\text{female}\}\times\{\text{age }15\text{--}29,\text{ age }30\text{--}59,\text{ age }\geq 60\}$. The baseline probability of infection for a given age group is determined using Bayesian updates of local public health data, under the guidance of local epidemiologists. We used publicly available epidemiological models from the Institute of Health Metrics and Evaluation (IHME) to estimate baseline infection rates in \redacted{San Luis Potosí}.\footnote{Estimated infection rates for SLP with IHME models can be found at their dashboard for different public behavior regimes: \url{https://covid19.healthdata.org/mexico/san-luis-potosi?view=infections-testing&tab=trend&test=infections}.} These estimates provided us with values for $\Pr[\text{infection}]$ for all individuals in the population, irrespective of their category. Furthermore, we estimated the probability that an individual belongs to a given category given an infection via official national data on testing results.\footnote{National testing aggregates can be found at \url{https://datos.covid-19.conacyt.mx} and \redacted{\url{https://covid19.healthdata.org/mexico/san-luis-potosi?view=infections-testing\&tab=trend\&test=infections}}.} These estimates provide us with values for $\Pr[\text{category} \mid \text{infected}]$ for each category. Finally, we used census data to compute the probability of membership to a given category at the state/national level.\footnote{Census data can be found at \url{https://www.inegi.org.mx/programas/ccpv/2020/}.}
This provides us with an estimate for $\Pr[\text{category}]$ for the population. With Bayes' rule, we compute the desired probability of infection per category as follows: $\Pr[\text{infection} \mid \text{category}] = \frac{\Pr[\text{category} \mid \text{infection}] \Pr[\text{infection}]}{\Pr[\text{category}]}$. The probabilities of being healthy were approximately 99.5\% for each group. The probabilities stayed constant throughout this trial, as it only ran for 4 weeks. If applied over a longer period, the health probabilities may also be updated.\footnote{If the random element of time spent off-site can be controlled for, Bayesian updating using test results may be preferred.}

\subsection{Randomization}
\label{appendix:randomization}

Sufficient separation of the treatment and control groups was important for evaluating the protocol, primarily to limit psychological spillovers between conditions. If control participants encountered treatment participants on-site, potential health spillovers were limited by the testing protocol: a negative qPCR result cleared treatment participants for a 72-hour window beginning at sample submission. Because results generally took approximately 24 hours, this translated into about 48 hours of actual on-site access. We could not control off-site social interactions.

Many individuals participating in our trial belong to a research `discipline', and within that discipline, to a working group. Similarly, staff in the administration belong to specific departments, e.g.,~accounting, which we label as their working group. Because only one working group from each discipline volunteered to participate, discipline and working group are analogous henceforth.

Randomization proceeded in two stages, and the resulting mixed design has direct implications for our standard-error choice in \cref{section:app:metrics-and-methods}. In the first stage, the initial volunteer cohort was cluster-randomized at the working-group level using the R function \texttt{cluster\_ra()} from the \texttt{randomizr} package (seed 957): whole working groups were assigned as units to either the treatment or control arm under complete cluster randomization. In the second stage, a small set of CNS late enrollers was randomized through a separate \texttt{cluster\_ra()} call (seed 1007), with each enroller treated as a singleton cluster. This procedure is equivalent to individual complete randomization of the late enrollers. The full sample therefore combines complete cluster randomization of the initial cohort with complete individual randomization of the late enrollers. We accommodate this mixed design through complementary cluster-robust standard-error specifications reported in \cref{sec:robustnessanalysis}: CR2 standard errors at the working-group level for the initial cluster-randomized cohort and CR2 standard errors using an \emph{effective cluster} variable for the full sample, following the design-based framework of \citet{abadie2023clustering}.

Treatment and control groups were spatially separated by different floors, offices, or even buildings. Crucial for this approach to be effective is that individuals across working disciplines are comparable. Given \redacted{IPICYT}'s reports about their staff, we know that staff, research students, and researchers are assigned to work/study in each team contingent only on their academic discipline, based on no individual characteristics. Hence, we may consider the assignment approximately random. Nevertheless, we further collect a number of covariates to conduct a balance analysis.

\begin{figure}[!htbp]
    \centering
    \begingroup
    \makeatletter
    \let\@makecaption\@maketablecaption
    \makeatother
    \caption{Kernel Density Estimates of Baseline Utility by Experimental Condition}
    \label{fig:utilities_bytreat}
    \endgroup
    \vspace{0.45em}
    \includegraphics[width=0.75\textwidth]{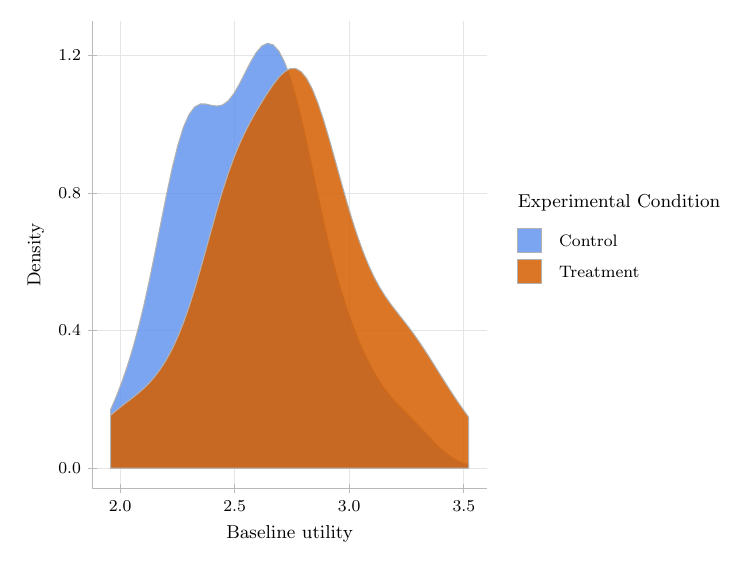}

    \par\vspace{1em}
    \begin{minipage}{\textwidth}
        \footnotesize
        \raggedright
        \emph{Notes.} Curves are Gaussian kernel density estimates, each
        normalized to integrate to one within its experimental condition;
        estimates use the 130 participants with nonmissing baseline utility data ($n=64$ control and $n=66$ treatment).
    \end{minipage}
\end{figure}

\paragraph{Covariate balance.} In \cref{tab:cov-balance} we provide a covariate balance analysis on the full analysis sample. Of the sixteen variables tested, three show imbalance at $p<0.05$: the baseline utility index ($\Delta = 0.169$, $p = 0.003$), its digital-resources component ($\Delta = 0.260$, $p = 0.004$), and baseline productivity ($\Delta = 0.27$, $p = 0.033$).\footnote{The Digital Resources Score is a computed measure of the need for digital and other resources found in the institute, and is one of the four components of the utility index.} Participants assigned to treatment had a higher mean baseline utility and a higher mean digital-resources score. Because each utility component was coded so that a higher value represents greater need for, or expected benefit from, institutional access, the treatment group had greater ex ante access-related need on average. \Cref{fig:utilities_bytreat} illustrates the overlap between the utility distributions by experimental condition; the treatment distribution exhibits greater concentration in the upper tail. Nevertheless, treatment participants received only conditional access under the testing protocol, whereas control participants received unrestricted access. In this access-related sense, evidence of mean equivalence is conservative: the protocol produced similar outcomes despite being imposed on the group with greater measured need for full access. We do not, however, assume that the baseline imbalance has an identical directional effect on every individual outcome. Our primary ANCOVA specifications therefore condition on the baseline value of each outcome and the full set of prespecified pretreatment covariates.

The baseline productivity imbalance motivates our choice of two complementary primary specifications. The ANCOVA level model conditions directly on the baseline value of each outcome (in addition to the full set of pre-treatment covariates), thereby adjusting for the baseline productivity imbalance; the change score (delta) model additionally eliminates all time-invariant individual heterogeneity. Covariate-adjusted treatment-effect estimates appear in \cref{tab:results-mentalhealth,tab:results-productivity}, and the main results are robust to the imbalance.

Based on the covariate balance discussion above, we do not treat the balance table as validating the randomization on its own; rather, these imbalances are addressed directly by our primary covariate-adjusted specifications (ANCOVA), which condition on the baseline value of each outcome and the full pre-treatment covariate set, and the main results are robust to them \citep{bruhn2009pursuit, altman1985comparability}.

\subsection{Scheduling preferences}
\label{section:app:mechanism}

Our mechanism allowed individuals in the treatment group to indicate their preferences over days on which they wished to access the institute. A negative pooled test on a given day allowed individuals in the pool to access the campus for 72 hours starting from sample submission, which meant 2 days after receiving their result in practice. In the web app, participants were given a set of 10 virtual tokens that they could distribute arbitrarily among all consecutive two-day windows (Monday \& Tuesday, Tuesday \& Wednesday, etc.) during which they wished to enter the institute. This distribution of tokens then expressed the agent's relative preferences. (Assigning more tokens to some two-day window indicated a stronger preference for these two days.) The individual's utility for each two-day block is then computed from baseline utilities as described in \cref{section:population-data} and their relative preference for the block.

\subsection{Outcomes and covariates}
\label{section:app:outcomes}

\paragraph{Mental health outcomes.}
\label{section:app:mentalhealth}
Mental health problems related to social isolation as a consequence of the COVID-19 pandemic have been documented for students and the general population \citep{martinez2021psychological}. We conjecture that putting in place a safe education protocol decreases stress levels among students, researchers, and staff, by increasing \emph{safe} sociability \citep{becchetti2017sociability} and modulating the perception of health risk in the institute \citep{shan2022effect}. Consequently, subjective well-being may also be positively affected. Stress is measured via the validated 4-item Perceived Stress Scale by Sheldon Cohen \citep{cohen1983perceived} and we use a variation of the European Quality of Life Survey measure of subjective well-being, using a `life/subject evaluation' approach \citep{oecd2013oecd}.\footnote{Note that we use baseline Stress and Subjective Well-being in our utilities' computation. On the other hand, we use endline Stress and Subjective Well-being to analyze between-group intervention effects. Further, in practice we compute the perceived-stress score as the mean of the available items after reverse-coding the two positively worded items. Partial non-response is rare (three participants at baseline and one at endline answered fewer than four items, and only one baseline participant answered a single item); we do not impose a minimum-item threshold, and the sole single-item case enters only as a baseline covariate, so it has negligible influence on any estimate. Scale reliability is acceptable for a four-item instrument (Cronbach's $\alpha = 0.67$ at baseline, $0.58$ at endline).}

\paragraph{Performance, productivity, and learning.}
\label{section:app:prodperflearning}
The pandemic has disrupted learning processes and decreased productivity of Mexican students \citep{limon2020social,martinez2021psychological} and female researchers \citep{king2021pandemic}. A significant portion of this downfall in productivity may be due to remote work with limited access to the necessary resources for work, research, and learning. We conjecture that our testing protocol improves (self-assessed) productivity and performance (in learning environments), and self-assessed learning experience when compared to a remote work policy. In the presence of an alternative reopening strategy---as is our case---we expect to see no difference between groups. That is, two competing opening strategies that allow all or some individuals in the population to socialize within the institutional premises should increase productivity. Since our mechanism imposes a greater logistical burden on subjects than the status quo reopening policy of \redacted{IPICYT}, no difference in performance, productivity and learning is an indication that a utility-maximizing approach to partial reopening is a successful strategy. 

Performance, productivity, and learning are analyzed as separate primary outcomes. Self-assessed performance is measured by survey question 5.1, productivity by question 5.3, and learning by question 5.2; the full wording and response options appear in \cref{section:app:survey}. The performance and productivity items are coded from 1 (high) to 5 (poor), so lower scores indicate better outcomes; only categories 1--4 are observed in the analysis sample. Learning is reported on a 1--10 scale, where 1 denotes a poor learning experience and 10 an excellent one.
Questions 5.4 and 5.5 measure achievement relative to supervisor-set and self-set goals, respectively, and enter the empirical analysis as secondary outcomes.

This outcome construction is distinct from the baseline utility calculation. For that calculation only, the point scores from performance, productivity, supervisor-goal achievement, and own-goal achievement (questions 5.1 and 5.3--5.5) were averaged to form the performance, productivity, and achievement component $u_i^{ppa}$:
\[
u_i^{ppa}
  = \frac{1}{|\mathcal Z_i^{ppa}|}
    \sum_{z\in\mathcal Z_i^{ppa}}P_{i,z}^{ppa},
\]
where $\mathcal Z_i^{ppa}\subseteq\{5.1,5.3,5.4,5.5\}$ is the set of these items answered by participant $i$. Learning did not enter this utility component.

\paragraph{Covariates.} 
\label{section:app:covariates}

Besides outcome variables, we collected additional socio-economic and psychosocial data of participants. These data are used in two ways. First, some of these features enter into the utility estimations needed for the testing algorithm. Second, we use relevant variables to assess balance between the experimental groups and as covariates in our primary and robustness specifications. We collect the following covariates:

\begin{enumerate}
    \item Socio-economic attributes: gender, age, ethnicity, educational affiliation,  perceived socio-economic status, financial dependents.
    \item Academic or job resources: internet access, access to job materials, need to collaborate in person, access to a dedicated working space outside of the office.
    \item Psychosocial attributes: Sociability, fear of the virus, subjective well-being.
\end{enumerate}
All covariates and their measurement strategy can be found in the baseline survey in \cref{section:app:survey}.

\subsection{Power and sample size}
\label{section:app:sample-size}

We summarize the trial's statistical power and complementary equivalence analysis. As a conventional individual-level benchmark, a two-sample, two-sided $t$-test calculation with 60 participants per arm, $\alpha=0.05$, and target power of 0.80 yields a minimum detectable standardized difference of approximately Cohen's $d=0.516$ (computed using \texttt{pwr.t.test()} in the R package \texttt{pwr}). This calculation does not account for the cluster-randomized portion of the trial and should therefore be interpreted as a descriptive benchmark rather than as a design-based power calculation for the realized mixed design.

The realized working-group intracluster correlations are near zero for the mental-health outcomes but larger for the work and study outcomes (\cref{tab:ICC}). Consequently, the individual-level benchmark is most informative for the mental-health outcomes, whereas power is lower and more dependent on the number and sizes of clusters for performance and learning. We account for the assignment structure in inference through the cluster-robust specifications reported in \cref{sec:robustnessanalysis}; we do not claim a single cluster-adjusted minimum detectable effect for all outcomes.

The five primary outcomes have different measurement scales. Perceived stress is a composite score formed by averaging the available Perceived Stress Scale items and can therefore take noninteger values. Life satisfaction and learning are recorded on 10-point ordinal rating scales. Performance and productivity are recorded using four ordered response categories. In the primary analysis, we treat these scores as approximately continuous and estimate linear models, which provide readily interpretable mean differences. As a robustness check, we estimate ordered logistic models for the four-category performance and productivity outcomes, together with the secondary goal-achievement outcomes, in \cref{sec:robustnessanalysis}.

Our central empirical question is whether the testing protocol preserves participants' well-being and work or study outcomes relative to unrestricted access. A nonsignificant difference test alone cannot establish this substantive null, because it does not distinguish evidence of equivalence from insufficient precision. We therefore complement the covariate-adjusted primary treatment-effect estimates with unadjusted two-sample TOST comparisons of the endline group means that directly assess whether effects large enough to be meaningful can be excluded \citep{lakens2017equivalence}. The preregistered analysis plan identified Cohen's $d=0.5$ as the benchmark for a moderate standardized effect in its power analysis, although it did not prespecify the formal equivalence procedure or designate a smallest effect size of interest. We use this preregistration-anchored benchmark as the TOST margin. For each outcome $j$, the raw-unit equivalence bounds are $\pm 0.5\widehat{\sigma}_{j,\mathrm{pooled}}$, where $\widehat{\sigma}_{j,\mathrm{pooled}}$ is the pooled endline standard deviation. Rejecting the TOST null hypothesis of non-equivalence therefore provides evidence that the treatment effect lies within this margin; failing to reject means that equivalence at this margin has not been established. Because the exact equivalence procedure was not prespecified, we interpret the TOST results as complementary evidence concerning the study's preregistered substantive null rather than as a confirmatory test based on a preregistered equivalence procedure.

\begin{table}[!htbp]
\centering
\begin{threeparttable}
\caption{TOST equivalence tests for primary outcomes.}
\label{tab:power-analysisTOST}
\small
\arxivtablefit{\footnotesize}{4.2pt}
    \begin{tabular}{lcccccc}
        \toprule
         & {Cohen's $d$} & {Lower bound} & {Upper bound} & {Equivalence $p$} & {NHST $p$} & {Conclusion} \\
         \midrule
         {Stress} & 0.500 & -0.370 & 0.370 & 0.079 & 0.184 & not established \\
         {Life satisfaction} & 0.500 & -1.024 & 1.024 & 0.007 & 0.812 & equivalence established \\
         {Performance} & 0.500 & -0.388 & 0.388 & 0.031 & 0.401 & equivalence established \\
         {Productivity} & 0.500 & -0.362 & 0.362 & 0.016 & 0.568 & equivalence established \\
         {Learning} & 0.500 & -0.783 & 0.783 & 0.018 & 0.544 & equivalence established\\
        \bottomrule
        \addlinespace
	    \end{tabular}
	    \begin{tablenotes}[flushleft]
    \footnotesize
	    \item[] \emph{Notes:} The TOST calculations compare unadjusted endline means between treatment and control using outcome-specific complete cases and equal-variance two-sample tests implemented with the R \texttt{TOSTER} package. Raw-unit equivalence margins are $\pm 0.5\widehat{\sigma}_{j,\mathrm{pooled}}$, where $\widehat{\sigma}_{j,\mathrm{pooled}}$ is calculated from the two endline groups. The equivalence $p$-value is the larger of the two one-sided $p$-values; the NHST $p$-value is from the conventional two-sided test of equal endline means. These calculations are distinct from, and complementary to, the covariate-adjusted ANCOVA and change-score models reported as the primary estimates.
    \end{tablenotes}
\end{threeparttable}
\end{table}

\Cref{tab:power-analysisTOST} reports the TOST results based on the unadjusted endline group means summarized in \cref{tab:TOST-input}. We fix the standardized benchmark at Cohen's $d=0.5$, independently of the observed treatment-effect estimates, and convert it to each outcome's scale using the pooled endline standard deviation, giving symmetric bounds $\pm 0.5\widehat{\sigma}_{\mathrm{pooled}}$.

Equivalence at the $d=0.5$ margin is established ($p<0.05$) for life satisfaction, performance, productivity, and learning. For stress, equivalence at this margin is not established ($p=0.079$): the TOST fails to reject the null hypothesis of non-equivalence. Stress has the largest observed standardized difference ($d \approx 0.24$), and the data are not sufficiently precise to exclude effects at the $d=0.5$ margin. Equivalence is, however, established at any margin $d \geq 0.6$, so the data rule out moderately large and large effects on stress. This is consistent with the differential-attrition and Lee-bounds analyses, which likewise single out stress (cf. \cref{tab:lee-bounds}). However, under Imbens--Manski bootstrapped confidence intervals, the intervals include zero, so the data provide no statistically significant evidence of an adverse effect.

\begin{table}[!htbp]
\centering
\begin{threeparttable}
\caption{Working-group ICCs for primary outcomes.}
\label{tab:ICC}
\small
    \begin{tabular}{lcccc}
        \toprule
         & {Full sample} & {Full sample} & {Cluster-RA cohort} & {Cluster-RA cohort} \\
         & {Baseline ICC} & {Endline ICC} & {Baseline ICC} & {Endline ICC} \\
         \midrule
         {Stress} & -0.204 & 0.011 & -0.183 & 0.005 \\
         {Life satisfaction} & -0.017 & 0.079 & -0.006 & 0.018 \\
         {Performance} & 0.199 & 0.100 & 0.273 & 0.143 \\
         {Productivity} & 0.012 & 0.090 & 0.090 & 0.105 \\
         {Learning} & 0.411 & 0.344 & 0.333 & 0.294 \\
        \bottomrule
        \addlinespace
	    \end{tabular}
	    \begin{tablenotes}[flushleft]
    \footnotesize
    \item[] \emph{Notes.} ICCs are ANOVA one-way random-effects estimates of $\rho$, computed using the R \texttt{fishmethods} package, for the five primary outcomes across two cluster definitions and two timepoints.
    \end{tablenotes}
\end{threeparttable}
\end{table}

The initial randomization relied on participants' affiliation to a working group, while late enrollers were randomized separately; the working-group variable is also the cluster used for the ICC calculation. \cref{tab:ICC} reports the working-group ICC (ANOVA one-way random-effects $\rho$ \citep{Donner-1986}, computed with the \texttt{fishmethods} R package) for the five primary outcomes under two cluster definitions: the effective cluster\footnote{Recall the effective cluster uses group id for the initial cohort, and user id for the late-enrollers.} on the full sample and the working group on the initial cluster-randomized cohort, and at both baseline and endline. The pattern is consistent across all four columns: ICCs are effectively zero for the mental-health outcomes (the small negative point estimates are finite-sample artifacts and should be read as zero), moderate for performance ($\approx 0.10$--$0.27$), and substantial for learning ($\approx 0.29$--$0.41$). Because these correlations are already present at baseline, the clustering in the work and study outcomes is structural rather than treatment-induced. The three \texttt{fishmethods} estimators (Lohr's $\rho$ \citep{lohr2021sampling}, an adjusted $R^2$, and the ANOVA $\rho$) agree to within roughly $0.02$, so we report the ANOVA $\rho$ only; the estimates are imprecise given the small number of clusters, so we treat them as a qualitative diagnostic and base inference on the cluster-robust CR2 standard errors and small-sample Satterthwaite tests reported in \cref{sec:robustnessanalysis,tab:cr2-satterthwaite}. These ICC estimates describe within-group outcome similarity but do not determine the appropriate design-based inference: treatment was assigned at the working-group level for the initial cohort regardless of the realized ICC. We therefore retain HC1 inference as the conventional full-sample specification but assess all five primary outcomes using the CR2 cluster-robust specifications and small-sample Satterthwaite tests reported in \cref{sec:robustnessanalysis,tab:cr2-satterthwaite}.

\paragraph{Attrition.} We observed two sources of missingness: non-systematic missing values across survey variables, for a total of 24 missing values spread across the dataframe at baseline, and 22 at endline. We further note that between baseline and endline, there was a total of 9 trial attriters, equivalent to 6.9 percent of our sample. We evaluated the relationship between attriters and experimental conditions using a logistic regression model. Moreover, attrition was asymmetric across experimental conditions: the treatment arm lost 8 of 67 participants (11.9\%) against 1 of 64 participants in control (1.6\%), and the treatment--attrition association is statistically significant at the $\alpha = 0.05$ threshold (logit $p = 0.046$). Because attrition is differential and plausibly induced by the protocol itself, we assess its impact non-parametrically with \citet{lee2009bounds} sharp bounds under monotone attrition, including 95\% Imbens--Manski confidence intervals from 2,000 arm-stratified bootstrap replications (\cref{tab:lee-bounds}). The point bounds exclude zero for stress ([0.037, 0.340]) and performance ([0.013, 0.231]) and straddle zero for the other three outcomes; however, once sampling uncertainty in the bounds is accounted for, the Imbens--Manski interval includes zero for every outcome. We therefore find no statistically distinguishable treatment effect on any primary outcome even under the worst-case selection pattern admitted by the observed attrition. The intervals are wide, reflecting the small sample and the differential attrition, and support the claim of no detectable adverse effect.

\subsection{Description of the data}
\label{section:app:describe-data}

A total of 141 volunteers responded to the recruitment call and were briefed on the trial requirements. By September 12, 131 participants had consented, completed the baseline survey, and entered the trial. Of these, 130 had the health-probability and utility data required for the testing allocation, and 122 completed the endline survey; thus, nine participants (6.9\%) did not complete the endline survey, with attrition statistically associated with treatment status. \Cref{tab:summarystatsbaseline} reports summary statistics at baseline and endline for all key variables. Of the 119 participants with valid sociability data at both baseline and endline (59 treatment and 60 control), 19 treatment and 20 control participants reported a decrease in in-person socializing, while 24 treatment and 31 control participants reported an increase. The endline summary in \cref{tab:summarystatsbaseline} contains 120 observations because the paired change calculation excludes one observation without valid sociability data at both waves. Counting unchanged-or-increased sociability as \emph{positive}, $n=40$ participants in each group exhibited positive sociability over the trial. Throughout the trial only one pooled test returned positive; on re-testing of the already-submitted saliva samples, a single infected individual was identified and isolated, and the remaining members of the pool were cleared.

\begin{table}[!htbp]
\centering
\begin{threeparttable}
\caption{Summary statistics at baseline and endline.}
\label{tab:summarystatsbaseline}
\small
\begin{tabular}[t]{lrrrrrrr}
\toprule
  & N & Mean & SD & Min & P25 & P75 & Max\\
\midrule
\addlinespace[0.3em]
\multicolumn{8}{l}{\textbf{Baseline}}\\
\hspace{1em}Treatment & 131 & 0.511 & 0.502 & 0.000 & 0.000 & 1.000 & 1.000\\
\hspace{1em}Baseline utility & 130 & 2.656 & 0.331 & 1.958 & 2.422 & 2.841 & 3.521\\
\hspace{1em}Health probability & 130 & 0.995 & 0.001 & 0.994 & 0.994 & 0.996 & 0.996\\
\hspace{1em}Age & 127 & 33.362 & 10.581 & 20.000 & 25.000 & 39.000 & 80.000\\
\hspace{1em}Dependants & 131 & 1.481 & 1.556 & 0.000 & 0.000 & 2.000 & 8.000\\
\hspace{1em}Household size & 131 & 3.015 & 1.741 & 0.000 & 2.000 & 4.000 & 8.000\\
\hspace{1em}Socio-economic class & 130 & 6.154 & 1.542 & 1.000 & 5.000 & 7.000 & 10.000\\
\hspace{1em}Socio-economic index & 131 & 2.828 & 0.662 & 1.333 & 2.250 & 3.250 & 4.500\\
\hspace{1em}Digital media index & 131 & 2.545 & 0.527 & 1.167 & 2.167 & 2.833 & 4.500\\
\hspace{1em}Sociability & 129 & 19.186 & 24.853 & 0.000 & 2.000 & 30.000 & 100.000\\
\hspace{1em}Fear & 131 & 2.985 & 1.008 & 1.000 & 2.000 & 4.000 & 5.000\\
\hspace{1em}Stress score & 131 & 2.366 & 0.696 & 1.000 & 2.000 & 3.000 & 4.000\\
\hspace{1em}Life satisfaction & 130 & 7.677 & 1.981 & 1.000 & 7.000 & 9.000 & 10.000\\
\hspace{1em}Institute satisfaction & 131 & 7.954 & 2.151 & 1.000 & 7.000 & 10.000 & 10.000\\
\hspace{1em}Psychosocial index & 131 & 2.988 & 0.565 & 1.750 & 2.604 & 3.333 & 4.417\\
\hspace{1em}Learning & 131 & 8.069 & 1.545 & 1.000 & 7.000 & 9.000 & 10.000\\
\hspace{1em}Productivity & 130 & 2.138 & 0.734 & 1.000 & 2.000 & 3.000 & 4.000\\
\hspace{1em}Performance & 130 & 2.023 & 0.762 & 1.000 & 1.000 & 3.000 & 4.000\\
\addlinespace[0.3em]
\multicolumn{8}{l}{\textbf{Endline}}\\
\hspace{1em}Treatment & 122 & 0.484 & 0.502 & 0.000 & 0.000 & 1.000 & 1.000\\
\hspace{1em}Baseline utility & 121 & 2.652 & 0.336 & 1.958 & 2.396 & 2.833 & 3.521\\
\hspace{1em}Health probability & 121 & 0.995 & 0.001 & 0.994 & 0.994 & 0.996 & 0.996\\
\hspace{1em}Sociability & 120 & 22.800 & 26.866 & 0.000 & 2.000 & 40.000 & 100.000\\
\hspace{1em}Fear & 121 & 2.736 & 1.109 & 1.000 & 2.000 & 4.000 & 5.000\\
\hspace{1em}Life satisfaction & 122 & 7.598 & 2.040 & 1.000 & 7.000 & 9.000 & 10.000\\
\hspace{1em}Institute satisfaction & 122 & 8.090 & 2.336 & 1.000 & 7.250 & 10.000 & 10.000\\
\hspace{1em}Learning & 119 & 8.277 & 1.562 & 2.000 & 8.000 & 9.000 & 10.000\\
\hspace{1em}Productivity & 120 & 2.133 & 0.721 & 1.000 & 2.000 & 3.000 & 4.000\\
\hspace{1em}Stress score & 122 & 2.331 & 0.742 & 1.000 & 1.750 & 2.750 & 4.250\\
\bottomrule
\addlinespace
\end{tabular}
\end{threeparttable}
\end{table}

\subsection{Metrics and methods}
\label{section:app:metrics-and-methods}

The analysis sample is the 122 trial participants who completed the endline survey (composition and balance are documented in \cref{section:app:describe-data,tab:cov-balance}; randomization details are in \cref{appendix:randomization}). Throughout we use a strict significance threshold of $p < 0.05$, and our core hypothesis is one of \emph{mean equivalence} rather than a positive treatment effect: because the control arm captures a fully unrestricted on-site reopening, a small or zero treatment effect on welfare and productivity outcomes is itself a substantive positive finding for the protocol.

\paragraph{Outcomes.} We denote the five primary outcomes as
$y_i \in \{s_i, w_i, l_i, p_i, pr_i\}$: stress score $s_i$ and life-satisfaction score $w_i$ (mental health), and self-assessed performance $p_i$, productivity $pr_i$, and learning $l_i$ (work and study).\footnote{We adapt a measure based on the fit of 10- and 5-point Likert scales, respectively, as per \citet{versteeg2019putting}.} Own-goal and supervisor-goal scores enter the analysis as secondary outcomes; they appear in the robustness tables but not in the equivalence, multiple-testing, or attrition-bound robustness checks.

\paragraph{Model specifications.} Three of the sixteen baseline covariates exhibit realized imbalance at $p < 0.05$ (baseline utility, digital-resources score, and baseline productivity; see \cref{tab:cov-balance}). Although random assignment ensures that the unadjusted difference in means is unbiased in expectation, covariate adjustment accounts for these realized baseline differences and can improve precision. We therefore use covariate-adjusted estimation for the primary inference. For each outcome we report two complementary linear models. The first is an ANCOVA level model: for each individual $i$,
\begin{equation}
    \label{equation1}
    Y_{i,\text{end}} = \alpha + \tau\, T_i + \gamma\, Y_{i,\text{base}}
                       + C_i'\boldsymbol{\delta} + \varepsilon_i,
\end{equation}
where $T_i$ is the treatment indicator, $Y_{i,\text{base}}$ is the baseline value of the outcome, and $C_i$ is a vector of pre-treatment covariates: gender, age, academic status, baseline utility, baseline sociability, baseline fear of COVID-19, baseline institutional satisfaction, baseline life satisfaction, baseline stress, recent illness, household size, digital-resources score, and socio-economic score. The treatment effect is the coefficient $\tau$. The second is a covariate-adjusted change score (delta) model:
\begin{equation}
    \label{equation2}
    \Delta Y_i = \alpha' + \tau'\, T_i + \tilde{C}_i'\boldsymbol{\delta}'
                 + \varepsilon_i',
\end{equation}
where $\Delta Y_i = Y_{i,\text{end}} - Y_{i,\text{base}}$ and $\tilde C_i$ is the six-element subset of $C_i$ that excludes time-invariant demographic covariates (already removed by the differencing) and the baseline outcome (built into the construction of $\Delta Y$): baseline utility, sociability, fear, institutional satisfaction, life satisfaction, and stress. ANCOVA conditions on baseline imbalance while reducing residual variance; the delta model eliminates time-invariant heterogeneity. We report both specifications because convergent inferences across them strengthen the qualitative conclusions \citep{Allison1990}.

\paragraph{Inference and robustness.} Standard errors on the primary specifications are heteroskedasticity-robust (HC1). To accommodate the mixed cluster and individually randomized design described in \cref{appendix:randomization}, we additionally report two cluster-robust variants in \cref{sec:robustnessanalysis}: CR2 standard errors at the working-group level, refit on the initial cluster-randomized cohort, and CR2 standard errors at an \emph{effective cluster} variable on the full sample (working group for initial-cohort members; individual identifier for late enrollers), following the design-based inference framework of
\citet{abadie2023clustering}. Cluster-robust hypothesis tests use Satterthwaite small-sample degrees of freedom, reported in \cref{tab:cr2-satterthwaite}. Separately from these adjusted models, equivalence between the unadjusted endline group means is assessed using two one-sided tests (TOST) \citep{lakens2017equivalence} in \cref{tab:power-analysisTOST}. Across the ten covariate-adjusted models for the five primary outcomes (5 outcomes $\times$ \{level, delta\}) we report Holm-Bonferroni (FWER) and Benjamini-Hochberg (FDR) multiple-testing corrections in \cref{tab:mt-correction}. Differential-attrition sensitivity is bounded non-parametrically via \citet{lee2009bounds} sharp bounds in \cref{tab:lee-bounds}. For the four-category Likert outcomes (performance, productivity, own goals, supervisor goals) we additionally estimate ordered logistic models as ordinal-form robustness.

\paragraph{ITT and compliance.} Our treatment coefficient estimates the intention-to-treat effect of assignment to the pooled-testing protocol. Participants assigned to treatment could decline an invitation to submit a saliva sample and therefore might not receive the testing exposure prescribed by the protocol; they nevertheless remain in the treatment arm for analysis. The allocation algorithm was rerun on the set of submitted samples, so non-submission changed the pools actually tested but did not change treatment assignment. Participants who did not submit a sample were unable to enter the premises until they subsequently obtained a negative result, but incurred no additional penalty. Because we do not instrument for testing participation or restrict the analysis to compliers, the estimates should not be interpreted as treatment effects among compliers.

\subsection{Results}
\label{section:app:results}

We present the regression analyses based on \cref{equation1,equation2} in \cref{tab:results-mentalhealth,tab:results-productivity}. Each table reports the covariate-adjusted level (ANCOVA) and change-score (delta) specifications side by side. No outcome shows a statistically significant treatment effect under either specification. This finding indicates an absence of statistically detectable differences in participant well-being and work or study outcomes, but does not by itself establish equivalence. The separate TOST analysis of unadjusted endline group means establishes equivalence at the $d=0.5$ margin for life satisfaction, performance, productivity, and learning (\cref{tab:power-analysisTOST}). For stress, equivalence at this margin is not established ($p=0.079$). The multiple-testing corrections preserve the conclusion that none of the ten adjusted estimates is statistically distinguishable from zero (\cref{tab:mt-correction}), while the Lee-bound confidence intervals likewise include zero but are too wide to establish equivalence (\cref{tab:lee-bounds}).

The change-score specifications corroborate the level ANCOVA results. The largest delta-side treatment estimate in magnitude is a coefficient of $-0.196$ on the change in productivity ($p = 0.155$). Because productivity is coded from 1 (high) to 5 (poor), lower scores indicate better self-assessed productivity. The negative estimate therefore directionally favors treatment. Descriptively, the raw arm means, calculated separately at each survey wave, moved from approximately $2.27$ to $2.17$ in treatment and from $2.00$ to $2.09$ in control. Note that these wave-specific means are not paired-person changes. The adjusted estimate remains statistically non-significant and is also non-significant after multiple-testing correction (Holm and Benjamini-Hochberg adjusted $p \geq 0.25$; \cref{tab:mt-correction}). All other delta-model treatment effects are smaller in magnitude and insignificant.

\paragraph{Mental health outcomes.} \cref{tab:results-mentalhealth} shows no statistically significant effect of the pooled-testing protocol on stress or subjective well-being. Under the ANCOVA level specification, treatment participants report a stress score $0.022$ points higher than control ($p = 0.86$); the small unadjusted treatment-control gap visible in the raw means disappears once we adjust for baseline stress and the standard pre-treatment covariate set. The corresponding change score (delta) coefficient on stress is $0.076$ ($p = 0.57$), indicating that treatment status induces little to no variation in stress between $t_0$ and $t_1$. For life satisfaction, treatment participants report a higher endline score ($\hat\tau = 0.293$, $p = 0.50$, ANCOVA) with a parallel delta-model coefficient of $0.322$ ($p = 0.46$). All four estimates remain non-significant under Holm-Bonferroni and Benjamini-Hochberg corrections across the ten primary tests (\cref{tab:mt-correction}), and the cluster-randomization-consistent robustness checks (\cref{tab:lm-mentalhealth-cov-cluster,tab:lm-mentalhealth-cov-effective}) yield qualitatively identical conclusions.

\begin{table}[!htbp]
\centering
\begin{threeparttable}
\caption{Treatment effect on mental-health outcomes.}
\label{tab:results-mentalhealth}
\small
\begin{tabular}[t]{lcccc}
\toprule
  & Stress & $\Delta$ Stress & Life sat. & $\Delta$ Life sat.\\
\midrule
Treatment & 0.022 & 0.076 & 0.293 & 0.322\\
 & (0.127) & (0.134) & (0.429) & (0.437)\\
Female & 0.104 &  & 0.035 & \\
 & (0.150) &  & (0.405) & \\
Prefer not to say & -1.254** &  & 1.441 & \\
 & (0.388) &  & (1.292) & \\
Age & -0.012 &  & 0.015 & \\
 & (0.008) &  & (0.019) & \\
Academic vs. staff & 0.148 &  & 0.684 & \\
 & (0.179) &  & (0.396) & \\
Utility & 0.400 & 0.120 & -0.017 & -0.222\\
 & (0.418) & (0.225) & (1.215) & (0.701)\\
Sociability & 0.002 & 0.000 & 0.001 & -0.006\\
 & (0.003) & (0.003) & (0.010) & (0.010)\\
Fear & 0.028 & -0.014 & 0.096 & 0.085\\
 & (0.076) & (0.076) & (0.188) & (0.225)\\
Institute satisfaction & -0.053 & -0.039 & -0.066 & -0.137\\
 & (0.039) & (0.037) & (0.101) & (0.111)\\
Life satisfaction & 0.026 & 0.167*** & 0.319 & \\
 & (0.049) & (0.042) & (0.196) & \\
Stress score & 0.420* &  & -0.320 & 0.738*\\
 & (0.164) &  & (0.414) & (0.321)\\
C19 recovered & -0.020 &  & -1.407 & \\
 & (0.348) &  & (1.159) & \\
Not recovered & -0.243 &  & -1.194 & \\
 & (0.335) &  & (1.085) & \\
Household size & -0.037 &  & 0.200 & \\
 & (0.039) &  & (0.146) & \\
Digital media score & 0.068 &  & -0.179 & \\
 & (0.137) &  & (0.534) & \\
Socio-economic score & -0.009 &  & -0.504 & \\
 & (0.182) &  & (0.588) & \\
Constant & 0.796 & -1.307 & 7.811* & -0.495\\
 & (0.859) & (0.767) & (3.521) & (2.086)\\
Observations & 115 & 119 & 115 & 119\\
$R^{2}$ & 0.364 & 0.161 & 0.203 & 0.084\\
Adjusted $R^{2}$ & 0.261 & 0.116 & 0.073 & 0.035\\
\bottomrule
\addlinespace
\end{tabular}
\begin{tablenotes}[flushleft]
\footnotesize
\item[] \emph{Notes.} Outcomes are stress and subjective well-being. $^* p < 0.05$, $^** p < 0.01$, $^*** p < 0.001$. Heteroskedasticity-robust (HC1) standard errors are in parentheses. Full endline sample ($n \approx 122$). Level columns are ANCOVA; delta columns regress the change in outcome on treatment plus baseline covariates. Exact treatment-effect estimates, robust SEs, raw $p$-values, and multiplicity-adjusted $p$-values are reported in \cref{tab:mt-correction}.
\end{tablenotes}
\end{threeparttable}
\end{table}

\begin{table}[!htbp]
\centering
\begin{threeparttable}
\caption{Treatment effect on work and study outcomes.}
\label{tab:results-productivity}
\small
\arxivtablefit{\footnotesize}{4.25pt}
\begin{tabular}[t]{lcccccc}
\toprule
  & Performance & $\Delta$ Performance & Productivity & $\Delta$ Productivity & Learning & $\Delta$ Learning\\
\midrule
Treatment & 0.070 & 0.038 & -0.134 & -0.196 & 0.338 & -0.041\\
 & (0.145) & (0.152) & (0.138) & (0.137) & (0.294) & (0.326)\\
Baseline outcome & 0.456*** &  & 0.367** &  & 0.269 & \\
 & (0.115) &  & (0.138) &  & (0.145) & \\
Female & 0.078 &  & 0.093 &  & 0.330 & \\
 & (0.135) &  & (0.136) &  & (0.286) & \\
Prefer not to say & -0.042 &  & -0.034 &  & 0.639 & \\
 & (0.347) &  & (0.351) &  & (0.898) & \\
Age & 0.000 &  & -0.004 &  & -0.003 & \\
 & (0.006) &  & (0.006) &  & (0.014) & \\
Academic vs. staff & 0.375* &  & 0.284 &  & -0.316 & \\
 & (0.174) &  & (0.177) &  & (0.357) & \\
Utility & 0.050 & -0.509 & 0.492 & -0.459* & 0.626 & 0.223\\
 & (0.493) & (0.257) & (0.514) & (0.212) & (1.000) & (0.620)\\
Sociability & 0.003 & -0.001 & 0.002 & -0.002 & 0.002 & -0.001\\
 & (0.003) & (0.002) & (0.003) & (0.003) & (0.007) & (0.007)\\
Fear & 0.015 & -0.008 & 0.028 & 0.035 & 0.190 & 0.296\\
 & (0.081) & (0.088) & (0.080) & (0.079) & (0.135) & (0.171)\\
Institute satisfaction & -0.094** & -0.076* & -0.065 & -0.029 & 0.048 & -0.159\\
 & (0.035) & (0.037) & (0.037) & (0.032) & (0.079) & (0.097)\\
Life satisfaction & -0.010 & -0.017 & 0.026 & 0.004 & 0.142 & 0.067\\
 & (0.054) & (0.050) & (0.056) & (0.047) & (0.155) & (0.167)\\
Stress score & -0.072 & -0.062 & 0.060 & 0.125 & -0.380 & -0.252\\
 & (0.130) & (0.129) & (0.124) & (0.121) & (0.244) & (0.269)\\
C19 recovered & -0.542* &  & -0.166 &  & -0.784 & \\
 & (0.264) &  & (0.354) &  & (0.917) & \\
Not recovered & -0.601* &  & -0.237 &  & -0.474 & \\
 & (0.257) &  & (0.343) &  & (0.915) & \\
Household size & -0.034 &  & -0.021 &  & 0.015 & \\
 & (0.032) &  & (0.032) &  & (0.118) & \\
Digital media score & -0.232 &  & -0.190 &  & -0.183 & \\
 & (0.183) &  & (0.168) &  & (0.446) & \\
Socio-economic score & 0.249 &  & -0.010 &  & -0.574 & \\
 & (0.166) &  & (0.178) &  & (0.504) & \\
Constant & 2.222 & 2.214* & 0.913 & 1.114 & 5.683 & 0.100\\
 & (1.180) & (1.111) & (1.020) & (0.843) & (3.027) & (2.572)\\
Observations & 112 & 116 & 113 & 117 & 112 & 116\\
$R^{2}$ & 0.406 & 0.075 & 0.350 & 0.071 & 0.300 & 0.059\\
Adjusted $R^{2}$ & 0.298 & 0.015 & 0.234 & 0.011 & 0.173 & -0.002\\
\bottomrule
\addlinespace
\end{tabular}
\begin{tablenotes}[flushleft]
\footnotesize
\item[] \emph{Notes.} Outcomes are performance, productivity, and learning. $^* p < 0.05$, $^** p < 0.01$, $^*** p < 0.001$. Heteroskedasticity-robust (HC1) standard errors are in parentheses. Full endline sample ($n \approx 122$). Level columns are ANCOVA; delta columns regress the change in outcome on treatment plus baseline covariates. Exact treatment-effect estimates, robust SEs, raw $p$-values, and multiplicity-adjusted $p$-values are reported in \cref{tab:mt-correction}.
\end{tablenotes}
\end{threeparttable}
\end{table}

\subsection{Robustness Analyses}
\label{sec:robustnessanalysis}

This section reports the robustness checks discussed in
\cref{section:app:metrics-and-methods}. The null finding, i.e. no statistically significant treatment effect on any outcome under our
pre-specified $p < 0.05$ threshold, is preserved quantitatively and qualitatively across a range of tests.

\paragraph{Ordered logistic regressions for categorical outcomes.} Performance, productivity, own-goal, and supervisor-goal scores are recorded using five ordered response categories. Performance and productivity use four observed categories in our sample, whereas the goal outcomes use all five. The covariate-adjusted ordered-logit estimates in \cref{tab:orderedlogit-cov} corroborate the linear-model findings: the treatment odds ratios are $1.170$ (performance), $0.606$ (productivity), $0.811$ (own goals), and $1.232$ (supervisor goals). Because higher productivity categories indicate worse outcomes, the productivity odds ratio below one directionally favors treatment; the performance odds ratio above one points in the opposite direction. Neither estimate is statistically distinguishable from one. The baseline outcome is the only consistently significant predictor across columns, as expected for an ANCOVA-style specification.

\paragraph{Cluster-randomization-consistent standard errors.}
The two cluster-robust specifications, CR2 at the working group with Satterthwaite small-sample tests for the treatment effect, restricted to the initial cluster-randomized cohort (\cref{tab:lm-mentalhealth-cov-cluster,tab:lm-performance-cov-cluster}), and CR2 at the \emph{effective cluster} on the full sample (\cref{tab:lm-mentalhealth-cov-effective,tab:lm-performance-cov-effective}), yield treatment coefficients essentially identical to the primary HC1 estimates. Standard errors widen modestly for the work and study outcomes, consistent with the small but non-zero intracluster correlations reported in \cref{tab:ICC}, and are essentially unchanged for the mental-health outcomes. No coefficient changes statistical inference under either cluster definition.\footnote{Note that the significance markers in the cluster regression tables use the model residual degrees of freedom; the small-sample inference for the treatment effect, using CR2 variance with Satterthwaite degrees of freedom, is reported in \cref{tab:cr2-satterthwaite}. The Satterthwaite degrees of freedom range from approximately $12.7$ to $14.4$ on the cluster-randomized cohort and from $15.7$ to $18.6$ on the full sample. Every treatment effect remains non-significant under this stricter test, so our conclusions do not rely on the more liberal table degrees of freedom.}

\paragraph{Multiple-testing correction.} We correct the ten primary hypothesis tests (five outcomes $\times$ two specifications) for multiplicity using Holm-Bonferroni (FWER) and Benjamini-Hochberg (FDR); results are reported in \cref{tab:mt-correction}. Adjusted $p$-values exceed our pre-specified $p < 0.05$ threshold for every outcome and specification, both across the full ten-test family and within the levels-only and deltas-only
subfamilies. The smallest unadjusted $p$-value belongs to a coefficient of $-0.196$ on the change in productivity ($p = 0.155$), which corresponds to a Holm/BH-adjusted $p \geq 0.25$.

\paragraph{Attrition bounds.} The Lee point-identified sets lie above zero for stress ($[0.037,0.340]$) and performance ($[0.013,0.231]$), while those for life satisfaction, productivity, and learning include zero (\cref{tab:lee-bounds}). Given the outcome coding, the stress and performance point bounds correspond to higher stress and poorer self-assessed performance under treatment assignment. However, the 95\% Imbens--Manski confidence intervals include zero for all five outcomes. The point bounds may therefore indicate adverse assignment effects on stress and performance under the maintained monotonicity assumption, but sampling uncertainty prevents concluding that either effect is statistically distinguishable from zero. These intervals are also too wide to establish equivalence.

\begin{table}[!htbp]
\centering
\begin{threeparttable}
\caption{Ordered-logit robustness for categorical outcomes.}
\label{tab:orderedlogit-cov}
\small
\begin{tabular}{@{}lcccc@{}}
\toprule
 & Performance & Productivity & Own goals & Sup. goals \\
\midrule
Treatment & 1.170 & 0.606 & 0.811 & 1.232 \\
 & (0.520) & (0.266) & (0.325) & (0.500) \\
Baseline outcome & 4.549*** & 3.465** & 2.452* & 1.373 \\
 & (1.745) & (1.406) & (0.855) & (0.494) \\
Female & 1.453 & 1.424 & 1.455 & 1.552 \\
 & (0.634) & (0.623) & (0.603) & (0.635) \\
Prefer not to say & 0.810 & 1.292 & 4.563 & 1.708 \\
 & (1.793) & (3.467) & (8.699) & (3.314) \\
Age & 0.998 & 0.989 & 1.005 & 0.993 \\
 & (0.020) & (0.021) & (0.020) & (0.020) \\
Academic vs. staff & 3.428* & 2.626 & 2.078 & 1.063 \\
 & (1.790) & (1.399) & (0.972) & (0.508) \\
\addlinespace
Utility & 1.421 & 6.803 & 0.784 & 29.743* \\
 & (2.056) & (10.075) & (1.223) & (49.370) \\
Sociability & 1.010 & 1.007 & 0.997 & 1.004 \\
 & (0.009) & (0.010) & (0.009) & (0.010) \\
Fear & 1.085 & 1.122 & 0.711 & 0.766 \\
 & (0.243) & (0.252) & (0.146) & (0.164) \\
Institute satisfaction & 0.728** & 0.799* & 0.878 & 0.914 \\
 & (0.081) & (0.088) & (0.084) & (0.092) \\
Life satisfaction & 0.977 & 1.111 & 1.197 & 1.292 \\
 & (0.161) & (0.184) & (0.187) & (0.199) \\
Stress score & 0.787 & 1.206 & 1.780 & 1.354 \\
 & (0.315) & (0.474) & (0.674) & (0.504) \\
\addlinespace
C19 recovered & 0.160 & 0.574 & 0.456 & 0.364 \\
 & (0.171) & (0.636) & (0.488) & (0.381) \\
Not recovered & 0.143 & 0.480 & 0.349 & 0.177 \\
 & (0.148) & (0.517) & (0.363) & (0.180) \\
Household size & 0.908 & 0.934 & 0.861 & 0.864 \\
 & (0.115) & (0.121) & (0.108) & (0.111) \\
Digital media & 0.432 & 0.486 & 1.161 & 0.761 \\
 & (0.242) & (0.260) & (0.649) & (0.424) \\
Socio-economic & 2.078 & 0.862 & 1.406 & 0.697 \\
 & (1.072) & (0.437) & (0.737) & (0.382) \\
\addlinespace
Observations & 112 & 113 & 112 & 113 \\
\bottomrule
\addlinespace
\end{tabular}
\begin{tablenotes}[flushleft]
\footnotesize
\item[] \emph{Notes.} Covariate-adjusted ordered-logit estimates for the four categorical Likert outcomes. Each model regresses the ordered endline outcome on treatment, its baseline value, and the same pre-treatment covariates as the primary ANCOVA specifications (see \cref{section:app:metrics-and-methods}). Entries are odds ratios; odds ratios greater than 1 indicate higher odds of reporting a higher outcome category. Default Hessian-based standard errors, transformed to the odds-ratio scale, are in parentheses. $^{*}\, p < 0.05$, $^{**}\, p < 0.01$, $^{***}\, p < 0.001$.
\end{tablenotes}
\end{threeparttable}
\end{table}

\begin{table}[!htbp]
\centering
\begin{threeparttable}
\caption{Mental-health outcomes with CR2 standard errors.}
\label{tab:lm-mentalhealth-cov-cluster}
\small
\begin{tabular}[t]{lcccc}
\toprule
  & Stress & Life satisfaction & $\Delta$ Stress & $\Delta$ Life satisfaction\\
\midrule
Treatment & 0.013 & 0.467 & 0.055 & 0.312\\
 & (0.133) & (0.345) & (0.113) & (0.379)\\
Female & 0.073 & 0.100 &  & \\
 & (0.140) & (0.360) &  & \\
Prefer not to say & -1.285*** & 1.950 &  & \\
 & (0.421) & (1.317) &  & \\
Age & -0.009 & 0.013 &  & \\
 & (0.007) & (0.020) &  & \\
Academic vs. staff & 0.207 & 0.513 &  & \\
 & (0.206) & (0.395) &  & \\
Utility & 0.289 & 0.030 & 0.155 & -0.259\\
 & (0.454) & (1.169) & (0.252) & (0.697)\\
Sociability & 0.002 & -0.001 & 0.000 & -0.008\\
 & (0.003) & (0.011) & (0.004) & (0.009)\\
Fear & 0.023 & 0.004 & -0.020 & 0.040\\
 & (0.091) & (0.229) & (0.082) & (0.278)\\
Institute satisfaction & -0.038 & -0.078 & -0.025 & -0.119\\
 & (0.043) & (0.094) & (0.040) & (0.102)\\
Life satisfaction & -0.005 & 0.443** & 0.170*** & \\
 & (0.050) & (0.187) & (0.040) & \\
Stress score & 0.386** & -0.156 &  & 0.849***\\
 & (0.192) & (0.382) &  & (0.317)\\
C19 recovered & 0.351 & -2.358* &  & \\
 & (0.257) & (1.252) &  & \\
Not recovered & 0.100 & -2.205* &  & \\
 & (0.243) & (1.258) &  & \\
Household size & -0.054 & 0.141 &  & \\
 & (0.045) & (0.143) &  & \\
Digital media score & 0.100 & -0.387 &  & \\
 & (0.145) & (0.599) &  & \\
Socio-economic score & 0.052 & -0.324 &  & \\
 & (0.259) & (0.753) &  & \\
Constant & 0.651 & 7.919** & -1.521 & -0.586\\
 & (0.960) & (3.623) & (0.921) & (2.304)\\
Observations & 100 & 100 & 104 & 104\\
$R^{2}$ & 0.360 & 0.252 & 0.169 & 0.108\\
Adjusted $R^{2}$ & 0.236 & 0.108 & 0.118 & 0.052\\
\bottomrule
\addlinespace
\end{tabular}
\begin{tablenotes}[flushleft]
\footnotesize
\item[] \emph{Notes.} Covariate-adjusted level and delta models for stress and life satisfaction, refit on the initial cluster-randomized cohort with CR2 cluster-robust standard errors at the working group. $n \approx 102$. $^* p < 0.1$, $^** p < 0.05$, $^*** p < 0.01$. Stars use model residual degrees of freedom; small-sample CR2/Satterthwaite treatment tests are reported in \cref{tab:cr2-satterthwaite}.
\end{tablenotes}
\end{threeparttable}
\end{table}

\begin{table}[!htbp]
\centering
\begin{threeparttable}
\caption{Work and study outcomes with CR2 standard errors.}
\label{tab:lm-performance-cov-cluster}
\small
\arxivtablefit{\footnotesize}{4.25pt}
\begin{tabular}[t]{lcccccc}
\toprule
  & Performance & Productivity & Learning & $\Delta$ Performance & $\Delta$ Productivity & $\Delta$ Learning\\
\midrule
Treatment & 0.121 & -0.104 & 0.192 & 0.084 & -0.155 & -0.238\\
 & (0.189) & (0.165) & (0.412) & (0.203) & (0.166) & (0.380)\\
Baseline outcome & 0.481*** & 0.345** & 0.339** &  &  & \\
 & (0.167) & (0.140) & (0.138) &  &  & \\
Female & 0.019 & 0.048 & 0.317 &  &  & \\
 & (0.138) & (0.149) & (0.278) &  &  & \\
Prefer not to say & -0.168 & -0.190 & 1.003 &  &  & \\
 & (0.301) & (0.388) & (0.873) &  &  & \\
Age & 0.000 & -0.003 & 0.004 &  &  & \\
 & (0.005) & (0.008) & (0.014) &  &  & \\
Academic vs. staff & 0.410** & 0.348 & -0.435 &  &  & \\
 & (0.190) & (0.223) & (0.430) &  &  & \\
Utility & -0.129 & 0.466 & -0.107 & -0.567** & -0.489*** & 0.337\\
 & (0.705) & (0.663) & (1.158) & (0.260) & (0.178) & (0.566)\\
Sociability & 0.003 & 0.003 & 0.000 & 0.000 & -0.002 & -0.002\\
 & (0.003) & (0.003) & (0.009) & (0.003) & (0.002) & (0.007)\\
Fear & -0.005 & 0.036 & 0.118 & -0.033 & 0.031 & 0.209\\
 & (0.097) & (0.072) & (0.106) & (0.096) & (0.069) & (0.131)\\
Institute satisfaction & -0.108*** & -0.068 & 0.053 & -0.080** & -0.027 & -0.128\\
 & (0.038) & (0.048) & (0.080) & (0.032) & (0.037) & (0.132)\\
Life satisfaction & -0.028 & -0.006 & 0.154 & -0.053 & -0.037 & 0.194\\
 & (0.063) & (0.084) & (0.148) & (0.065) & (0.055) & (0.152)\\
Stress score & -0.093 & 0.020 & -0.282 & -0.124 & 0.057 & -0.110\\
 & (0.167) & (0.186) & (0.287) & (0.200) & (0.159) & (0.316)\\
C19 recovered & -0.666 & 0.101 & -0.467 &  &  & \\
 & (0.465) & (0.437) & (1.169) &  &  & \\
Not recovered & -0.754 & 0.017 & -0.283 &  &  & \\
 & (0.456) & (0.392) & (1.144) &  &  & \\
Household size & -0.044 & -0.020 & -0.076 &  &  & \\
 & (0.037) & (0.052) & (0.070) &  &  & \\
Digital media score & -0.264 & -0.218 & 0.043 &  &  & \\
 & (0.289) & (0.263) & (0.388) &  &  & \\
Socio-economic score & 0.299 & -0.018 & -0.017 &  &  & \\
 & (0.260) & (0.257) & (0.395) &  &  & \\
Constant & 3.107*** & 1.126 & 4.749 & 2.880** & 1.622* & -1.432\\
 & (1.142) & (1.029) & (3.674) & (1.336) & (0.929) & (2.305)\\
Observations & 97 & 98 & 97 & 101 & 102 & 101\\
$R^{2}$ & 0.444 & 0.363 & 0.343 & 0.104 & 0.073 & 0.087\\
Adjusted $R^{2}$ & 0.324 & 0.228 & 0.201 & 0.037 & 0.004 & 0.018\\
\bottomrule
\addlinespace
\end{tabular}
\begin{tablenotes}[flushleft]
\footnotesize
\item[] \emph{Notes.} Covariate-adjusted level and delta models for performance, productivity, and learning, refit on the initial cluster-randomized cohort with CR2 cluster-robust standard errors at the working group. $n \approx 102$. $^* p < 0.1$, $^** p < 0.05$, $^*** p < 0.01$. Stars use model residual degrees of freedom; small-sample CR2/Satterthwaite treatment tests are reported in \cref{tab:cr2-satterthwaite}.
\end{tablenotes}
\end{threeparttable}
\end{table}

\paragraph{Distributions of outcome variables.}
\label{section:app:outcomedistribution}

We evaluate the testing allocation protocol with linear models, accepting that some outcomes are not strictly continuous, on the understanding that our interest is in mean equivalence between
experimental conditions rather than in the magnitude of the coefficients. Stress (a mean of four $1$--$5$ items, including fractions) and life satisfaction ($1$--$10$ Likert) are continuous or near-continuous and approximately normally distributed in our sample. Learning is coded on a $1$--$10$ Likert scale and is similarly close to normal. Performance, productivity, own goals, and supervisor goals, by contrast, are coded on $1$--$5$ ordinal Likert scales. In practice, the top category (5) was not selected for performance or productivity, so these two primary outcomes take observed values in $\{1, 2, 3, 4\}$; the secondary own-goal and supervisor-goal outcomes use the full $\{1, 2, 3, 4, 5\}$ support. The covariate-adjusted ordered-logit robustness check in \cref{tab:orderedlogit-cov} accommodates this categorical structure, and its treatment odds ratios corroborate the ANCOVA primary findings.

\begin{table}[!htbp]
\centering
\begin{threeparttable}
\caption{Small-sample CR2 tests of treatment effects.}
\label{tab:cr2-satterthwaite}
\small
\arxivtablefit{\small}{5.5pt}
    \begin{tabular}{llccccc}
        \toprule
         & {Specification} & {Cluster} & {Estimate} & {CR2 SE} & {$df$ (Satt.)} & {$p$-value} \\
         \midrule
         {Stress} & level & group\_id (cluster-RA) & 0.013 & 0.133 & 12.693 & 0.925 \\
         {Stress} & delta & group\_id (cluster-RA) & 0.055 & 0.113 & 13.790 & 0.631 \\
         {Stress} & level & effective cluster (full) & 0.022 & 0.122 & 15.672 & 0.862 \\
         {Stress} & delta & effective cluster (full) & 0.076 & 0.115 & 17.748 & 0.519 \\
         {Life satisfaction} & level & group\_id (cluster-RA) & 0.467 & 0.345 & 12.693 & 0.200 \\
         {Life satisfaction} & delta & group\_id (cluster-RA) & 0.312 & 0.379 & 13.905 & 0.424 \\
         {Life satisfaction} & level & effective cluster (full) & 0.293 & 0.347 & 15.672 & 0.411 \\
         {Life satisfaction} & delta & effective cluster (full) & 0.322 & 0.367 & 18.170 & 0.391 \\
         {Performance} & level & group\_id (cluster-RA) & 0.121 & 0.189 & 13.377 & 0.532 \\
         {Performance} & delta & group\_id (cluster-RA) & 0.084 & 0.203 & 14.403 & 0.685 \\
         {Performance} & level & effective cluster (full) & 0.070 & 0.169 & 16.478 & 0.685 \\
         {Performance} & delta & effective cluster (full) & 0.038 & 0.181 & 18.569 & 0.834 \\
         {Productivity} & level & group\_id (cluster-RA) & -0.104 & 0.165 & 12.802 & 0.538 \\
         {Productivity} & delta & group\_id (cluster-RA) & -0.155 & 0.166 & 14.184 & 0.367 \\
         {Productivity} & level & effective cluster (full) & -0.134 & 0.144 & 15.865 & 0.366 \\
         {Productivity} & delta & effective cluster (full) & -0.196 & 0.149 & 18.396 & 0.203 \\
         {Learning} & level & group\_id (cluster-RA) & 0.192 & 0.412 & 13.033 & 0.649 \\
         {Learning} & delta & group\_id (cluster-RA) & -0.238 & 0.380 & 13.717 & 0.542 \\
         {Learning} & level & effective cluster (full) & 0.338 & 0.370 & 16.068 & 0.373 \\
         {Learning} & delta & effective cluster (full) & -0.041 & 0.389 & 17.936 & 0.917 \\
        \bottomrule
        \addlinespace
    \end{tabular}
\begin{tablenotes}[flushleft]
\footnotesize
\item[] \emph{Notes.} Tests use CR2 cluster-robust standard errors with Satterthwaite small-sample degrees of freedom under both cluster definitions: working-group ID for the initial cluster-randomized cohort and effective cluster---working-group ID for initial-cohort members and user ID for late enrollers---for the full sample. The Satterthwaite degrees of freedom range from $12.693$ to $14.403$ for the initial cluster-randomized cohort and from $15.672$ to $18.569$ for the full sample; they vary by outcome and specification. Every treatment effect remains non-significant at $p<0.05$.
\end{tablenotes}
\end{threeparttable}
\end{table}

\subsection{Additional tables}

\begin{table}[!htbp]
    \centering
    \begin{threeparttable}
    \caption{TOST inputs: unadjusted endline summaries by experimental condition.}
    \label{tab:TOST-input}
    \small
    \begin{tabular}{lccccccc}
        \toprule
         & {Mean 1(t)} & {Mean 2(c)} & {SD 1(t)} & {SD 2(c)} & {$N$ 1(t)} & {$N$ 2(c)} & {Pooled SD} \\
         \midrule
         {Stress} & 2.424 & 2.245 & 0.755 & 0.725 & 59 & 63 & 0.740 \\
         {Life satisfaction} & 7.644 & 7.556 & 2.041 & 2.054 & 59 & 63 & 2.048 \\
         {Performance} & 2.103 & 1.984 & 0.872 & 0.671 & 58 & 61 & 0.775 \\
         {Productivity} & 2.172 & 2.097 & 0.798 & 0.646 & 58 & 62 & 0.723 \\
         {Learning} & 8.368 & 8.194 & 1.508 & 1.618 & 57 & 62 & 1.566 \\
        \bottomrule
        \addlinespace
    \end{tabular}
\begin{tablenotes}[flushleft]
\footnotesize
\item[] \emph{Notes.} Unadjusted endline means, standard deviations, and outcome-specific complete-case sample sizes for treatment (t) and control (c). The pooled standard deviations are used to convert the standardized $d=0.5$ margin into the raw-unit bounds reported in \cref{tab:power-analysisTOST}. These summaries are not covariate-adjusted estimates.
\end{tablenotes}
\end{threeparttable}
\end{table}

\begin{table}[!htbp]
\centering
\begin{threeparttable}
\caption{Covariate balance analysis.}
\label{tab:cov-balance}
\small
\begin{tabular}[t]{lcclc}
\toprule
Covariate & Difference (mean) & Method & $p$-value & Status\\
\midrule
Utility & 0.1693 & Welch t-test & 0.0031 & unbalanced\\
Gender & N/A & Fisher exact & 0.7933 & balanced\\
Age & -1.2273 & Welch t-test & 0.5163 & balanced\\
Role (Staff vs. Academics) & N/A & Fisher exact & 0.3034 & balanced\\
C19 recovered & N/A & Fisher exact & 0.6320 & balanced\\
Household size & 0.3048 & Welch t-test & 0.3196 & balanced\\
Socio-economic status & 0.1199 & Welch t-test & 0.3020 & balanced\\
Sociability & -0.7532 & Welch t-test & 0.8636 & balanced\\
Stress (baseline) & 0.1054 & Welch t-test & 0.3864 & balanced\\
Learning & 0.0121 & Welch t-test & 0.9643 & balanced\\
Life satisfaction & -0.0824 & Welch t-test & 0.8132 & balanced\\
Productivity & 0.2727 & Welch t-test & 0.0333 & unbalanced\\
Institutional satisfaction & 0.0326 & Welch t-test & 0.9314 & balanced\\
Digital resources score & 0.2602 & Welch t-test & 0.0042 & unbalanced\\
Own goals (achieving) & 0.2438 & Welch t-test & 0.1445 & balanced\\
Supervisor goals (achieving) & 0.2959 & Welch t-test & 0.0773 & balanced\\
\bottomrule
\addlinespace
\end{tabular}
\begin{tablenotes}[flushleft]
\footnotesize
\item[] \emph{Notes.} Continuous covariates use Welch's two-sample t-test; categorical covariates use Fisher's exact test. Status `unbalanced' indicates $p < 0.05$.
\end{tablenotes}
\end{threeparttable}
\end{table}

\begin{table}[!htbp]
\centering
\begin{threeparttable}
\caption{Treatment-effect summary with multiplicity corrections.}
\label{tab:mt-correction}
\small
\arxivtablefit{\small}{4.5pt}
    \begin{tabular}{lcccccccc}
        \toprule
         & {Family} & {Estimate} & {SE} & {Raw $p$} & {Holm all} & {BH all} & {Holm w-fam} & {BH w-fam} \\
         \midrule
         {Stress} & level & 0.022 & 0.127 & 0.865 & 1.000 & 0.900 & 1.000 & 0.865 \\
         {Life satisfaction} & level & 0.293 & 0.429 & 0.496 & 1.000 & 0.900 & 1.000 & 0.788 \\
         {Performance} & level & 0.070 & 0.145 & 0.630 & 1.000 & 0.900 & 1.000 & 0.788 \\
         {Productivity} & level & -0.134 & 0.138 & 0.335 & 1.000 & 0.900 & 1.000 & 0.788 \\
         {Learning} & level & 0.338 & 0.294 & 0.252 & 1.000 & 0.900 & 1.000 & 0.788 \\
         {$\Delta$ Stress} & delta & 0.076 & 0.134 & 0.573 & 1.000 & 0.900 & 1.000 & 0.900 \\
         {$\Delta$ Life satisfaction} & delta & 0.322 & 0.437 & 0.462 & 1.000 & 0.900 & 1.000 & 0.900 \\
         {$\Delta$ Performance} & delta & 0.038 & 0.152 & 0.801 & 1.000 & 0.900 & 1.000 & 0.900 \\
         {$\Delta$ Productivity} & delta & -0.196 & 0.137 & 0.155 & 1.000 & 0.900 & 0.774 & 0.774 \\
         {$\Delta$ Learning} & delta & -0.041 & 0.326 & 0.900 & 1.000 & 0.900 & 1.000 & 0.900 \\
        \bottomrule
        \addlinespace
    \end{tabular}
\begin{tablenotes}[flushleft]
\footnotesize
\item[] \emph{Notes.} Covariate-adjusted specifications use ANCOVA for levels and change score with baseline covariates for deltas, with HC1 robust inference. The table reports the treatment-effect estimate, robust standard error (SE), raw $p$-value, and multiplicity-adjusted $p$-values across the ten primary tests. Holm = Holm--Bonferroni family-wise error correction; BH = Benjamini--Hochberg false discovery rate correction.
\end{tablenotes}
\end{threeparttable}
\end{table}

\begin{table}[!htbp]
\centering
\begin{threeparttable}
\caption{Mental-health outcomes with effective-cluster CR2 standard errors.}
\label{tab:lm-mentalhealth-cov-effective}
\small
\begin{tabular}[t]{lcccc}
\toprule
  & Stress & Life satisfaction & $\Delta$ Stress & $\Delta$ Life satisfaction\\
\midrule
Treat vs. control & 0.022 & 0.293 & 0.076 & 0.322\\
 & (0.122) & (0.347) & (0.115) & (0.367)\\
Female & 0.104 & 0.035 &  & \\
 & (0.128) & (0.372) &  & \\
Prefer not to say & -1.254*** & 1.441 &  & \\
 & (0.369) & (1.263) &  & \\
Age & -0.012 & 0.015 &  & \\
 & (0.008) & (0.021) &  & \\
Academic vs. staff & 0.148 & 0.684* &  & \\
 & (0.190) & (0.406) &  & \\
Utility & 0.400 & -0.017 & 0.120 & -0.222\\
 & (0.413) & (1.194) & (0.243) & (0.653)\\
Sociability & 0.002 & 0.001 & 0.000 & -0.006\\
 & (0.003) & (0.010) & (0.004) & (0.009)\\
Fear & 0.028 & 0.096 & -0.014 & 0.085\\
 & (0.080) & (0.222) & (0.076) & (0.284)\\
Institute satisfaction & -0.053 & -0.066 & -0.039 & -0.137\\
 & (0.043) & (0.105) & (0.038) & (0.106)\\
Life satisfaction & 0.026 & 0.319* & 0.167*** & \\
 & (0.046) & (0.168) & (0.037) & \\
Stress score & 0.420** & -0.320 &  & 0.738**\\
 & (0.172) & (0.355) &  & (0.316)\\
C19 recovered & -0.020 & -1.407 &  & \\
 & (0.394) & (1.260) &  & \\
Not recovered & -0.243 & -1.194 &  & \\
 & (0.376) & (1.215) &  & \\
Household size & -0.037 & 0.200 &  & \\
 & (0.044) & (0.153) &  & \\
Digital media score & 0.068 & -0.179 &  & \\
 & (0.129) & (0.540) &  & \\
Socio-economic score & -0.009 & -0.504 &  & \\
 & (0.207) & (0.692) &  & \\
Constant & 0.796 & 7.811** & -1.307 & -0.495\\
 & (0.900) & (3.558) & (0.875) & (2.155)\\
Observations & 115 & 115 & 119 & 119\\
$R^{2}$ & 0.364 & 0.203 & 0.161 & 0.084\\
Adjusted $R^{2}$ & 0.261 & 0.073 & 0.116 & 0.035\\
\bottomrule
\addlinespace
\end{tabular}
\begin{tablenotes}[flushleft]
\footnotesize
\item[] \emph{Notes.} Covariate-adjusted level and delta models for stress and life satisfaction, refit on the full sample with CR2 cluster-robust standard errors at the effective cluster. Effective cluster is defined as working group for initial-cohort members and user id for late enrollers. $n \approx 122$. $^* p < 0.1$, $^** p < 0.05$, $^*** p < 0.01$. Stars use model residual degrees of freedom; small-sample CR2/Satterthwaite treatment tests are reported in \cref{tab:cr2-satterthwaite}.
\end{tablenotes}
\end{threeparttable}
\end{table}

\begin{table}[!htbp]
\centering
\begin{threeparttable}
\caption{Work and study outcomes with effective-cluster CR2 standard errors.}
\label{tab:lm-performance-cov-effective}
\small
\arxivtablefit{\footnotesize}{4.25pt}
\begin{tabular}[t]{lcccccc}
\toprule
  & Performance & Productivity & Learning & $\Delta$ Performance & $\Delta$ Productivity & $\Delta$ Learning\\
\midrule
Treatment & 0.070 & -0.134 & 0.338 & 0.038 & -0.196 & -0.041\\
 & (0.169) & (0.144) & (0.370) & (0.181) & (0.149) & (0.389)\\
Baseline outcome & 0.456*** & 0.367*** & 0.269* &  &  & \\
 & (0.154) & (0.110) & (0.137) &  &  & \\
Female & 0.078 & 0.093 & 0.330 &  &  & \\
 & (0.136) & (0.139) & (0.262) &  &  & \\
Prefer not to say & -0.042 & -0.034 & 0.639 &  &  & \\
 & (0.299) & (0.343) & (0.840) &  &  & \\
Age & 0.000 & -0.004 & -0.003 &  &  & \\
 & (0.006) & (0.007) & (0.013) &  &  & \\
Academic vs. staff & 0.375* & 0.284 & -0.316 &  &  & \\
 & (0.192) & (0.208) & (0.395) &  &  & \\
Utility & 0.050 & 0.492 & 0.626 & -0.509** & -0.459*** & 0.223\\
 & (0.586) & (0.485) & (1.186) & (0.247) & (0.172) & (0.581)\\
Sociability & 0.003 & 0.002 & 0.002 & -0.001 & -0.002 & -0.001\\
 & (0.003) & (0.003) & (0.009) & (0.002) & (0.002) & (0.008)\\
Fear & 0.015 & 0.028 & 0.190 & -0.008 & 0.035 & 0.296*\\
 & (0.088) & (0.064) & (0.125) & (0.089) & (0.067) & (0.165)\\
Institute satisfaction & -0.094*** & -0.065 & 0.048 & -0.076** & -0.029 & -0.159\\
 & (0.035) & (0.043) & (0.068) & (0.033) & (0.035) & (0.127)\\
Life satisfaction & -0.010 & 0.026 & 0.142 & -0.017 & 0.004 & 0.067\\
 & (0.048) & (0.063) & (0.124) & (0.052) & (0.047) & (0.157)\\
Stress score & -0.072 & 0.060 & -0.380 & -0.062 & 0.125 & -0.252\\
 & (0.134) & (0.143) & (0.240) & (0.162) & (0.134) & (0.273)\\
C19 recovered & -0.542 & -0.166 & -0.784 &  &  & \\
 & (0.334) & (0.395) & (1.083) &  &  & \\
Not recovered & -0.601* & -0.237 & -0.474 &  &  & \\
 & (0.320) & (0.375) & (1.073) &  &  & \\
Household size & -0.034 & -0.021 & 0.015 &  &  & \\
 & (0.033) & (0.042) & (0.122) &  &  & \\
Digital media score & -0.232 & -0.190 & -0.183 &  &  & \\
 & (0.249) & (0.217) & (0.435) &  &  & \\
Socio-economic score & 0.249 & -0.010 & -0.574 &  &  & \\
 & (0.187) & (0.167) & (0.525) &  &  & \\
Constant & 2.222** & 0.913 & 5.683* & 2.214* & 1.114 & 0.100\\
 & (1.061) & (0.952) & (3.301) & (1.163) & (0.844) & (2.425)\\
Observations & 112 & 113 & 112 & 116 & 117 & 116\\
$R^{2}$ & 0.406 & 0.350 & 0.300 & 0.075 & 0.071 & 0.059\\
Adjusted $R^{2}$ & 0.298 & 0.234 & 0.173 & 0.015 & 0.011 & -0.002\\
\bottomrule
\addlinespace
\end{tabular}
\begin{tablenotes}[flushleft]
\footnotesize
\item[] \emph{Notes.} Covariate-adjusted level and delta models for performance, productivity, and learning, refit on the full sample with CR2 cluster-robust standard errors at the effective cluster. Effective cluster is defined as working group for initial-cohort members and user id for late enrollers. $n \approx 122$. $^* p < 0.1$, $^** p < 0.05$, $^*** p < 0.01$. Stars use model residual degrees of freedom; small-sample CR2/Satterthwaite treatment tests are reported in \cref{tab:cr2-satterthwaite}.
\end{tablenotes}
\end{threeparttable}
\end{table}

\begin{table}[!htbp]
\centering
\begin{threeparttable}
  \caption{Lee sharp bounds under differential attrition.}
  \label{tab:lee-bounds} 
  \small
\setlength{\tabcolsep}{4pt}
\arxivtablefit{\footnotesize}{2pt}
\begin{tabular}{@{}lccccccr@{}} 
\toprule
 & Observed difference & Lower Bound & Upper Bound & CI Lower & CI Upper & Trim frac & Arm trimmed \\ 
\midrule
Stress & $0.179$ & $0.037$ & $0.340$ & $-0.203$ & $0.570$ & $0.105$ & control \\ 
Life Satisfaction & $0.089$ & $-0.481$ & $0.394$ & $-1.126$ & $1.067$ & $0.105$ & control \\ 
Performance & $0.120$ & $0.013$ & $0.231$ & $-0.241$ & $0.484$ & $0.105$ & control \\ 
Productivity & $0.076$ & $-0.064$ & $0.191$ & $-0.307$ & $0.437$ & $0.105$ & control \\ 
Learning & $0.175$ & $-0.213$ & $0.405$ & $-0.657$ & $0.917$ & $0.105$ & control \\ 
\bottomrule
\addlinespace
\end{tabular} 
\begin{tablenotes}[flushleft]
\footnotesize
\item[] \emph{Notes.} \citet{lee2009bounds} bounds on the effect of treatment assignment for the always-observed principal stratum, accounting for differential attrition. The maintained monotonicity assumption is that assignment to treatment can only weakly reduce endline retention, $S_i(1)\leq S_i(0)$, where $S_i(t)$ indicates whether participant $i$'s outcome would be observed under assignment $t$. Because observed retention was lower in treatment, the control outcome distribution is trimmed to equalize retention rates across arms. Columns report the 95\% Imbens--Manski confidence interval for this effect, with bound standard errors computed using $2{,}000$ arm-stratified bootstrap replications.
\end{tablenotes}
\end{threeparttable}
\end{table}

\clearpage

\subsection{Survey}\label{section:app:survey}
\begin{enumerate}
    \item \textbf{Socio-demographic attributes}
\end{enumerate}

\begin{enumerate}[label=1.\arabic*]
 
 \item Identifier \vskip0.5em
  Please write down your \redacted{IPICYT} ID number: \Qline{4cm}  \vskip0.5em
 
  \item Role \vskip0.5em
  What is your role at the university? \vskip0.5em
  \QO{} Taught student \vskip0.2em
  \QO{} Research student \vskip0.2em
  \QO{} Researcher \vskip0.2em
  \QO{} Staff (Administration, maintenance, other employees of \redacted{IPICYT}) \Qline{2.5cm} \vskip0.5em
  
 \item Affiliation [Only for students and researchers] \vskip0.5em
  Which department are you affiliated with? \vskip0.5em
  \QO{} Maths and Computer Science \vskip0.2em
  \QO{} Natural Sciences \vskip0.2em
  \QO{} Other: \Qline{5.5cm} \vskip0.5em
 
  \item Gender \vskip0.5em
  Which gender do you identify yourself with? \vskip0.5em
  \QO{} Female \vskip0.2em
  \QO{} Male \vskip0.2em
  \QO{} Other: \Qline{5.5cm} \vskip0.2em
  \QO{} Prefer not to say \vskip0.5em
  
  \item Age \vskip0.5em
  Please indicate your age in two digits: \Qline{1cm} \vskip0.5em
  
  \item Ethnicity \vskip0.5em
  Which ethnic group do you identify most with? \vskip0.5em
  \QO{} White \vskip0.2em
  \QO{} Indigenous \vskip0.2em
  \QO{} Mestizo \vskip0.2em
  \QO{} Afrolatino \vskip0.2em
  \QO{} Other: \Qline{5.5cm}\vskip0.5em

\end{enumerate}

\bigskip

\begin{enumerate}\addtocounter{enumi}{1}
    \item \textbf{Family, work, and socio-economic features}
\end{enumerate}

\begin{enumerate}[label=2.\arabic*]
    \item (se) How many dependants do you have? \vskip0.5em
    This could be children, children and partner, other relatives, etc.
    \vskip0.5em
    \QO{} Answer: \Qline{5.5cm}\vskip0.5em
    \QO{} 0--1 [1pt.]\vskip0.2em
    \QO{} 2   [2pt.]\vskip0.2em
    \QO{} 3   [3pt.]\vskip0.2em
    \QO{} 4   [4pt.]\vskip0.2em
    \QO{} 5+  [5pt.]\vskip0.5em
    \item (se) How many people live in the same household as you? \vskip0.5em
    This could be children, children and partner, siblings, other relatives, housemates etc. \vskip0.5em
    \QO{} Answer: \Qline{5.5cm}\vskip0.5em
    \QO{} 0--1 [1pt.]\vskip0.2em
    \QO{} 2 [2pt.]\vskip0.2em
    \QO{} 3 [3pt.]\vskip0.2em
    \QO{} 4  [4pt.]\vskip0.2em
    \QO{} 5+ [5pt.]\vskip0.5em
    \item (dr) How much of your time during a normal workday do you spend working on a computer? \vskip0.5em
    \QO{} 0--10\% [5pt.]\vskip0.2em
    \QO{} 11--30\% [4pt.]\vskip0.2em
    \QO{} 31--50\% [3pt.]\vskip0.2em
    \QO{} 51--70\%  [2pt.]\vskip0.2em
    \QO{} 71--100\% [1pt.]\vskip0.5em
    \item (dr) How much of your time during a normal workday do you spend on communication with colleagues? \vskip0.5em
    \QO{} 0--10\% [1pt.]\vskip0.2em
    \QO{} 11--30\% [2pt.]\vskip0.2em
    \QO{} 31--50\% [3pt.]\vskip0.2em
    \QO{} 51--70\% [4pt.]\vskip0.2em
    \QO{} 71--100\% [5pt.] \vskip0.5em
    \item (dr) How much of your time during a normal workday do you spend working in a team? \vskip0.5em
    \QO{} 0--10\% [1pt.]\vskip0.2em
    \QO{} 11--30\% [2pt.]\vskip0.2em
    \QO{} 31--50\% [3pt.]\vskip0.2em
    \QO{} 51--70\% [4pt.]\vskip0.2em
    \QO{} 71--100\% [5pt.]\vskip0.5em
    \item Socio-economic class \vskip0.5em
    (se) Look at the image of the ladder below. Imagine this ladder pictures how Mexican society is set up: 
    \begin{figure}[!htbp]
        \centering
        \begingroup
        \makeatletter
        \let\@makecaption\@maketablecaption
        \makeatother
        \caption{MacArthur ladder for perceived socio-economic status.}
        \label{fig:macarthur-ladder}
        \endgroup
        \vspace{0.45em}
        \includegraphics[width=4cm]{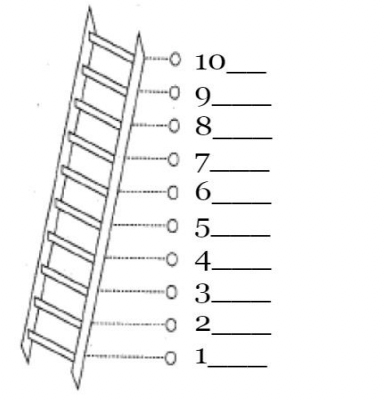}
    \end{figure}
    \begin{itemize}
        \item[·]At the top of the ladder are the people that are best off - they have the most money, the highest amount of schooling, and the jobs that bring the most respect.
        \item[·] At the bottom are the people who are the worst off - they have the least money, little or no education, no job or jobs that no one wants or respects.
    \end{itemize}
    Now think of your family, please tell us where you think your family would be on this ladder: \Qline{1.5cm} \vskip0.5em
    \QO{} 9--10 [1pt.]\vskip0.2em
    \QO{} 7--8  [2pt.]\vskip0.2em
    \QO{} 5--6  [3pt.]\vskip0.2em
    \QO{} 3--4  [4pt.]\vskip0.2em
    \QO{} 1--2  [5pt.]\vskip0.5em
    
     \item (se) Perceived socio-economic status \vskip0.5em
  People sometimes describe themselves as belonging to the working class, the middle class, or the upper or lower class. Would you describe yourself as belonging to the \vskip0.5em
 \QO{} Upper class [1pt.]\vskip0.5em
 \QO{} Upper middle class [2pt.]\vskip0.5em
 \QO{} Lower middle class [3pt.]\vskip0.5em
 \QO{} Working class [4pt.]\vskip0.5em
 \QO{} Lower class [5pt.]\vskip0.5em
 \QO{} Prefer not to answer [0pt.]\vskip0.5em
\end{enumerate}

\bigskip
  
\begin{enumerate}\addtocounter{enumi}{2}
    \item \textbf{Using digital media}
\end{enumerate}

\begin{enumerate}[label=3.\arabic*]
    \item (dr) How much of your work time do you spend using the internet?
    \vskip0.5em
    \QO{} 0--10\%   [5pt.]\vskip0.2em
    \QO{} 11--30\%  [4pt.]\vskip0.2em
    \QO{} 31--50\%  [3pt.]\vskip0.2em
    \QO{} 51--70\%  [2pt.]\vskip0.2em
    \QO{} 71--100\% [1pt.]\vskip0.5em
    \item (dr) How much of your leisure time do you spend using the internet?
    \vskip0.5em
    \QO{} 0--10\%   [5pt.]\vskip0.2em
    \QO{} 11--30\%  [4pt.]\vskip0.2em
    \QO{} 31--50\%  [3pt.]\vskip0.2em
    \QO{} 51--70\%  [2pt.]\vskip0.2em
    \QO{} 71--100\% [1pt.]\vskip0.5em
    \item (dr) How do you access the internet from home most of the time?
    \vskip0.5em
    \QO{} Through laptop + Wi-Fi             [1pt.]\vskip0.2em
    \QO{} Through laptop + mobile connection [2pt.]\vskip0.2em
    \QO{} Through phone + Wi-Fi              [3pt.]\vskip0.2em
    \QO{} Through phone + mobile connection  [4pt.]\vskip0.2em
    \QO{} N/A                                [5pt.]\vskip0.5em
\end{enumerate}

\bigskip

\begin{enumerate}\addtocounter{enumi}{3}
    \item \textbf{Psychosocial features}
\end{enumerate}
\begin{enumerate}[label=4.\arabic*] 
    \item (psy) Sociability \vskip0.5em
    Please write down the percentage of individuals (in your social circle) who would agree with the following statement about yourself: `I spend a lot of time visiting friends' \Qline{1.5cm}\vskip0.2em
    
    \QO{} 0--10\%   [5pt.]\vskip0.2em
    \QO{} 11--30\%  [4pt.]\vskip0.2em
    \QO{} 31--50\%  [3pt.]\vskip0.2em
    \QO{} 51--70\%  [2pt.]\vskip0.2em
    \QO{} 71--100\% [1pt.]\vskip0.5em
    \item Fear \vskip0.5em
    Please rate the extent to which you experience the following feelings at this moment: Fear because of the COVID-19 disease/ the SARS-CoV-2 virus. \vskip0.5em
    \QO{} Not at all \vskip0.2em
    \QO{} Not really \vskip0.2em
    \QO{} Neutral    \vskip0.2em
    \QO{} Somewhat   \vskip0.2em
    \QO{} Very much  \vskip0.5em
    \item (psy) Perceived Stress Scale \vskip0.5em
    \begin{table}[!htbp]
    \centering
    \begin{threeparttable}
    \caption{Perceived Stress Scale response options.}
    \label{tab:StressScale}
    \small
    \begin{tabular}{c|c|c|c|c}
        \textbf{1} & \textbf{2} & \textbf{3} & \textbf{4} & \textbf{5} \\
        \hline
         Never & Almost never & Sometimes & Fairly often & Very often
    \end{tabular}
    \end{threeparttable}
    \end{table}   
    Based on the scale above, where 1 indicates never experiencing that situation and 5 indicates experiencing that situation very often, please rate the following statements:\\
    \begin{itemize}
        \item[·] In the last month, how often have you felt that you were unable to control the important things in your life?\Qline{1.5cm}
        \item[·] In the last month, how often have you felt confident about your ability to handle your personal problems?\Qline{1.5cm}
        \item[·] In the last month, how often have you felt that things were going your way?\Qline{1.5cm}
        \item[·] In the last month, how often have you felt difficulties were piling up so high that you could not overcome them?\Qline{1.5cm}
    \end{itemize}
    [For scoring, the responses to the first and fourth items retain their original $1$--$5$ coding, while the positively worded second and third items are reverse-coded as $6-x$. The perceived-stress score is the mean of the available coded responses, with higher values indicating greater perceived stress.] \vskip0.5em
    \item (psy) Subjective well-being \vskip0.5em
    All things considered, how satisfied would you say you are with your life these days? Please tell me on a scale of 1 to 10, where 1 means very dissatisfied and 10 means very satisfied: \Qline{1.5cm} \vskip0.5em
    \QO{} 9--10 [1pt.]\vskip0.2em
    \QO{} 7--8  [2pt.]\vskip0.2em
    \QO{} 5--6  [3pt.]\vskip0.2em
    \QO{} 3--4  [4pt.]\vskip0.2em
    \QO{} 1--2  [5pt.]\vskip0.5em
    \item Subjective well-being \vskip0.5em
    Taking all things together on a scale of 1 to 10, how satisfied are you about \redacted{IPICYT}’s efforts to keep you safe in the institute throughout the pandemic?
    \Qline{1.5cm}\vskip0.5em
\end{enumerate}
\begin{enumerate}\addtocounter{enumi}{4}
    \item \textbf{Performance self-assessment}
\end{enumerate}
\begin{enumerate}[label=5.\arabic*] 
    \item (ppa) Self-assessment of performance \vskip0.5em
    How would you rate your overall performance for your job or degree in the past 4 weeks? \vskip0.5em
    \QO{} Poor          [5pt.]\vskip0.2em
    \QO{} Below average [4pt.]\vskip0.2em
    \QO{} Average       [3pt.]\vskip0.2em
    \QO{} Above average [2pt.]\vskip0.2em
    \QO{} High          [1pt.]\vskip0.2em
    \item Self-assessment of learning 
    \vskip0.5em
    After the COVID-19 pandemic began, the way we learn and interact with our peers drastically changed. How would you say your learning experience has been in the past 4 weeks?\vskip0.5em
  
    Please rate your learning process and experience between 1 and 10, where 1 is  poor and 10 is excellent:  \Qline{1.5cm}\vskip0.5em
    \item (ppa) Self-assessment of productivity \vskip0.5em
    How would you rate your day-to-day productivity in your work in the past 4 weeks? \vskip0.5em
    \QO{} Poor          [5pt.]\vskip0.2em
    \QO{} Below average [4pt.]\vskip0.2em
    \QO{} Average       [3pt.]\vskip0.2em
    \QO{} Above average [2pt.]\vskip0.2em
    \QO{} High          [1pt.]\vskip0.2em
    
    \item (ppa) Self-assessment of achievement (supervisor goals) \vskip0.5em
    Considering again the work for your job or degree during the past 4 weeks, please select the statement that fits your situation best. \vskip0.5em
    \QO{} I have struggled to achieve the goals set by my supervisor/employer/course teachers          [5pt.]\vskip0.2em
    \QO{} I have managed to achieve some of the goals set by my supervisor/employer/course teachers          [4pt.]\vskip0.2em
    \QO{} I have achieved many of the goals set by my supervisor/employer/course teachers          [3pt.]\vskip0.2em
    \QO{} I have achieved most of the goals set by my supervisor/employer/course teachers          [2pt.]\vskip0.2em
    \QO{} I have achieved all or exceeded the goals set by my supervisor/employer/course teachers          [1pt.]\vskip0.2em
    
    \item (ppa) Self-assessment of achievement (own goals) \vskip0.5em
    Considering again the work for your job or degree during the past 4 weeks, please select the statement that fits your situation best. \vskip0.5em
    \QO{} I have struggled to achieve the goals I set for myself [5pt.]\vskip0.2em
    \QO{} I have managed to achieve some of the goals I set for myself [4pt.]\vskip0.2em
    \QO{} I have achieved many of the goals I set for myself [3pt.]\vskip0.2em
    \QO{} I have achieved most of the goals I set for myself [2pt.]\vskip0.2em
    \QO{} I have achieved all or exceeded the goals I set for myself [1pt.]\vskip0.2em
\end{enumerate}

\subsection{Consent form}
\label{section:app:consent_form}
\paragraph{Note on the reproduced consent text.}
The consent text below reflects the study protocol and schedule as originally planned: the trial was expected to take place in August 2022, with control participants remaining subject to the institute's remote-working policy. Following the federally mandated reopening of educational institutions, the trial was implemented in September 2022, and the control group received unrestricted on-site access, as described in \cref{section:design}. The consent text is reproduced here in its original form. [Participants were informed of this change before the trial began, and the modification received the required institutional and ethical approval.]

You are invited to take part in a research project conducted by researchers from \redacted{the University of Oxford, Harvard University, and the United Nations University} in conjunction with \redacted{IPICYT}. This project is funded by \redacted{IPICYT}. In accordance with international standards in the practice of randomized studies, this project has received ethical approval from the Research Ethics Committee at \redacted{IPICYT} and \redacted{The Central University Research Ethics Committee at Oxford University, ethics reference R81719/RE001.}

We ask that you read this form carefully prior to deciding to participate in the study. If you decide you do not want to participate, you may leave at any time without providing a reason and without penalty.

\noindent \textbf{Purpose:} The purpose of this study is to understand how the implementation of an algorithm-based safe-education protocol affects the well-being and productivity of students and staff during a pandemic.

\noindent \textbf{What happens during the study:} This study requires you to follow one of two protocols. 

If you are selected to be part of the treatment group, you will participate in COVID-19 pooled testing. Throughout the course of the study, you may receive emails inviting you to submit a saliva sample, which will be pooled with other samples and tested at the \redacted{LANBAMA} laboratory at \redacted{IPICYT}. If your pooled test is negative, no SARS-CoV-2 viral material was detected in the pooled sample, and everyone in your pool is permitted to enter the institute for 48 hours.\footnote{A negative result reduces but does not eliminate the possibility that a pool member is infected.} If your pooled test is positive, the result indicates that at least one sample in the pool may contain SARS-CoV-2 viral material. You and the other individuals in your pool are then not permitted to enter the institute until you are selected for re-testing and the next test result is negative. At no point are you obliged to submit a saliva sample, or to enter the building.

If you are selected to be part of the control group, you will be asked to follow the same remote working policy that is currently in place at \redacted{IPICYT}. If you would like to access the institute, you must contact the head of your department for permission.

We also ask all participants to respond to a short survey at the beginning and end of the trial, approximately one month apart. The survey includes sociodemographic and psychological questions. You are not required to answer any questions that you may find uncomfortable. Furthermore, for the purpose of COVID-19 testing, you may be asked to provide a saliva sample to the technicians at \redacted{LANBAMA} if you are selected for pooled testing. Each sample will be used on the day of receipt and destroyed after being processed for a qPCR test; samples will not be stored. You will be informed of the result of every pooled test containing your sample.

\noindent \textbf{Participation:} The trial is expected to run for a month, throughout August 2022, during which participants in the treatment group will receive free COVID-19 testing. Participants are asked to fill in a survey at the beginning and end of the study. In addition, participants in the treatment group are able to indicate their preference for which days they wish to be tested. Throughout the course of the month, the principal investigators will link health data (i.e. COVID-19 test results) to survey data (collected at the beginning and end of the trial). However, at the end of the trial all gathered data will be anonymized. If you wish to withdraw consent on the use of your data at any point during the study, please contact \redacted{\href{mailto:mail@c-sef.com}{mail@c-sef.com}}.
You always have the option of stopping your participation in the study and you may leave at any time during the study (4 weeks from the start of the trial) without providing a reason and without penalty. If you decide to leave, the data you have provided up to this point will be anonymized immediately and deleted after attrition analysis.

\noindent \textbf{Potential risks:}  If you choose not to participate in the study, or you participate and are selected into the control group, you will not be exposed to any additional risk. If you choose to participate and are selected into the treatment group, there is a risk that you will be infected if you are permitted to enter the institute and decide to do so. This risk is small, as all individuals must test negative in order to enter the institute. In particular, \redacted{the C-SEF protocols} are much safer than reopening without monitoring for infections. While the probability of infection can be minimized and contained, it is not guaranteed to be zero. There is always a small risk of infection when participating in social activities, and COVID-19 can have minor or major consequences, including fever, cough, loss of taste or smell, respiratory problems, and, in some cases, death.

Your survey responses are strictly confidential and will only be accessible to the researchers. Below, we describe the steps we are taking to protect your privacy. In addition, your decision on whether to participate will not adversely affect your relationship with \redacted{IPICYT} or any other institution to which the researchers are affiliated.

\noindent \textbf{Benefits:} Participating in this study means that you are aiding further development of science. Additionally, a successful trial would allow \redacted{IPICYT} to reformulate the institutional policy regarding work and study during the current and future waves of the pandemic into one that gives you more social interactions and flexibility with a minimized risk of contagion.

\noindent \textbf{Data protection and privacy:}
The information collected during the study will be kept private. In accordance with the \redacted{UK General Data Protection Regulation and Data Protection Act of 2018, the University of Oxford} is the data controller with respect to your personal data, and as such will determine how your personal data is used in the research. The University will process your personal data for the purpose of the research outlined above. Research is a task that is performed in the public interest. Further information about your rights with respect to your personal data is available at \redacted{\href{https://compliance.admin.ox.ac.uk/individual-rights}{https://compliance.admin.ox.ac.uk/individual-rights}}.

Responsible members of \redacted{the University of Oxford}
and \redacted{IPICYT} may be given access to data for monitoring and/or auditing of the study to ensure that we comply with the guidelines or as otherwise required by law. Moreover, in accordance with the signed Memorandum of Understanding, the \redacted{Potosinian Institute of Scientific Research and Technology (IPICYT)} will store and anonymize the original data in a secure server. During the trial, no one other than the head of the \redacted{IPICYT} Supercomputing Centre and responsible members of \redacted{the University of Oxford and the United Nations University (Maastricht)} will have access to any records of this trial. The data will be stored in electronic form, encrypted and password protected. At the conclusion of the trial, all data will be anonymized, and none of the records will identify you. A copy of the anonymized data will be provided to the principal investigators of the trial. The data that we collect from you may be transferred to, and stored or processed at a destination outside Mexico. Archived/stored data, once anonymized, is available for research purposes upon request (primarily for peer-review replication processes). By submitting your personal data, you agree to this transfer, storing, or processing. After completion of the study, you cannot withdraw your personal information. Your individual privacy will be maintained in all publications or presentations resulting from this study. No information about you provided by you during this research will be disclosed to others without your written permission, except:

\begin{itemize}
    \item[-] if necessary to protect your rights or welfare (for example, if you are injured and need emergency care); or

    \item[-] if required by law.
\end{itemize}

\noindent \textbf{Additional information:} If you are interested in receiving additional information about the results of the study, please contact the study authors.

\noindent \textbf{Concerns:} If you have any questions or concerns about any aspect of this project, you can contact the study authors at \redacted{\href{mailto:mail@c-sef.com}{mail@c-sef.com}}, who will do their best to answer your query. The researchers should acknowledge receipt of your concern within 10 working days and give you an indication of how they intend to address it. If you fail to receive a response, are dissatisfied with the response you receive, or desire to report an aspect of how the study is being conducted, please contact the relevant chair of the Research Ethics Committee at the
\redacted{University of Oxford}:

Chair, Social Sciences \& Humanities Inter-Divisional Research Ethics Committee;\\
Email: \redacted{ethics@socsci.ox.ac.uk}\\
Address: \redacted{Research Services, University of Oxford, Wellington Square, Oxford OX1 2JD} 

The Chair will seek to resolve the matter in a reasonably expeditious manner. 
\vskip2em

\newpage

\begin{center}
\textbf{\Large Consent form}
\end{center}\vskip0.8em

\begin{center}
\textbf{Please confirm the following by marking each of the boxes next to the statements.}
\end{center}\vskip1em

\begin{flushright}
\textit{Please Mark Each Box}
\end{flushright}

\begin{multicols}{2}
\begin{itemize}
    \item I confirm that I have read and understand the information for the above study and have had the opportunity to properly consider the information provided. \hskip0.4cm \QO{}
    \item I understand that my participation is voluntary and that I am free to withdraw at any time, without giving any reason and without any adverse consequences. \hskip0.4cm \QO{}
    \item I understand the risks associated with participating in this study as explained in the information sheet.\hskip0.4cm \QO{}
    \item I understand that a saliva sample will be taken during the study and that this sample will be tested for COVID-19. I understand that the sample will be destroyed after completion of this test or if I withdraw my consent.\hskip0.4cm \QO{}
    \item I consider these samples a gift to \redacted{University of Oxford}
    and the \redacted{LANBAMA} laboratory and I understand I will not gain any direct personal benefit from this.\hskip0.4cm \QO{}
    \item I understand that research data collected during the study may be looked at by designated individuals from \redacted{the University of Oxford} and \redacted{IPICYT} where it is relevant to my taking part in this study. I give permission for these individuals to access my data. I give permission for anonymized data to be made publicly available at the end of the research. \hskip0.4cm \QO{}
    \item I understand that this project has been reviewed by, and received ethics clearance through, the Research Ethics Committee at \redacted{IPICYT} and \redacted{the Central University Research Ethics Committee at Oxford University.}
    \hskip0.4cm \QO{}
    \item I understand who will have access to the personal data provided, how the data will be stored, and what will happen to the data at the end of the project. \hskip0.4cm \QO{}
    \item I understand how this research will be written up and published. \hskip0.4cm \QO{}
    \item I understand how to raise a concern or make a complaint. \hskip0.4cm \QO{}
    \item I agree to take part in the study. \hskip0.4cm \QO{}

\end{itemize}
\end{multicols}
\vskip1em

By selecting “Yes, I agree to participate” below, you signify that you have read and understood the information above and agree to the processing of the data you provide during the study as described above.\vskip1em
\QO{} Yes, I agree to participate
\bigskip

\QO{} No, I do not agree to participate

\section{Additional Elements: Preregistration and Open Source Code}
\label{section:preregistration}

You can find the preregistered experiment on the American Economic Association RCT Registry using the following DOI: \redacted{\url{https://doi.org/10.1257/rct.9466-2.0}}
    \bigskip

\noindent If you would like to install and run the web application used in the experiment, the open-source code is available on GitHub: \redacted{\url{https://github.com/edwinlock/csef}}

\end{document}